\documentclass[twocolumn,tighten]{aastex63}

\usepackage{graphicx}
\usepackage{amsmath}
\usepackage{comment}
\usepackage{array,multirow,color,tablefootnote}
\usepackage{xcolor}
\usepackage[maxfloats=256]{morefloats}
\usepackage{subfigure}

\newcommand{\tess}{TESS}

\shorttitle{Year One TESS phase curves}
\shortauthors{Wong et al.}

\begin{document}
\title{Systematic Phase Curve Study of Known Transiting Systems from Year One of the TESS Mission} 
\correspondingauthor{Ian Wong}
\email{iwong@mit.edu}

\author[0000-0001-9665-8429]{Ian Wong}
\altaffiliation{51 Pegasi b Fellow}
\affil{Department of Earth, Atmospheric and Planetary Sciences, Massachusetts Institute of Technology,
Cambridge, MA 02139, USA}

\author[0000-0002-1836-3120]{Avi Shporer}
\affil{Department of Physics and Kavli Institute for Astrophysics and Space Research, Massachusetts Institute of Technology, Cambridge, MA 02139, USA}

\author[0000-0002-6939-9211]{Tansu Daylan}
\altaffiliation{Kavli Fellow}
\affiliation{Department of Physics and Kavli Institute for Astrophysics and Space Research, Massachusetts Institute of Technology, Cambridge, MA 02139, USA}

\author[0000-0001-5578-1498]{Bj{\" o}rn Benneke}
\affiliation{Department of Physics and Institute for Research on Exoplanets, Universit{\' e} de Montr{\' e}al, Montr{\' e}al, QC, Canada}

\author{Tara Fetherolf}
\affiliation{Department of Physics and Astronomy, University of California, Riverside, CA 92521, USA}

\author[0000-0002-7084-0529]{Stephen R. Kane}
\affiliation{Department of Earth and Planetary Sciences, University of California, Riverside, CA 92521, USA}

\author[0000-0003-2058-6662]{George R. Ricker}
\affiliation{Department of Physics and Kavli Institute for Astrophysics and Space Research, Massachusetts Institute of Technology, Cambridge, MA 02139, USA}

\author[0000-0001-6763-6562]{Roland Vanderspek}
\affiliation{Department of Physics and Kavli Institute for Astrophysics and Space Research, Massachusetts Institute of Technology, Cambridge, MA 02139, USA}

\author[0000-0001-9911-7388]{David W. Latham}
\affiliation{Center for Astrophysics ${\rm \mid}$ Harvard {\rm \&} Smithsonian, 60 Garden Street, Cambridge, MA 02138, USA}


\author[0000-0002-4265-047X]{Joshua N. Winn}
\affiliation{Department of Astrophysical Sciences, Princeton University, Princeton, NJ 08544, USA}

\author[0000-0002-4715-9460]{Jon M. Jenkins}
\affiliation{NASA Ames Research Center, Moffett Field, CA 94035, USA}

\author[0000-0003-0442-4284]{Patricia T. Boyd}
\affiliation{Astrophysics Science Division, NASA Goddard Space Flight Center, Greenbelt, MD 20771, USA}

\author[0000-0002-5322-2315]{Ana Glidden}
\affiliation{Department of Earth, Atmospheric and Planetary Sciences, Massachusetts Institute of Technology, Cambridge, MA 02139, USA}
\affiliation{Department of Physics and Kavli Institute for Astrophysics and Space Research, Massachusetts Institute of Technology, Cambridge, MA 02139, USA}

\author[0000-0003-1748-5975]{Robert F. Goeke}
\affiliation{Department of Physics and Kavli Institute for Astrophysics and Space Research, Massachusetts Institute of Technology, Cambridge, MA 02139, USA}

\author[0000-0001-5401-8079]{Lizhou Sha}
\affiliation{Department of Physics and Kavli Institute for Astrophysics and Space Research, Massachusetts Institute of Technology, Cambridge, MA 02139, USA}




\author[0000-0002-8219-9505]{Eric B. Ting}
\affiliation{NASA Ames Research Center, Moffett Field, CA 94035, USA}

\author[0000-0003-4755-584X]{Daniel Yahalomi}
\affiliation{Center for Astrophysics ${\rm \mid}$ Harvard {\rm \&} Smithsonian, 60 Garden Street, Cambridge, MA 02138, USA}

\begin{abstract}
We present a systematic phase curve analysis of known transiting systems observed by the Transiting Exoplanet Survey Satellite (TESS) during year one of the primary mission. Using theoretical predictions for the amplitude of the planetary longitudinal atmospheric brightness modulation, stellar ellipsoidal distortion and Doppler boosting, as well as brightness considerations to select targets with likely detectable signals, we applied a uniform data processing and light-curve modeling framework to fit the full-orbit phase curves of 22 transiting systems with planet-mass or brown dwarf companions, including previously published systems. Statistically significant secondary eclipse depths and/or atmospheric brightness modulation amplitudes were measured for HIP 65A, WASP-18, WASP-19, WASP-72, WASP-100, WASP-111, WASP-121, and WASP-122/KELT-14. For WASP-100b, we found marginal evidence that the brightest region of the atmosphere is shifted eastward away from the substellar point. We detected significant ellipsoidal distortion signals in the light curves of HIP 65A, TOI-503, WASP-18, and WASP-30, with HIP 65A, TOI-503 and WASP-18 also exhibiting Doppler boosting. The measured amplitudes of these signals agree with the predictions of theoretical models. Combining the optical secondary eclipse depths with previously published Spitzer 3.6 and 4.5 $\mu$m measurements, we derived dayside brightness temperatures and visible-light geometric albedos for a subset of the analyzed systems. We also calculated updated transit ephemerides combining the transit timings from the \tess\ light curves with previous literature values.
\end{abstract}

\section{Introduction}\label{sec:intro}

Since the start of science observations on 2018 July 25, the Transiting Exoplanet Survey Satellite (TESS) has been observing most of the sky in search of new transiting exoplanets around bright, nearby stars. In addition to the thousands of planet candidates and several dozen confirmed planets that the mission has detected to date, hundreds of previously discovered exoplanet systems have been observed, providing nearly continuous broadband visible photometry spanning at least one month for every target. This treasury of light curves has proven to be an invaluable resource for time-domain astronomy of known exoplanet systems.

\tess\ has been especially fruitful for the study of orbital phase curves. Long-baseline photometric monitoring of transiting systems can reveal the secondary eclipse, when the orbiting companion is occulted by the host star, as well as photometric variations phased to the orbital period. Short-period systems are expected to be tidally locked \citep[e.g.,][]{mazeh2008}, with fixed dayside and nightside hemispheres that may differ greatly in temperature. The changing viewing phase of the orbiting companion results in a periodic modulation of the observed atmospheric brightness with maxima and minima near mid-eclipse (superior conjunction) and mid-transit (inferior conjunction), respectively (see \citealt{parmentier2017} for a review of this phase curve component).

The depth of the secondary eclipse corresponds to the relative brightness of the companion's dayside hemisphere. At visible wavelengths, the eclipse depth contains contributions from both thermal emission and reflected starlight. The addition of secondary eclipse measurements at infrared wavelengths breaks the degeneracy between reflected light and thermal emission, yielding direct constraints on the optical geometric albedo, an important quantity for inferring the presence of clouds and hazes on the dayside hemisphere. When combined with measurements of the amplitude of the atmospheric brightness modulation, one can deduce the dayside and nightside temperatures. Meanwhile, a detected phase shift in the atmospheric brightness modulation indicates an offset in the region of maximum brightness relative to the substellar point, which may be caused by inhomogeneous clouds \citep[e.g.,][]{shporer2015} or an eastward-shifted dayside hotspot due to super-rotating equatorial winds \citep[e.g.,][]{perna2012,perezbecker}.

For massive orbiting companions, gravitational interactions can cause variations in the host star's brightness that are detectable in precise visible-light phase curves. First, the Doppler boosting signal is produced when the radial velocity (RV) modulation of the star induced by the gravitational pull of the orbiting companion leads to periodic blue- and red-shifting of the stellar spectrum as well as modulations in photon emission rate in the observer's direction \citep[e.g.,][]{shakura1987,loeb2003,zucker2007}. Second, the companion's gravity raises tidal bulges on the star, with the long dimension aligned with the star--companion axis \citep[e.g.,][]{morris1985,morris1993,pfahl2008}. This produces a modulation in the star's sky-projected area and apparent flux that comes to maximum at the quadratures, resulting in a phase curve signal with a leading term at the first harmonic of the orbital phase. A detailed overview of the astrophysics of visible-light phase curves is provided in \citet{shporer2017}. 

Analyses of individual high signal-to-noise phase curves from the \tess\ mission have been published for several systems, including WASP-18 \citep{shporer2019}, WASP-19 \citep{wong2019wasp19}, WASP-100 \citep{jansen2020}, WASP-121 \citep{bourrier2019,daylan2019}, and KELT-9 \citep{wong2019kelt9}. These studies have reported robust detections of phase curve signals attributed to all of the aforementioned processes. Building upon these previous studies, as well as the legacy of analogous works from the CoRoT and Kepler eras \citep[e.g.,][]{mazeh2010,esteves2013,esteves2015,angerhausen2015}, we seek to expand the search for phase curve signals in \tess\ photometry to cover all confirmed star--planet systems. By extending our analysis to systems with lower signal-to-noise datasets, we will maximize the science yield of the \tess\ mission in the realm of phase curves.

In this paper, we present a systematic phase curve study of known transiting systems observed during the first year of the \tess\ mission. We consider both planetary-mass companions and brown dwarfs, and include targets that were discovered prior to the \tess\ mission as well as newly confirmed systems discovered by \tess. Special attention is given to utilizing a uniform data processing and phase curve modeling framework and applying a consistent treatment of instrumental systematics across all datasets, analogous to the techniques used in our previously published studies.

The paper is organized as follows. Section \ref{sec:ana} describes the \tess\ observations and data processing techniques used to produce the light curves for our fits. The target selection criteria for filtering out systems with phase curve signals that are likely to be undetectable are detailed in Section \ref{sec:selec}. The results of the phase curve analyses are presented in Section \ref{sec:res}. In Section \ref{sec:dis}, we discuss these results in the context of theoretical predictions of the gravitational phase curve amplitudes; we also combine previously published Spitzer secondary eclipse measurements with our \tess-band eclipse depths to calculate the dayside brightness temperatures and optical geometric albedos for a subset of the analyzed systems. We summarize the main results of this work in Section \ref{sec:con}.

\section{Light Curves and Data Analysis}\label{sec:ana}

\subsection{\tess\ Photometry}\label{subsec:obs}

During year one of the \tess\ primary mission (2018 July 25 to 2019 July 18), the spacecraft observed most of the southern ecliptic hemisphere. \tess\ has four identical wide-field cameras, each with an effective aperture diameter of 10 cm. The combined field of view of $24^{\circ}\times96^{\circ}$ is oriented with the long axis along a line of constant ecliptic longitude. In latitude, the field of view begins at $-6^{\circ}$ and reaches $12^{\circ}$ past the southern ecliptic pole. The Southern Sky was divided into 13 sectors; each sector was observed for 27.4 days, corresponding to two geocentric spacecraft orbits, with an interruption in data collection between orbits during perigee for data downlink.

Each of the four cameras consists of four CCDs with a total on-sky area of $4096\times4096$ pixels. \tess\ utilizes a red-optical bandpass spanning 600--1000 nm, centered on the Cousins \textit{I} band ($\lambda=786.5$ nm). The entire array is read out at 2 s intervals, with individual frames combined on board into $11\times11$ pixel stamps at 2 minute cadence and full-frame images at 30 minute cadence prior to downlink. The targets for which 2-minute data are compiled have been selected from the \tess\ Input Catalog \citep{stassun2018} and include almost all of the bright, known transiting exoplanet systems within the \tess\ sectors.

The downlinked pixel stamps are passed through the Science Processing Operations Center (SPOC) pipeline \citep{jenkins2016}. After the optimal photometric extraction apertures are determined, two types of light curves are produced: simple aperture photometry (SAP) and pre-search data conditioning (PDC) light curves. To construct the PDC light curves, the raw aperture photometry is detrended for common-mode instrumental systematics using co-trending basis vectors empirically calculated from other sources on the corresponding detector \citep{smith2012,stumpe2014}. The PDC light curves are also corrected for flux contamination from nearby stars. Both the SAP and PDC light curves are publicly released and hosted on the Mikulski Archive for Space Telescopes (MAST). For most of the systems studied in this paper, we found the PDC light curves to be cleaner than the SAP light curves, showing significantly fewer short-timescale flux variations and reduced scatter. The exceptions are WASP-19 and WASP-121, for which the systematics correction process led to noisier photometry and increased red noise; for these two systems, we utilized the SAP light curves instead. Analyses of the same target using PDC and SAP light curves yielded statistically consistent parameter values in all cases.

Momentum dumps were scheduled two to four times per spacecraft orbit in order to reset the onboard reaction wheels. These events often lead to discontinuities in the flux time series, as well as occasional flux ramps before or after, lasting up to one day. To adequately model the residual instrumental systematics in our light-curve fits, we followed previous work \citep{wong2019wasp19,wong2019kelt9} and split each orbit's time series into discrete segments, separated by the momentum dumps. Each of these segments is assigned its own systematics model in the joint fits (see Sections \ref{subsec:model} and \ref{subsec:fit} for details). 

In cases where discernible flux ramps are present, we chose to trim these short-timescale features from the time series, because retaining them would necessitate significantly higher-order systematics detrending functions and may lead to biases in the fitted astrophysical phase curve amplitudes. All of the trimmed time intervals were set to multiples of 0.25 day. 

Periods of abnormal spacecraft operation and significant scattered light on the detector are automatically flagged by the SPOC pipeline, and we removed all flagged points from the time series. Prior to fitting, we applied a 16-point wide moving median filter to the light curve (excluding regions near primary transits) and trimmed $>$3$\sigma$ outliers. The flagged point and outlier trimming process typically removed less than $5\%$ of the data points.

A full list of the data segments considered in this work is provided in Appendix A. We did not include any data segment that spans less than one day (much shorter than the orbital period of most of our targets). In a handful of cases, severe systematic artifacts (e.g., sharp, short-term flux variations and periods of significantly increased scatter) were present in individual segments; because such features are not readily removed using typical systematics detrending methods, we discarded these segments in their entirety prior to fitting.

\subsection{Full Phase Curve Model}\label{subsec:model}
The light-curve modeling in this work is largely identical to the methods used in previous papers \citep{shporer2019,wong2019wasp19,wong2019kelt9}. The core computational framework for our analysis is the ExoTEP pipeline \citep[e.g.,][]{benneke2019,wong2019} --- a modular, Python-based tool for data extraction and light-curve fitting.

The transit and secondary eclipse light curves --- $\lambda_{t}(t)$ and $\lambda_{e}(t)$, respectively --- are modeled using \texttt{batman} \citep{kreidberg2015}. The out-of-eclipse phase curve variation is appropriately divided into terms describing variations in the orbiting companion's flux $\psi_{p}(t)$ and those attributed to the host star's flux $\psi_{*}(t)$. Defining the orbital phase as $\phi\equiv 2\pi(t-T_0)/P$, where $T_0$ is the mid-transit time, and $P$ is the orbital period, the component photometric signals are expressed as
\begin{align}
\label{planet}\psi_{p}(t) &= \bar{f_{p}} - A_{\mathrm{atm}} \cos(\phi+\delta),\\
\label{star}\psi_{*}(t) &= 1-A_{\mathrm{ellip}}\cos(2\phi)+A_{\mathrm{Dopp}}\sin(\phi).
\end{align}
Here, $\bar{f_{p}}$ is the average relative brightness of the orbiting companion, and $A_{\mathrm{atm}}$ and $\delta$ are the semi-amplitude and phase shift of the object's atmospheric brightness modulation, respectively. The parameter $\delta$ is defined such that a positive value denotes an eastward shift in the region of maximum brightness. $A_{\mathrm{ellip}}$ and $A_{\mathrm{Dopp}}$ are the semi-amplitudes of the ellipsoidal distortion and Doppler beaming phase curve modulations, respectively. The sign convention in Equations \eqref{planet} and \eqref{star} is chosen so as to yield positive amplitudes, assuming the expected behavior for the associated physical processes \citep[e.g.,][]{shporer2017}. In the case where both $\bar{f_{p}}$ and $A_{\mathrm{atm}}$ are robustly detected, the secondary eclipse depth and nightside flux are, by definition, $D_{d}=\bar{f_{p}}-A_{\mathrm{atm}}\cos(\pi+\delta)$ and $D_{n}=\bar{f_{p}}-A_{\mathrm{atm}}\cos(\delta)$, respectively.

The astrophysical phase curve model described in Equations \eqref{planet} and \eqref{star} includes several simplifying assumptions, which we describe and validate below. First, the atmospheric brightness modulation measured in visible light contains contributions from both the thermal emission from the orbiting companion and any reflected starlight off the dayside hemisphere. Given the relatively low geometric albedos and high dayside temperatures of the targets with detectable visible-light phase curve signals (see Section \ref{subsec:temp}), the thermal emission component is expected to be dominant, even in the \tess\ bandpass. While the thermal emission component is well described by a cosine term (see, for example, the numerous Spitzer phase curve analyses in the literature; e.g., \citealt{wong2016}, \citealt{beatty2019}), the form of the reflection component can vary based on the assumed scattering properties of the atmosphere.

A common prescription used for visible-light phase curves is Lambertian scattering \citep[e.g.,][]{esteves2015}. When assuming Lambertian scattering, the reflection component deviates from the simple cosine variation characteristic of geometric scattering; as described in \citet{faigler2015}, this modulation can be represented as a leading-order term at the cosine of the orbital period and a second-order term at the first harmonic of the cosine (analogous to the ellipsoidal distortion component), with an amplitude less than $20\%$ of the leading-order term. Note that the relative amplitude of the second-order term is with respect to the \textit{reflection-only} component of the overall atmospheric brightness modulation, so in the emission-dominated overall atmospheric brightness modulation, the predicted relative contribution of this second-order reflection signal is significantly less than $20\%$.

In our analysis of targets for which significant atmospheric brightness modulation is detected (and no ellipsoidal distortion), we experimented with including the first harmonic of the cosine in the phase curve model. No significant amplitudes were measured, indicating that the single cosine model in Equation \eqref{planet} is generally sufficient in describing the atmospheric brightness modulation in these light curves. This of course does not mean that the reflection component is not consistent with Lambertian scattering; instead, the signal-to-noise of the photometry does not allow us to discern between geometric scattering and Lambertian scattering. With the added time baseline provided by the extended mission, we expect our analysis to become sensitive to small discrepancies in the reflection component for some of the brightest systems (e.g., WASP-18 and WASP-121).

The second simplification in the phase curve model relates to our treatment of ellipsoidal distortion. The photometric signal that arises from the tidal distortion of the host star is formally expressed as a series of cosine terms \citep[e.g.,][]{morris1985,morris1993}, with the leading-order term at the first harmonic of the orbital period, as presented in Equation \eqref{star}. The second-order term varies at the second harmonic of the orbital period (i.e., $\cos(3\phi)$) and has a typical amplitude that is more than an order-of-magnitude smaller than that of the leading-order term. The higher-order terms of the ellipsoidal distortion modulation are detectable in high signal-to-noise photometry, such as stellar binary light curves from the Kepler mission \citep[e.g.,][]{wong2019koi964}. In the case of the \tess\ targets, however, the predicted amplitude of the second-order ellipsoidal distortion modulation is well within the uncertainties. Nevertheless, for the targets in our current analysis with detected ellipsoidal distortion, we experimented with fitting for the amplitudes of the higher-order cosine terms. In all cases, no significant amplitudes were detected, and for the results presented in this paper, we fit the light curves using the prescription in Equation \eqref{star}, which includes only the leading-order term of the ellipsoidal distortion modulation.

Following our previous work on \tess\ phase curves, we renormalized the full astrophysical phase curve model such that the average combined star and companion brightness is unity:
\begin{equation}
    \psi(t) = \frac{\psi_{*}(t)\lambda_t(t)+\lambda_e(t)\psi_{p}(t)}{1+\bar{f_p}}.
\end{equation}

All remaining temporal variations in the light-curve segments (e.g., from residual uncorrected instrumental systematics or stellar variability) are described by generalized polynomial functions in time,
\begin{equation}\label{systematics}
    S_N^{\lbrace i\rbrace}(t) = \sum\limits_{j=0}^{N}c_j^{\lbrace i\rbrace}(t-t_0)^j,
\end{equation}
where $t_0$ is the time of the first data point in segment $i$, and $N$ is the order of the detrending polynomial. The full phase curve and systematics model is
\begin{equation}\label{fita}
    f(t) = S_N^{\lbrace i\rbrace}(t)|_{i=1-6}\times\psi(t).
\end{equation}

The optimal polynomial order for each segment was determined by carrying out full phase curve fits to the individual segment light curves. When selecting the orders, we considered both the Bayesian information criterion $\mathrm{BIC}\equiv \gamma\log m -2 \log L$ and the Akaike information criterion $\mathrm{AIC}\equiv 2\gamma -2 \log L$, where $\gamma$ is the number of free parameters in the fit, $m$ is the number of data points in the segment, and $L$ is the maximum log-likelihood. For the majority of data segments, minimization of the BIC and AIC yielded the same optimal polynomial order; for cases in which the AIC preferred a higher order than the BIC, we conservatively chose the order that minimized the BIC in order to reduce the number of free systematics parameters in the fit. Using the higher orders did not incur any significant changes to the astrophysical parameter values from the overall joint fits.
The optimal polynomial orders for all data segments considered in our analysis are listed in Appendix A. Appendix B contains a compilation of the raw and systematics-corrected light curves. The systematics-corrected light curves show no significant anomalous time-correlated signals, demonstrating the efficacy of our systematics modeling.

\subsection{Model Fitting}\label{subsec:fit}

In the joint fits, we allowed the transit depth (parameterized by the planet--star radius ratio $R_{p}/R_{*}$), transit ephemeris (mid-transit time $T_{0}$ and orbital period $P$), and transit shape parameters (impact parameter $b$ and scaled orbital semimajor axis $a/R_{*}$) to vary freely. In particular, we did not apply priors on $T_{0}$ and $P$ from previous measurements, and as such, we obtained independent transit timing measurements for all targets, which we use in Section \ref{subsec:ephem} to derive updated transit ephemerides. Likewise, in many cases, the constraints we derived for $b$ and $a/R_{*}$ are comparable to or more precise than available literature values, and we did not place priors on these parameters in the fits. For all of the systems analyzed in this paper, the available data are consistent with a circular orbit, so we fixed the orbital eccentricity $e$ to zero during our fitting procedure. In addition to the astrophysical parameters, we simultaneously fit for the systematics model coefficients $\left\lbrace c_{j}^{\lbrace i\rbrace}\right\rbrace$ (see Section \ref{wasp100} for the exception to this rule: WASP-100).

We used a quadratic limb-darkening model and fit for linear combinations of the limb-darkening coefficients $\gamma_{1}$ and $\gamma_{2}$, following \citet{holman2006}, in order to break the degeneracy that arises when fitting for the quadratic limb-darkening coefficients $u_{1}$ and $u_{2}$ directly: $\gamma_{1}\equiv 2u_{1}+u_{2}$ and $\gamma_{2}\equiv u_{1}-2u_{2}$. To achieve maximally conservative constraints on the transit depth and other system parameters, given the uncertainties in theoretical limb-darkening models, we allowed $\gamma_{1}$ and $\gamma_{2}$ to vary freely, pursuant to physically motivated constraints: a monotonically increasing, concave-down brightness profile from limb to center of disk. An exception to this rule was HIP 65A, which has a grazing transit ($b>1$): because the stellar limb-darkening profile is largely unconstrained in such instances, we instead placed Gaussian priors on the limb-darkening coefficients derived from the tabulated values in \citet{claret2018}.

For each target, we ran a suite of joint fits with various subsets of the phase curve parameters $\bar{f_{p}}$, $A_{\mathrm{atm}}$, $\delta$, $A_{\mathrm{ellip}}$, and $A_{\mathrm{Dopp}}$. When selecting the final set of results to present in this paper, we considered both the BIC and AIC, generally choosing the fit that minimizes both the AIC and the BIC. By incurring a penalty for each additional free parameter included in the phase curve model, scaled by the length of the time series, this process robustly determines which phase curve components show statistically significant signals in the data. 

Given the inherent degeneracy between Doppler boosting and a phase shift in the atmospheric brightness modulation, the parameters $A_{\mathrm{Dopp}}$ and $\delta$ were never included together as unconstrained fit parameters. For the two systems (HIP 65A and WASP-18) where both Doppler boosting and atmospheric brightness modulation were detected, even marginally, we chose to fit for $A_{\mathrm{Dopp}}$ and then place constraints on the phase shift in the atmospheric brightness component by extrapolating from the predicted Doppler boosting amplitude (see Section \ref{subsec:grav}). 

For all targets, we utilized the AIC and BIC to determine whether any of the gravitationally induced phase curve signals could be robustly detected from the photometry. In cases where neither ellipsoidal distortion nor Doppler boosting could be retrieved, we applied Gaussian priors on both of the corresponding amplitudes based on the theoretical predictions and literature values for the star--companion mass ratio $M_{p}/M_{*}$ (see Section \ref{sec:selec}). To check for any remaining periodic signals in the photometry (e.g., from possible phase shifts in the ellipsoidal distortion signal or unexpected higher harmonics of the atmospheric brightness modulation), we examined the Lomb--Scargle periodogram of the residual array from the best-fit model to ensure that no significant periodicities were unaccounted for.

ExoTEP utilizes the affine-invariant Markov Chain Monte Carlo (MCMC) routine \texttt{emcee} \citep{emcee} to simultaneously compute the posterior distributions of all free parameters. In each fit, we set the number of walkers to four times the number of free parameters and initiated each chain near the best-fit parameter values from the corresponding individual segment fits. We typically set the chain lengths to 30,000--40,000 steps and discarded a burn-in equal to $60\%$ of each chain prior to calculating the posterior distributions. As a test for convergence, we checked that the Gelman--Rubin statistic $\hat{R}$ is below 1.1 \citep{gelmanrubin}.

As an empirical consistency test, particularly in cases where there is significant stellar variability in the raw light curves, we carried out phase curve fits on each spacecraft orbit's worth of photometry (i.e., two per sector) separately. Comparing the astrophysical fit parameter values across these orbit-wide fits, we found that they were mutually consistent at better than the $2\sigma$ level in all cases.

\subsection{Red Noise}\label{subsec:rednoise}

The \tess\ light curves show significant time-correlated noise (i.e., red noise), even after systematics correction, and we addressed its effect on the fitted astrophysical parameters in two different ways. The first method directly estimated and corrected for the red noise contribution through a two-step (MCMC) analysis. In the first step, we included an additional parameter: a uniform per-point uncertainty $\sigma_i$ for each data segment. Allowing this parameter to vary freely ensures that the resultant reduced $\chi^2$ value is unity. This process accounts for both white noise and any red noise at timescales comparable to the cadence of the observations (i.e., 2 minutes), yielding self-consistently adjusted uncertainties on the astrophysical parameters. Nevertheless, there is residual red noise on timescales longer than the time sampling of the \tess\ light curves.

\begin{figure}
    \centering
    \includegraphics[width=\linewidth]{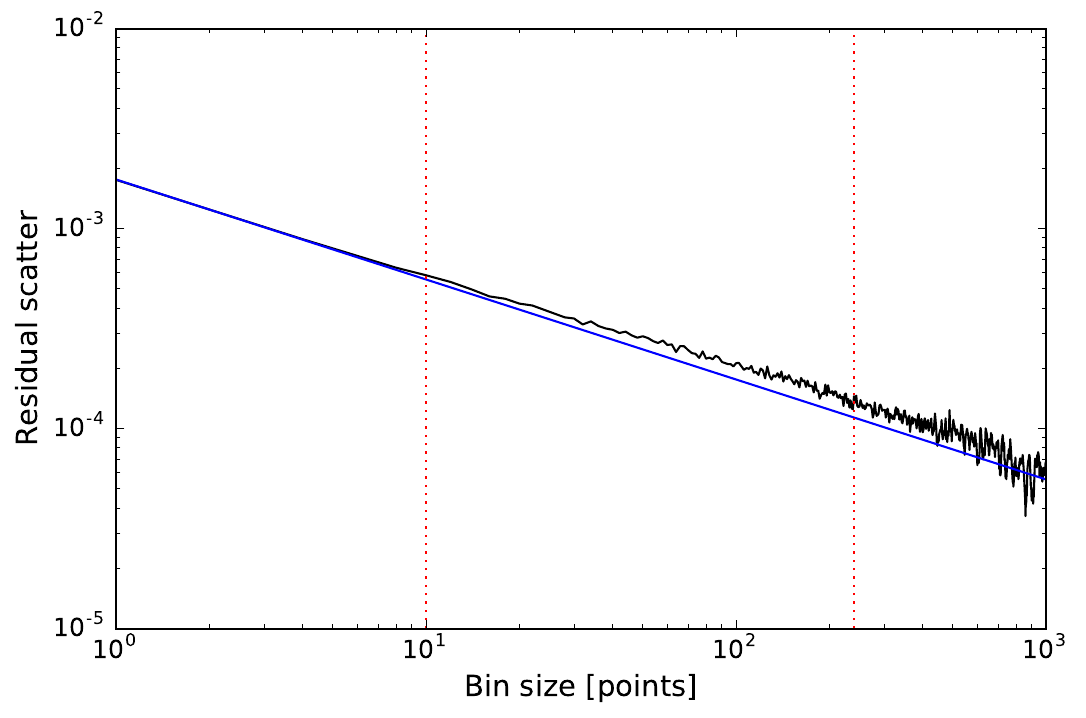}
    \caption{The scatter in the residuals from the best-fit model to the HATS-24 light curve, binned at various intervals (black curve). The blue line indicates the theoretical scaling law assuming pure white noise, with the zero-point set to the per-point uncertainty that ensures a reduced $\chi^2$ value of unity. The positive deviation of the black curve indicates the presence of time-correlated red noise at the corresponding timescales. In the final fits presented in this paper, we computed the average ratio between the measured and expected scatter for bin sizes spanning the range 10--240 and inflated the per-point uncertainty in the light curve by that ratio in order to propagate the additional red noise contribution to the resultant parameter uncertainties.}
    \label{fig:rms}
\end{figure}

To quantify the red noise contribution, we followed the technique described in \citet{pont2006} and used extensively in the analysis of Spitzer phase curves \citep[e.g.,][]{cowan2012,wong2016}. We binned the residuals from the initial joint fit using various bin sizes $n$ and plotted the curve alongside the $1/\sqrt{n}$ scaling law expected for pure white noise. Because the first joint fit already inflated the per-point uncertainty to account for any non-white noise on a per-point basis, the two curves are aligned at $n=1$. Figure \ref{fig:rms} shows an example of such a plot for the HATS-24 light curve. The actual scatter in the binned residuals deviates from the $1/\sqrt{n}$ curve, indicating the presence of additional red noise at various timescales. To include this additional contribution in the final fit, we computed the ratio $\beta$ of the two curves and took the average between bin sizes $n=10$ and $n=240$. These correspond to time intervals of 20 minutes and 8 hr, respectively, i.e., timescales relevant to the eclipse ingress/egress and phase curve variations, respectively. Across the targets we analyze in this paper, $\beta$ ranged from 1.04 to 1.52. In the second step of the MCMC analysis, we inflated the best-fit per-point uncertainties from the initial fits with $\beta$ and reran the joint fits with the uniform per-point uncertainties fixed to the new values.

The second method we utilized was ``prayer bead'' (PB) residual permutation (see, for example, \citealt{gillon2009}). For each light curve, the residual array was obtained from the best-fit combined astrophysical and systematics model in the initial MCMC fit, as described above. We then shifted the entire residual array by 5000 equal intervals, each time adding the cyclically permuted residuals back to the initial astrophysical model and deriving the best-fit parameters using Levenberg--Marquardt least-squares optimization. After all 5000 residual permutations were completed, we constructed the distribution of values for each parameter and calculated the median and $1\sigma$ uncertainties. In almost all cases, the PB method yielded smaller error bars than the uncertainty-inflated MCMC analysis (by $\sim$5--30\%). The notable exception is the mid-transit time, for which the PB method generally produced significantly larger uncertainties. For the results presented in this paper, we provide values and uncertainties for all parameters that were derived from the uncertainty-inflated MCMC analysis; we additionally include the transit time estimates from the PB analysis.

\section{Target Selection}\label{sec:selec}

The input database for our target selection included all known transiting planetary systems as well as new confirmed \tess\ discoveries, published or submitted as of 2019 December 1. We also searched through the catalog of known transiting brown dwarf systems, as compiled in \citet{carmichael2019} and \citet{mireles2020}.
To select targets with potentially detectable phase curve signals and secondary eclipses, we considered both photometric precision and theoretical predicted values for the various signals.  Extrapolating from the experience of our previous \tess\ phase curve studies \citep{shporer2019,wong2019wasp19,wong2019kelt9}, we limited our focus to systems with apparent \tess-band magnitudes brighter than 12.5 mag; this benchmark corresponds to a scaled 1 hr combined differential photometric precision of roughly 1000 ppm \citep{sullivan2015,stassun2017}. We also only considered systems for which 2 minute cadence data from the SPOC pipeline are available, in order to adequately resolve the ingress and egress of individual transits and secondary eclipses.

Systems displaying significant stellar variability other than the phased photometric modulations studied here (originating from, for example, pulsations or starspots) are challenging for phase curve analyses. The problems are particularly severe when the characteristic timescale of the variability is shorter than the orbital period, because techniques for detrending such additional photometric modulation can strongly bias the resultant measured astrophysical phase curve signals, or even remove them altogether. In this work, we did not analyze systems that show discernible short-term photometric variability on timescales shorter than or comparable to the orbital period. Several otherwise promising systems were rejected due to this variability constraint, including WASP-87A, WASP-120, WASP-167, and K2-237. The raw light curves for these systems are included in Appendix B, from which the variability is clearly discernible.

For transiting planetary-mass companions, the phase curve feature with the largest relative amplitude is typically the secondary eclipse. We calculated the predicted secondary eclipse depth using basic flux balance considerations, with the inclusion of reflected starlight \citep[e.g.,][]{esteves2013,esteves2015,shporer2017}:
\begin{equation}\label{eclipse}
D'_{d} = \left(\frac{R_{p}}{R_{*}}\right)^{2}\frac{\int B_{\lambda}(T_{p})\tau(\lambda) d\lambda}{\int B_{\lambda}(T_{*})\tau(\lambda) d\lambda}+A_{g}\left(\frac{R_{p}}{a}\right)^{2}.
\end{equation}
Here, for simplicity, we have approximated the star and planet flux as blackbodies $B_{\lambda}$ with temperatures $T_{p}$ and $T_{*}$, respectively. The fluxes are integrated over the \tess\ bandpass, with the associated transmission function $\tau(\lambda)$. The contribution of reflected light to the eclipse depth is parameterized by the geometric albedo in the \tess\ band: $A_{g}$. 

To compute the maximum limiting case for the planet's dayside temperature, we stipulated zero heat distribution across the planet's surface (i.e., instant reradiation). We assumed Lambertian scattering when relating Bond albedo to geometric albedo: $A_{B}\equiv \frac{3}{2}A_{g}$. It follows that the planet's dayside temperature can be expressed as \citep[e.g.,][]{esteves2013,esteves2015}
\begin{equation}\label{planettemp}
T_{p} = T_{*}\sqrt{\frac{R_{*}}{a}}\left\lbrack\frac{2}{3}\left(1-\frac{3}{2}A_{g}\right)\right\rbrack^{1/4}.
\end{equation}
The orbiting companions with potentially detectable secondary eclipses at visible wavelengths are almost all hot Jupiters and brown dwarfs. Both published geometric albedo measurements \citep[e.g.,][]{hengalbedo,esteves2015} and atmospheric models \citep[e.g.,][]{mayorga2019} indicate very low typical reflectivities for these objects, particularly in the red-optical. 

Given the manner in which the \tess\ sectors were arranged, most of ecliptic Southern Sky was only observed during one sector. Adjacent sectors overlapped at low ecliptic latitude, with a continuous viewing zone  (CVZ) near the ecliptic pole that was observed in all 13 sectors during the first year of the \tess\ mission. For targets observed in one sector, we established a selection criterion of $D'_{d}>100 \mathrm{ppm}$, assuming $A_{g}=0.1$. This benchmark value was appropriately adjusted for multi-sector targets. 

The list of known transiting systems in the ecliptic Southern Sky without previously published phase curves that satisfy the aforementioned constraints on brightness, stellar variability, and predicted secondary eclipse depth is as follows: HATS-24 \citep{hats24}, WASP-4 \citep{wasp4}, WASP-5 \citep{wasp5}, WASP-36 \citep{wasp36}, WASP-43 \citep{wasp43}, WASP-46 \citep{wasp46}, WASP-64 \citep{wasp6472}, WASP-72 \citep{wasp6472}, WASP-77A \citep{wasp77}, WASP-78 \citep{wasp78}, WASP-82 \citep{wasp82}, WASP-111 \citep{wasp111}, WASP-122/KELT-14 \citep{wasp122,kelt14}, WASP-142 \citep{wasp142}, and WASP-173A \citep{wasp173}. We also include WASP-100 \citep{wasp100}, the phase curve of which has been published by \citet{jansen2020}; in this paper, we present our independent analysis of the \tess\ light curve. In addition, we reanalyzed the WASP-18, WASP-19, and WASP-121 light curves in order to extend the identical data processing and analysis framework to those previously published phase curves and obtain a full sample of uniformly derived phase curve fits for year one of the \tess\ mission.

For the most massive short-period planets and brown dwarfs, the gravitationally induced phase curve signals --- ellipsoidal distortion and Doppler boosting --- may be detectable with \tess, as was the case with WASP-18 \citep{shporer2019} and KELT-9 \citep{wong2019kelt9}.
The leading term of the ellipsoidal distortion signal has a semi-amplitude of \citep[e.g.,][]{morris1985,shporer2017}
\begin{equation}\label{ellip}
A'_{\mathrm{ellip}}= \alpha_{\mathrm{ellip}}\frac{M_{p}}{M_{*}}\left(\frac{R_{*}}{a}\right)^{3}\sin^2 i,
\end{equation}
where $M_{p}$ and $M_{*}$ are the planet and stellar masses, respectively, $i$ is the orbital inclination, and the pre-factor $\alpha_{\mathrm{ellip}}$ is related to the linear limb-darkening and gravity-darkening coefficients $u$ and $g$ for the host star as follows:
\begin{equation}\label{alphaellip}
\alpha_{\mathrm{ellip}}=\frac{3}{20}\frac{(u+15)(g+1)}{3-u}.
\end{equation}
Tabulated values of $u$ and $g$ calculated for the \tess\ bandpass can be found in \citet{claret2017}.

The Doppler boosting semi-amplitude is related to the system's RV semi-amplitude $K_{\mathrm{RV}}$ by \citep[e.g.,][]{loeb2003,shporer2017}
\begin{equation}\label{doppler}
A'_{\mathrm{Dopp}}= \alpha_{\mathrm{Dopp}}\frac{K_{\mathrm{RV}}}{c},
\end{equation}
where $c$ is the speed of light, and the beaming factor $\alpha_{\mathrm{Dopp}}$ is of the order of unity and depends on the logarithmic derivative of the host star's flux with respect to wavelength, integrated over the \tess\ bandpass:
\begin{equation}\label{alphadopp}
\alpha_{\mathrm{Dopp}} = 3-\left\langle\frac{d \log F_{\nu}}{d \log \nu }\right\rangle_{\mathrm{TESS}}.
\end{equation}

Using published system parameters for targets that satisfy the aforementioned brightness and variability constraints, we calculated predicted values for $A'_{\mathrm{ellip}}$ and $A'_{\mathrm{Dopp}}$ and selected systems for which one or both of these amplitudes exceed 25 ppm. From this, we obtained three additional systems to include in our analysis --- HIP 65A \citep{hip65}, WASP-30 \citep{wasp30}, and TOI-503 \citep{subjak2019} --- for a total of 22 targets.

\section{Results}\label{sec:res}

For each of the 22 targets selected for detailed analysis, we fit the combined light curve (i.e., all utilized segments, as listed in Appendix A) to the full phase curve and systematics model in Equation \eqref{fita}. We ran a series of fits with different combinations of phase curve components, comparing the AIC and BIC of each run with the corresponding values of a fit with no phase curve components (secondary eclipse or sinusoidal terms), which we hereafter refer to as the null case. Systems for which the null case has the lowest BIC are considered non-detections or marginal detections; they are discussed briefly in Section \ref{subsec:marginal}. The 10 targets for which statistically robust signals were measured are discussed individually in the subsequent subsections.

\begin{deluxetable}{cccccc}
\tablewidth{0pc}
\tabletypesize{\scriptsize}
\tablecaption{
    Marginal Detections and Non-detections
    \label{tab:bad}
}
\tablehead{
    \colhead{Target} &
    \colhead{Sector}                     &
    \colhead{\textit{T}\tablenotemark{\scriptsize$\mathrm{a}$}} &
    \colhead{$D_{d,\mathrm{pred}}$\tablenotemark{\scriptsize$\mathrm{b}$}}  &
    \colhead{$D_{d,\mathrm{meas}}$\tablenotemark{\scriptsize$\mathrm{b}$}} &
    \colhead{$A_{\mathrm{atm}}$\tablenotemark{\scriptsize$\mathrm{c}$}}}
\startdata
HATS-24  & 13 & 12.2 & 280 & $290_{-110}^{+130}$ & $160\pm70$\\
WASP-4  & 2 & 11.8 & 230 & $120_{-70}^{+80}$ & $<70$\\ 
WASP-5 & 2 & 11.5 & 100 & $31_{-53}^{+72}$ & $<70$\\
WASP-36 & 8 & 12.1 & 120 & $90_{-70}^{+100}$ & $<100$\\
WASP-43 & 9 & 11.0 & 170 & $170\pm70$ & $52^{+27}_{-28}$\\
WASP-46  & 1 & 12.4 & 140 & $230_{-110}^{+140}$& $80\pm60$ \\
WASP-64  & 6,7 & 12.0 & 100 & $230_{-110}^{+130}$ & $89\pm53$\\
WASP-77A & 4 & 9.5 & 130 & $53_{-22}^{+32}$& $29\pm15$\\
WASP-78  & 5 & 11.8 & 160 & $210_{-90}^{+100}$ & $<80$ \\
WASP-82  & 5 & 9.5 & 130 & $72_{-39}^{+37}$& $<40$ \\
WASP-142 & 8 & 12.3 & 150 & $200_{-120}^{+160}$& $<140$\\
WASP-173A  & 2 & 10.9 & 140 & $150_{-60}^{+70}$& $72\pm34$\\
\enddata
\textbf{Notes.}
\vspace{-0.25cm}\tablenotetext{\textrm{a}}{Apparent magnitude in the \tess\ bandpass.}
\vspace{-0.25cm}\tablenotetext{\textrm{b}}{Predicted and measured secondary eclipse depths, in parts-per-million. Predictions assume $A_{g}=0.1$ and zero heat transport to the nightside.}
\vspace{-0.25cm}\tablenotetext{\textrm{c}}{Measured semi-amplitude of the atmospheric brightness modulation, in parts-per-million. In some cases, $2\sigma$ upper limits are provided.}
\end{deluxetable}

\subsection{Marginal Detections and Non-detections}\label{subsec:marginal}

Out of the 22 targets analyzed in this paper, 12 showed no robust phase curve signals of any kind. For most of these systems, the highest signal-to-noise phase curve component is the secondary eclipse. These marginal detections and non-detections tend to occur in relatively faint systems, illustrating the important limiting role that photometric precision plays in phase curve detectability. 

We list the marginal detections and non-detections in Table \ref{tab:bad}. The secondary eclipse depths $D_{d}$ and atmospheric brightness modulation semi-amplitudes $A_{\mathrm{atm}}$ derived from phase curve fits including both parameters are also given. We also include the corresponding predicted values, computed following the prescription described in Section \ref{sec:selec}. The measured secondary eclipse depths have formal statistical significances ranging from $<$$1\sigma$ to $\sim$$2.6\sigma$, with the fitted values broadly in agreement with the predictions. For all of these systems, the BIC of the null case was lower than the next-lowest case by a margin of at least $\Delta\mathrm{BIC}=2.2$. Even though these secondary eclipse detections fail the BIC test, we utilize some of them to place constraints on the visible geometric albedo in Section \ref{subsec:temp}.

For the systems without significant secondary eclipse or phase curve signal detections, we carried out transit-only fits of the \tess\ light curves (i.e., fixing the out-of-transit light curve to a flat line). The results are listed in Table \ref{tab:transitfit}. From the light-curve fit parameters $b$, $a/R_{*}$, $\gamma_{1}$, and $\gamma_2$, we derive the inclination $i$ and the quadratic limb-darkening coefficients $u_{1}$ and $u_{2}$. We also utilize the stellar radius reported in the respective discovery papers to obtain the orbital semimajor axis $a$ and the radius of the orbiting companion $R_{p}$. In all cases, the new measurements agree with literature values at better than the $2\sigma$ level, with the vast majority of the results consistent to well within $1\sigma$. The systematics-corrected, phase-folded transit light curves from these fits are compiled in Figure \ref{fig:transits}. For completeness, the full-orbit light curves from these transit-only fits are provided in Appendix C.

\begin{splitdeluxetable*}{lllllllllBlllllllllBlllllllll}
\tablewidth{0pc}
\tabletypesize{\scriptsize}
\tablecaption{
    Results from Transit-only Light-curve Fits
    \label{tab:transitfit}
}
\vspace{-0.3cm}
\tablehead{& \multicolumn{2}{c}{\underline{HATS-24}} \vspace{-0.2cm}& \multicolumn{2}{c}{\underline{WASP-4}} &
\multicolumn{2}{c}{\underline{WASP-5}} &  \multicolumn{2}{c}{\underline{WASP-36}} & & \multicolumn{2}{c}{\underline{WASP-43}} &  \multicolumn{2}{c}{\underline{WASP-46}} &
\multicolumn{2}{c}{\underline{WASP-64}} &
\multicolumn{2}{c}{\underline{WASP-77A}} & & 
\multicolumn{2}{c}{\underline{WASP-78}} &  \multicolumn{2}{c}{\underline{WASP-82}} &
\multicolumn{2}{c}{\underline{WASP-142}} &
\multicolumn{2}{c}{\underline{WASP-173A}}\\
    \colhead{Parameter} \vspace{-0.1cm}&
    \colhead{Value}                     &
    \colhead{Error}  &
    \colhead{Value}                     &
    \colhead{Error}  & 
    \colhead{Value}                     &
    \colhead{Error}  &  
    \colhead{Value}                     &
    \colhead{Error}  &
    \colhead{Parameter} &
    \colhead{Value}                     &
    \colhead{Error}  &
    \colhead{Value}                     &
    \colhead{Error}  & 
    \colhead{Value}                     &
    \colhead{Error} &
    \colhead{Value}                     &
    \colhead{Error}  &
    \colhead{Parameter} &
    \colhead{Value}                     &
    \colhead{Error}  &
    \colhead{Value}                     &
    \colhead{Error}  & 
    \colhead{Value}                     &
    \colhead{Error} &
    \colhead{Value}                     &
    \colhead{Error} 
}
\startdata
\multicolumn{2}{l}{\textit{Fitted parameters}} & & & & & & & & \multicolumn{2}{l}{\textit{Fitted parameters}}& &  & & & & & & \multicolumn{2}{l}{\textit{Fitted parameters}}\\
$R_p/R_*$     & 0.1342 & $_{-0.0022}^{+0.0018}$ & 0.1523 & 0.0010 & 0.1104 & $_{-0.0015}^{+0.0014}$  &  0.1346 & $_{-0.0033}^{+0.0029}$ & $R_p/R_*$   & 0.1595 & $_{-0.0020}^{+0.0015}$    & 0.1385 & $_{-0.0032}^{+0.0022}$ & 0.1208 & $_{-0.0018}^{+0.0022}$  &  0.1193 & $_{-0.0014}^{+0.0008}$ & $R_p/R_*$     & 0.0857 & $_{-0.0010}^{+0.0011}$ & 0.07726 & $_{-0.00046}^{+0.00062}$ & 0.1020 & $_{-0.0027}^{+0.0024}$ & 0.11273 & $_{-0.00088}^{+0.00085}$ \\
$T_{0,\mathrm{MCMC}}$\tablenotemark{\scriptsize a}  & 667.45838 & $_{-0.00021}^{+0.00019}$ & 365.890364 & $_{-0.000083}^{+0.000086}$ & 366.90713 & 0.00018 &  526.19222 & $_{-0.00021}^{+0.00020}$ & $T_{0,\mathrm{MCMC}}$ & 555.805663 & $_{-0.000047}^{+0.000048}$   & 337.45196 & $_{-0.00019}^{+0.00018}$ & 493.18907 & 0.00024 & 427.305069 & $_{-0.000048}^{+0.000047}$ & $T_{0,\mathrm{MCMC}}$ & 449.07749 & $_{-0.00057}^{+0.00048}$ & 447.08367 & $_{-0.00022}^{+0.00023}$ & 526.90343 & $_{-0.00068}^{+0.00073}$ & 367.67562 & $_{-0.00012}^{+0.00013}$ \\
$T_{0,\mathrm{PB}}$\tablenotemark{\scriptsize a}  & 667.45848 & $_{-0.00032}^{+0.00037}$ & 365.89038 & 0.00015 & 366.90697 & $_{-0.00038}^{+0.00044}$ &  526.19223 & $_{-0.00035}^{+0.00037}$ & $T_{0,\mathrm{PB}}$ & 555.80567 & $_{-0.00011}^{+0.00010}$   & 337.45186 & $_{-0.00033}^{+0.00039}$ & 493.18895 & $_{-0.00034}^{+0.00042}$  & 427.30509 & $_{-0.00013}^{+0.00011}$ & $T_{0,\mathrm{PB}}$ & 449.07700 & $_{-0.00073}^{+0.00075}$ & 447.08368 & $_{-0.00041}^{+0.00038}$ & 526.9036 & $_{-0.0011}^{+0.0010}$ & 367.67559 & $_{-0.00023}^{+0.00026}$ \\
$P$ (days)    & 1.348503 & 0.000033 & 1.338233 & $_{-0.000014}^{+0.000015}$ & 1.628411 & $_{-0.000044}^{+0.000039}$ &  1.537389 & 0.000039      & $P$ (days) & 0.8134722 & $_{-0.0000056}^{+0.0000060}$ & 1.430343 & $_{-0.000034}^{+0.000035}$ & 1.573253 & $_{-0.000027}^{+0.000028}$ & 1.3600266 & $_{-0.0000087}^{+0.0000081}$ &  $P$ (days) & 2.17538 & 0.00014 & 2.705838 & $_{-0.000081}^{+0.000086}$ & 2.05287 & $_{-0.00018}^{+0.00017}$ & 1.386658 & 0.000024 \\
$b$           & 0.30 & $_{-0.16}^{+0.11}$ & 0.02 & 0.13 & 0.37 & $_{-0.13}^{+0.10}$ &  0.651 & $_{-0.060}^{+0.054}$    & $b$  & 0.688 & $_{-0.018}^{+0.014}$         & 0.666 & $_{-0.057}^{+0.041}$ & 0.04 & $_{-0.31}^{+0.30}$ & 0.316 & $_{-0.096}^{+0.050}$ & $b$ & $-$0.05 & $_{-0.20}^{+0.25}$ & 0.20 & $_{-0.13}^{+0.12}$ & 0.750 & $_{-0.072}^{+0.044}$ & 0.07 & $_{-0.21}^{+0.19}$ \\
$a/R_*$       & 4.65 & $_{-0.16}^{+0.13}$ & 5.438 & $_{-0.057}^{+0.044}$ & 5.36 & $_{-0.22}^{+0.21}$ & 5.76 & $_{-0.27}^{+0.26}$     & $a/R_*$   & 4.734 & $_{-0.053}^{+0.054}$ & 6.17 & $_{-0.24}^{+0.28}$ & 5.53 & $_{-0.25}^{+0.14}$ & 5.162 & $_{-0.08}^{+0.12}$ & $a/R_*$   & 3.778 & $_{-0.098}^{+0.060}$ & 4.38 & $_{-0.14}^{+0.06}$ & 4.63 & $_{-0.32}^{+0.45}$ &  5.13 & $_{-0.12}^{+0.06}$ \\
$\gamma_{1}$\tablenotemark{\scriptsize b}  & 0.70 & $_{-0.13}^{+0.15}$ & 0.972 & $_{-0.067}^{+0.062}$ & 0.72 & 0.13 & 1.06 & $_{-0.27}^{+0.26}$ & $\gamma_{1}$ & 0.99 & $_{-0.14}^{+0.12}$  & 0.94 & $_{-0.25}^{+0.26}$ & 0.97 & $_{-0.14}^{+0.15}$ & 0.900 & $_{-0.037}^{+0.051}$ & $\gamma_{1}$ & 0.72 & 0.15 & 0.748 & $_{-0.053}^{+0.054}$ & 0.95 & 0.34 & 0.661 & $_{-0.086}^{+0.088}$ \\
$\gamma_{2}$\tablenotemark{\scriptsize b}  & $-$0.45 & $_{-0.66}^{+0.54}$ & $-$0.07 & $_{-0.38}^{+0.34}$ & $-$0.35 & $_{-0.58}^{+0.47}$ & $-$0.56 & $_{-0.67}^{+0.68}$  & $\gamma_{2}$  & $-$0.33 & $_{-0.83}^{+0.63}$ & $-$0.54 & $_{-0.70}^{+0.67}$ & $-$0.77 & $_{-0.56}^{+0.66}$ & $-$0.01 & $_{-0.31}^{+0.27}$ & $\gamma_{2}$ & $-$0.20 & $_{-0.51}^{+0.37}$ & 0.03 & $_{-0.34}^{+0.24}$ & $-$0.43 & $_{-0.72}^{+0.63}$ & $-$0.13 & $_{-0.45}^{+0.31}$ \\
\multicolumn{2}{l}{\textit{Derived parameters}} & & & & & & &  & \multicolumn{2}{l}{\textit{Derived parameters}} & &  & & & & & & \multicolumn{2}{l}{\textit{Derived parameters}} \\
$i$ ($^{\circ}$)      & 86.3 & $_{-1.5}^{+2.0}$ & 89.8 & 1.4 & 86.1 & $_{-1.3}^{+1.5}$ & 83.52 & $_{-0.87}^{+0.84}$   & $i$ ($^{\circ}$)   & 81.65 & $_{-0.25}^{+0.29}$   & 83.80 & $_{-0.64}^{+0.75}$ & 89.6 & 3.2 & 86.5 & $_{-0.6}^{+1.1}$ & $i$ ($^{\circ}$)   & 90.8 & $_{-3.8}^{+3.1}$ & 87.4 & 1.8 & 80.7 & $_{-1.3}^{+1.6}$ & 89.2 & $_{-2.2}^{+2.3}$ \\
$u_{1}$  & 0.18 & 0.12 & 0.374 & $_{-0.072}^{+0.065}$ & 0.21 & $_{-0.12}^{+0.11}$ & 0.30 & $_{-0.20}^{+0.21}$ & $u_{1}$  & 0.32 & $_{-0.20}^{+0.16}$ & 0.25 & $_{-0.17}^{+0.21}$ & 0.24 & $_{-0.14}^{+0.13}$ &  0.357 & $_{-0.054}^{+0.053}$ & $u_{1}$ & 0.24 & $_{-0.12}^{+0.10}$ & 0.304 & $_{-0.072}^{+0.053}$ & 0.26 &  $_{-0.18}^{+0.25}$ & 0.236 & $_{-0.091}^{+0.067}$ \\
$u_{2}$  & 0.32 & $_{-0.22}^{+0.28}$ & 0.22 & $_{-0.14}^{+0.16}$ & 0.28 & $_{-0.19}^{+0.24}$ & 0.30 & $_{-0.27}^{+0.26}$ & $u_{2}$    & 0.33 & $_{-0.24}^{+0.32}$ & 0.40 & $_{-0.26}^{+0.29}$ & 0.51 & $_{-0.27}^{+0.23}$ & 0.18 & $_{-0.11}^{+0.13}$ & $u_{2}$ & 0.23 & $_{-0.15}^{+0.20}$ & 0.14 & $_{-0.10}^{+0.14}$ & 0.35 & $_{-0.24}^{+0.30}$ & 0.18 & $_{-0.13}^{+0.19}$ \\
$a$ (au) & 0.0253 & 0.0011 & 0.02369 & 0.00091 & 0.0256 & 0.0018 & 0.0255 & 0.0013 & $a$ (au) & 0.01321 & 0.00078  & 0.0263 & 0.0014 & 0.0272 & 0.0012 & 0.02293 & 0.00058 & $a$ (au) & 0.0387 & 0.0023 & 0.0444 & 0.0017 & 0.0353 & 0.0034 & 0.0265 & 0.013 \\
$R_{p}$ ($R_{\mathrm{Jup}}$) & 1.531 & 0.049 & 1.389 & 0.053 & 1.102 & 0.064 & 1.246 & 0.037 & $R_{p}$ ($R_{\mathrm{Jup}}$)& 0.931 & 0.055  & 1.236 & 0.045 & 1.244 & 0.036 & 1.11 & 0.02 &$R_{p}$ ($R_{\mathrm{Jup}}$) & 1.84 & 0.10 &  1.639 & 0.050 & 1.628 & 0.089 & 1.218 & 0.056\\
\enddata
\textbf{Notes.}
\vspace{-0.25cm}\tablenotetext{\textrm{a}}{Mid-transit times are given in units of $\mathrm{BJD}_{\mathrm{TDB}}-2458000$. The two transit times are derived from the MCMC and prayer bead (PB) analyses.}
\vspace{-0.25cm}\tablenotetext{\textrm{b}}{Modified limb-darkening parameters $\gamma_{1}\equiv 2u_{1}+u_{2}$ and $\gamma_{2}\equiv u_{1}-2u_{2}$.}
\end{splitdeluxetable*}

\begin{figure*}
\includegraphics[width=0.95\linewidth]{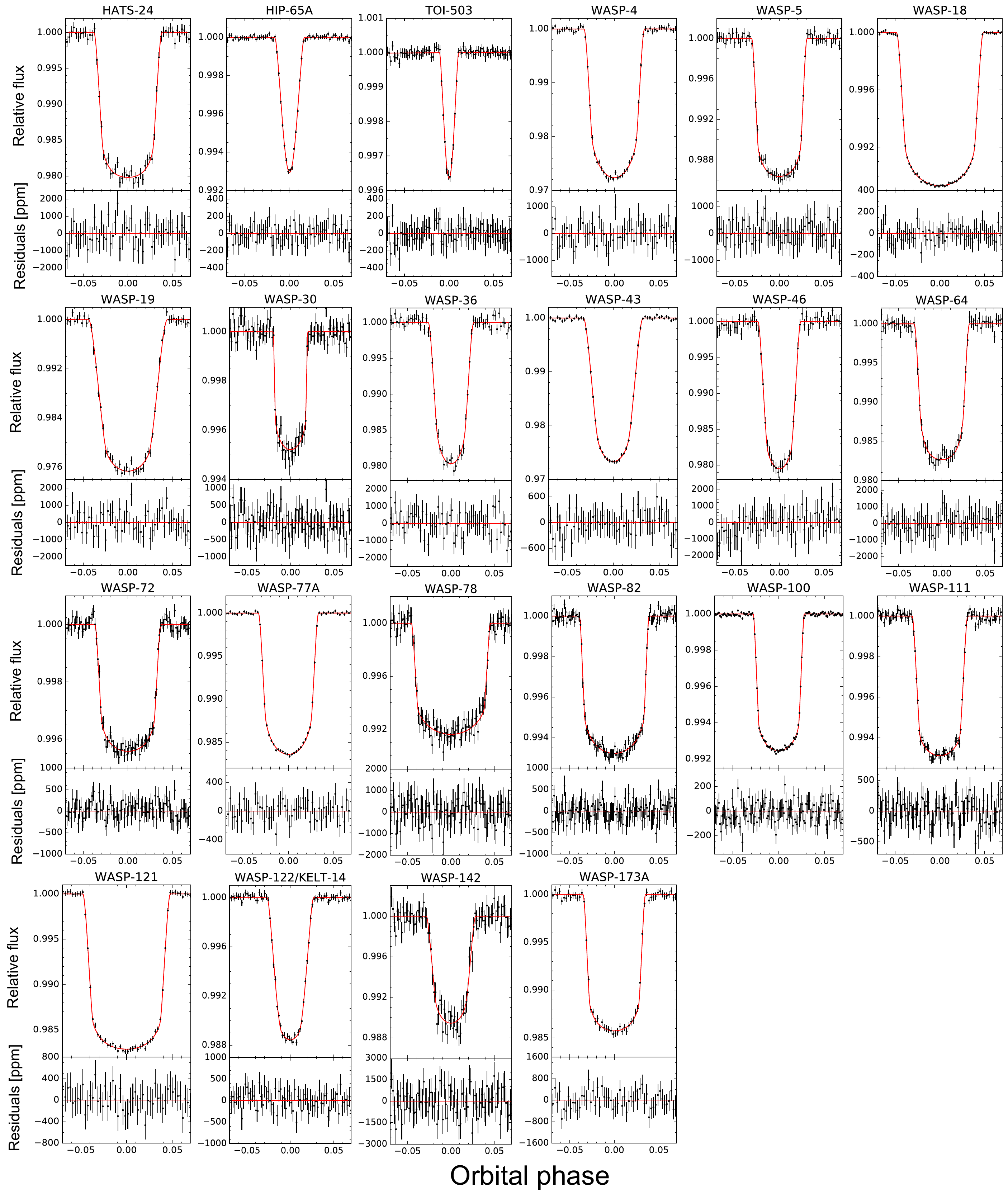}
\caption{Compilation plot of the binned and phase-folded light curves in the vicinity of the primary transit for the 22 systems analyzed in this work. The bottom panels show the residuals from the best-fit model. All systematics and phase curve signals have been removed. The light curves of systems with orbital periods in the ranges $P<1$ day, $1\ge P\ge 3$ days, and $P>3$ days are binned in 3, 5, and 10 minute intervals, respectively.}
\label{fig:transits}
\end{figure*}

\begin{splitdeluxetable*}{lllllllllllBlllllllllll}
\tablewidth{0pc}
\tabletypesize{\scriptsize}
\tablecaption{
    Results from Phase Curve Fits
    \label{tab:fit}
}
\vspace{-0.3cm}
\tablehead{& \multicolumn{2}{c}{\underline{HIP 65A}} \vspace{-0.2cm}& \multicolumn{2}{c}{\underline{TOI-503}} &
\multicolumn{2}{c}{\underline{WASP-30}} &  \multicolumn{2}{c}{\underline{WASP-72}} &  \multicolumn{2}{c}{\underline{WASP-100}} & & \multicolumn{2}{c}{\underline{WASP-111}} &  \multicolumn{2}{c}{\underline{WASP-122/KELT-14}} &
\multicolumn{2}{c}{\underline{WASP-18}\tablenotemark{\scriptsize f}} &
\multicolumn{2}{c}{\underline{WASP-19}\tablenotemark{\scriptsize f}} &
\multicolumn{2}{c}{\underline{WASP-121}\tablenotemark{\scriptsize f}} \\
    \colhead{Parameter} \vspace{-0.1cm}&
    \colhead{Value}                     &
    \colhead{Error}  &
    \colhead{Value}                     &
    \colhead{Error}  & 
    \colhead{Value}                     &
    \colhead{Error}  &  
    \colhead{Value}                     &
    \colhead{Error}  &
    \colhead{Value}                     &
    \colhead{Error}  &
    \colhead{Parameter} &
    \colhead{Value}                     &
    \colhead{Error}  &
    \colhead{Value}                     &
    \colhead{Error}  & 
    \colhead{Value}                     &
    \colhead{Error} &
    \colhead{Value}                     &
    \colhead{Error}  &
    \colhead{Value}                     &
    \colhead{Error}  
}
\startdata
\multicolumn{2}{l}{\textit{Fitted parameters}} & & & & & & & & & & \multicolumn{2}{l}{\textit{Fitted parameters}} & & & & & & & & &\\
$R_p/R_*$     & 0.30 & $_{-0.10}^{+0.12}$ & 0.085 & $_{-0.013}^{+0.051}$ & 0.0653 & 0.0010  &  0.06446 & $_{-0.00089}^{+0.00078}$    & 0.08385 & $_{-0.00039}^{+0.00032}$  & $R_p/R_*$  & 0.08188 & $_{-0.00055}^{+0.00049}$ & 0.1102 & $_{-0.0012}^{+0.0014}$  & 0.09774 & $_{-0.00039}^{+0.00035}$ & 0.1522 & $_{-0.0021}^{+0.0015}$ & 0.12409 & $_{-0.00043}^{+0.00045}$ \\
$T_{0,\mathrm{MCMC}}$\tablenotemark{\scriptsize a}  & 354.55211 & $_{-0.00009}^{+0.00010}$ & 503.08554 & $_{-0.00036}^{+0.00037}$ & 369.43329 & $_{-0.00060}^{+0.00064}$ &  412.21664 & 0.00034 & 509.105435 & $_{-0.000072}^{+0.000077}$ & $T_{0,\mathrm{MCMC}}$  & 337.13712 & $_{-0.00017}^{+0.00019}$ & 503.53225 & $_{-0.00016}^{+0.00017}$ & 374.228432 & 0.000033 & 555.45472 & 0.00012 & 503.473091 & $_{-0.000094}^{+0.000088}$ \\
$T_{0,\mathrm{PB}}$\tablenotemark{\scriptsize a}  & 354.5523 & $_{-0.0014}^{+0.0009}$ & 503.08580 & $_{-0.00081}^{+0.00097}$ & 369.4316 & $_{-0.0014}^{+0.0019}$ &  412.21665 & $_{-0.00062}^{+0.00065}$ & 509.10542 & 0.00014 & $T_{0,\mathrm{PB}}$  & 337.13711 & $_{-0.00033}^{+0.00036}$ & 503.53227 & 0.00021 & 374.228437 & $_{-0.000067}^{+0.000070}$ & 555.45469 & $_{-0.00019}^{+0.00021}$ & 503.47309 & $_{-0.00016}^{+0.00012}$ \\
$P$ (days)    & 0.9809761 & $_{-0.0000045}^{+0.0000041}$  & 3.67745 & $_{-0.00016}^{+0.00017}$ & 4.15699 & $_{-0.00033}^{+0.00034}$ &  2.216789 & $_{-0.000054}^{+0.000041}$     & 2.8493822 & $_{-0.0000020}^{+0.0000023}$ & $P$ (days) & 2.310946 & $_{-0.000054}^{+0.000052}$ & 1.710019 & 0.000038 & 0.9414532 & 0.0000020 & 0.788846 & $_{-0.000012}^{+0.000013}$ & 1.274929 & 0.000016 \\
$b$           & 1.22 & $_{-0.12}^{+0.13}$ & 0.973 & $_{-0.033}^{+0.071}$ & 0.17 & $_{-0.12}^{+0.17}$ &  0.659 & $_{-0.070}^{+0.046}$    & 0.558 & $_{-0.017}^{+0.016}$ & $b$         & 0.689 & $_{-0.024}^{+0.020}$ & 0.845 & $_{-0.018}^{+0.013}$ & 0.348 & $_{-0.036}^{+0.029}$ & 0.648 & $_{-0.031}^{+0.024}$ & 0.081 & $_{-0.057}^{+0.077}$ \\
$a/R_*$       & 5.180 & $_{-0.084}^{+0.089}$ & 7.25 & $_{-0.59}^{+0.61}$ & 8.42 & $_{-0.35}^{+0.16}$ & 3.74 & $_{-0.18}^{+0.24}$       & 5.389 & 0.064 & $a/R_*$ & 4.39 & $_{-0.10}^{+0.11}$ & 4.39 & $_{-0.09}^{+0.10}$ & 3.539 & $_{-0.035}^{+0.039}$ & 3.582 & $_{-0.067}^{+0.074}$ & 3.815 & $_{-0.032}^{+0.018}$ \\
$\bar{f_p}$ (ppm)    & $\left(220\right)$\tablenotemark{\scriptsize b}& $\left(_{-150}^{+170}\right)$ & $\left(200\right)$ & $\left(_{-120}^{+200}\right)$& $\left(60\right)$& $\left(_{-43}^{+64}\right)$ & 41 & $_{-27}^{+37}$     & 47 & $_{-17}^{+16}$ & $\bar{f_p}$ (ppm) & 102 & $_{-37}^{+38}$ & 101 & $_{-57}^{+61}$ & 157 & $_{-19}^{+20}$ & 160 & $_{-100}^{+120}$ & 272 & $_{-52}^{+53}$ \\
$A_{\mathrm{atm}}$ (ppm)   & 29.6 & $_{-9.1}^{+8.3}$ & $\left(<20\right)$ & $\dots$& $\left(34\right)$& $\left(_{-26}^{+29}\right)$&  70 & $_{-16}^{+15}$  & 48.5 & $_{-5.4}^{+5.7}$ & $A_{\mathrm{atm}}$ (ppm) &$\left(<31\right)$& $\dots$& 64 & 17 & 181.7 & $^{+8.2}_{-8.1}$ & 311 & $_{-54}^{+57}$ & 214 & $_{-26}^{+25}$ \\
$\delta$ ($^{\circ}$) & $\dots$& $\dots$& $\dots$& $\dots$& $\dots$& $\dots$& $\left(<25\right)$ &  $\dots$ & 12.0 & $_{-5.7}^{+6.3}$ & $\delta$ ($^{\circ}$) & $\dots$& $\dots$& $\left(<34\right)$& $\dots$ & $\dots$& $\dots$ & $\left(<17\right)$& $\dots$ & $\left(-9.7\right)$& $\left(5.7\right)$ \\
$A_{\mathrm{ellip}}$ (ppm)  & 28.4 & $_{-8.0}^{+9.6}$ & 61.6 & $_{-8.1}^{+9.4}$ & 97 & $_{-25}^{+24}$ & $\lbrack18\rbrack$\tablenotemark{\scriptsize b} & $\lbrack6\rbrack$  & $\lbrack11\rbrack$ & $\lbrack5\rbrack$& $A_{\mathrm{ellip}}$ (ppm) & $\lbrack13\rbrack$ & $\lbrack2\rbrack$& $\lbrack12\rbrack$ & $\lbrack2\rbrack$ & 181.9 & $_{-8.4}^{+8.6}$ & $\lbrack27\rbrack$ & $\lbrack5\rbrack$ & $\lbrack18\rbrack$ & $\lbrack1\rbrack$ \\
$A_{\mathrm{Dopp}}$ (ppm)  & 18.7 & $_{-8.4}^{+9.1}$ & 30 & 10 & $\left(<62\right)$& $\dots$& $\lbrack2.0\rbrack$ & $\lbrack0.1\rbrack$  & $\lbrack2.1\rbrack$ & $\lbrack0.2\rbrack$ & $A_{\mathrm{Dopp}}$ (ppm) & $\lbrack2.2\rbrack$ & $\lbrack0.2\rbrack$& $\lbrack2.0\rbrack$ & $\lbrack0.1\rbrack$ & 20.0 & $_{-6.7}^{+6.4}$ & $\lbrack3.0\rbrack$ & $\lbrack0.3\rbrack$ & $\lbrack1.8\rbrack$ & $\lbrack0.1\rbrack$ \\
$\gamma_{1}$\tablenotemark{\scriptsize c}  & 1.17 & $_{-0.11}^{+0.13}$ & 0.98 & $_{-0.48}^{+0.36}$ & 0.72 & $_{-0.17}^{+0.19}$ & 0.67 & $_{-0.13}^{+0.14}$  & 0.652 & $_{-0.053}^{+0.050}$ & $\gamma_{1}$ & 0.59 & 0.11 & 0.65 & $_{-0.29}^{+0.31}$ & 0.750 & 0.022 & 0.81 & 0.16 & 0.643 & 0.041 \\
$\gamma_{2}$\tablenotemark{\scriptsize c}  & 0.16 & $_{-0.12}^{+0.11}$ & $-$0.16 & $_{-0.72}^{+0.46}$ & $-$0.07 & $_{-0.54}^{+0.31}$ & $-$0.45 & $_{-0.47}^{+0.54}$   & $-$0.15 & $_{-0.36}^{+0.33}$ & $\gamma_{2}$ & $-$0.17 & $_{-0.44}^{+0.36}$ & $-$0.23 & $_{-0.54}^{+0.41}$ & $-$0.03 & 0.21 & $-$0.58 & $_{-0.59}^{+0.66}$ & $-$0.03 & $_{-0.27}^{+0.21}$ \\
\multicolumn{2}{l}{\textit{Derived parameters}} & & & & & & &  & & & \multicolumn{2}{l}{\textit{Derived parameters}} & & & & & & & \\
$D_{d}$ (ppm)\tablenotemark{\scriptsize d}  & $\left(250\right)$& $\left(_{-150}^{+170}\right)$& $\left(200\right)$ & $\left(_{-120}^{+200}\right)$& (97) & $\left(_{-51}^{+70}\right)$& 113  & $_{-31}^{+39}$  & 94 & 17 & $D_{d}$ (ppm) & 102 & $_{-37}^{+38}$ & 165 & $_{-60}^{+63}$ & 339 & 21 & 470 & $_{-110}^{+130}$ & 486 & 59  \\
$D_{n}$ (ppm)\tablenotemark{\scriptsize d}   & $\left(190\right)$& $\left(_{-150}^{+170}\right)$& $\dots$& $\dots$& (27) & $\left(_{-48}^{+64}\right)$&  $-$28 & $_{-31}^{+38}$  & 0 & 18 & $D_{n}$ (ppm) & $\dots$& $\dots$& 38 & $_{-58}^{+63}$ & $-$25 & $_{-20}^{+21}$ & $-$150 & $_{-110}^{+130}$ & 58 & 59 \\
$i$ ($^{\circ}$)      & 76.3 & $_{-1.3}^{+1.5}$ & 82.3 & $_{-1.3}^{+0.9}$ & 88.8 & $_{-1.2}^{+0.8}$ & 79.9 & $_{-1.3}^{+1.6}$      & 84.06 & 0.25 & $i$ ($^{\circ}$)   & 80.97 & $_{-0.48}^{+0.54}$ & 78.91 & $_{-0.41}^{+0.47}$ & 84.36 & $_{-0.53}^{+0.64}$ & 79.59 & $_{-0.59}^{+0.67}$ & 88.8 & $_{-1.2}^{+0.9}$ \\
$u_{1}$  & 0.504 & $_{-0.051}^{+0.053}$ & 0.31 & $_{-0.22}^{+0.26}$ & 0.26 & $_{-0.13}^{+0.11}$ & 0.16 & $_{-0.12}^{+0.15}$  & 0.230 & $_{-0.089}^{+0.084}$ & $u_{1}$ & 0.20 & $_{-0.12}^{+0.11}$ & 0.17 & $_{-0.12}^{+0.21}$ & 0.293 & $_{-0.043}^{+0.044}$ & 0.20 & $_{-0.14}^{+0.17}$ & 0.251 &  $_{-0.057}^{+0.045}$  \\
$u_{2}$  & 0.168 & $_{-0.046}^{+0.053}$ & 0.24 & $_{-0.18}^{+0.30}$ & 0.17 & $_{-0.12}^{+0.21}$ & 0.31 & $_{-0.19}^{+0.18}$    & 0.19 & $_{-0.12}^{+0.14}$ & $u_{2}$ & 0.18 & $_{-0.13}^{+0.16}$ & 0.21 & $_{-0.14}^{+0.23}$ & 0.162 & $_{-0.081}^{+0.082}$ & 0.38 & $_{-0.25}^{+0.24}$ & 0.14 & $_{-0.08}^{+0.11}$ \\
$a$ (au) & 0.01745 & 0.00036& 0.05732 & 0.0050 & 0.0507 & 0.0017 & 0.0344 & 0.0046 & 0.0501 & 0.0075 & $a$ (au) & 0.0378 & 0.0022 & 0.0279 & 0.0017 & 0.0200 & 0.0010 & 0.01566 & 0.00073 & 0.02587 & 0.00056 \\
$R_{p}$ ($R_{\mathrm{Jup}}$) & 2.11 & 0.78 & 1.41 & 0.53 & 0.823 & 0.017 & 1.24 & 0.15 & 1.63 & 0.24 & $R_{p}$ ($R_{\mathrm{Jup}}$) & 1.474 & 0.080 & 1.467 & 0.085 & 1.157 & 0.058 & 1.392 & 0.061 & 1.761 & 0.037\\
$M_{p}$ ($M_{\mathrm{Jup}}$)\tablenotemark{\scriptsize e} & 2.64 & 0.71 & 28--53 & $\dots$ & 60 & 17 & $\dots$ & $\dots$ & $\dots$ & $\dots$ & $M_{p}$ ($M_{\mathrm{Jup}}$) & $\dots$ & $\dots$ & $\dots$ & $\dots$ & 9.8 & 1.1 & $\dots$& $\dots$ & $\dots$ & $\dots$\\
\enddata
\textbf{Notes.}
\vspace{-0.25cm}\tablenotetext{\textrm{a}}{Mid-transit times are given in units of $\mathrm{BJD}_{\mathrm{TDB}}-2458000$. The two transit times were derived from the MCMC and PB analyses.}
\vspace{-0.25cm}\tablenotetext{\textrm{b}}{For statistical non-detections, values or $2\sigma$ upper limits are provided in parentheses. Values in square brackets were constrained by Gaussian priors.}
\vspace{-0.25cm}\tablenotetext{\textrm{c}}{Modified limb-darkening parameters $\gamma_{1}\equiv 2u_{1}+u_{2}$ and $\gamma_{2}\equiv u_{1}-2u_{2}$. For the HIP 65A phase curve fit, these parameters were constrained by priors derived from values tabulated in \citet{claret2018}: $(\gamma_1,\gamma_2) = (1.16\pm0.11,0.18\pm0.11)$.}
\vspace{-0.25cm}\tablenotetext{\textrm{d}}{$D_{d}$ and $D_{n}$ are the secondary eclipse depth (i.e., dayside flux) and nightside flux, respectively, as defined in Section \ref{subsec:model}.}
\vspace{-0.25cm}\tablenotetext{\textrm{e}}{Companion masses derived from the measured ellipsoidal distortion and Doppler boosting semi-amplitudes (see Section \ref{subsec:grav}).}
\vspace{-0.25cm}\tablenotetext{\textrm{f}}{Reanalysis of phase curves previously published in \citet{shporer2019}, \citet{wong2019wasp19}, \citet{bourrier2019}, and \citet{daylan2019}.}
\end{splitdeluxetable*}

\subsection{HIP 65A}\label{hip65}
The discovery of this system was reported in \citet{hip65}. The system, observed by \tess\ during sectors 1 and 2, consists of a 3.23 $M_{\mathrm{Jup}}$ planet on a grazing 0.98 day orbit around an active 0.79 $M_{\Sun}$ K dwarf. Due to the small occulted area of the planet during superior conjunction, as well as the relatively low stellar temperature, the predicted depth of the secondary eclipse is very small, and indeed, we did not measure any significant secondary eclipse in our fits. The full list of fitted astrophysical parameters is given in Table \ref{tab:fit}. Due to the extremely grazing nature of the transit, we placed Gaussian priors on the quadratic limb-darkening coefficients. We interpolated the tabulated coefficient values from \citet{claret2018} to the effective temperature and surface gravity measurements reported in \citet{hip65} ($T_{\mathrm{eff}}=4590$ K, $\log g=4.611$) and set the width of the Gaussian to 0.05 for both coefficients: $(u_1,u_2) = (0.50\pm0.05,0.16\pm0.05)$, corresponding to $(\gamma_1,\gamma_2) = (1.16\pm0.11,0.18\pm0.11)$. 

\begin{figure}
\includegraphics[width=\linewidth]{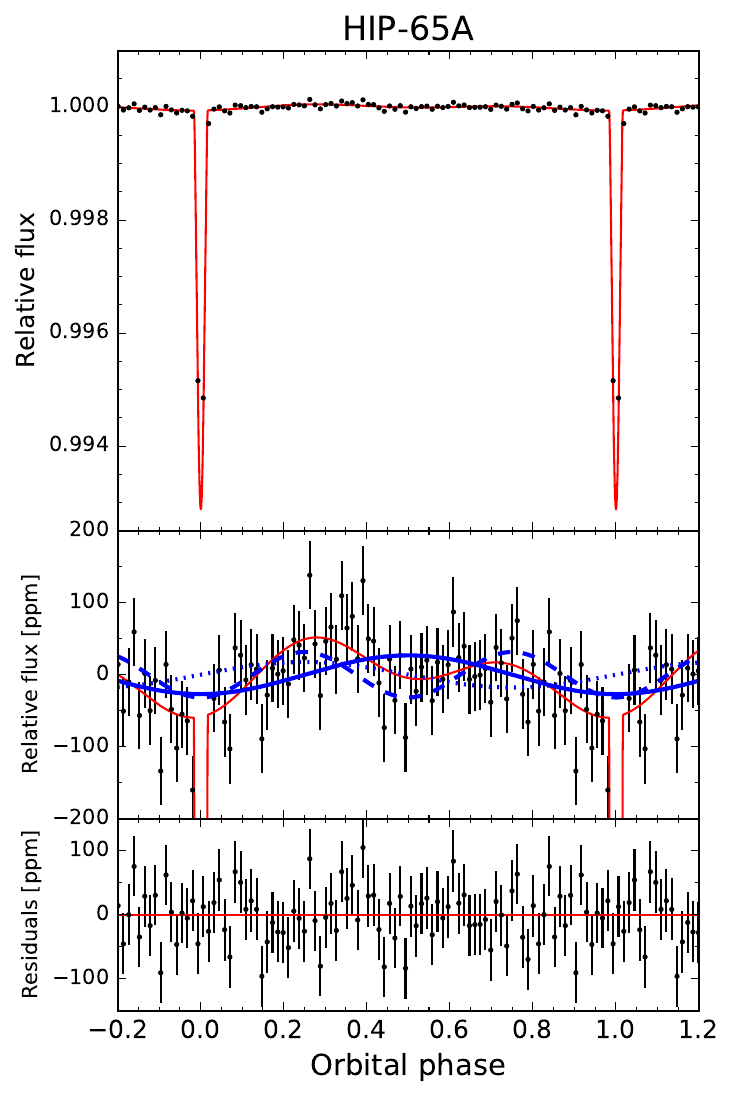}
\caption{Top: systematics-removed, phase-folded \tess\ light curve of HIP 65A, binned in 18 minute intervals (black points), along with the best-fit full phase curve model (red curve). Middle: zoomed-in view to show the orbital phase curve variations, relative to unity. We detected signals corresponding to the atmospheric brightness modulation ($A_{\mathrm{atm}}=29.6_{-9.1}^{+8.3}$ ppm), ellipsoidal distortion ($A_{\mathrm{ellip}}=28.4^{+9.6}_{-8.0}$ ppm), and Doppler boosting ($A_{\mathrm{Dopp}}=18.7^{+9.1}_{-8.4}$ ppm). These signals are shown separately by the solid, dashed, and dotted blue curves, respectively. The full results from our joint light-curve fit are listed in Table \ref{tab:fit}. Bottom: corresponding residuals from the best-fit phase curve model.}
\label{fig:hip65}
\end{figure}

The phase-folded, systematics-removed light curve and best-fit phase curve model are shown in Figure \ref{fig:hip65}. For this and all subsequent plots, the binning interval is chosen so that roughly 75 bins span the orbital period. In addition, the transit light curves, with all phase curve signals and systematics trends removed, are presented in the compilation plot in Figure \ref{fig:transits}. We report strong detections of phase curve amplitudes corresponding to the atmospheric brightness modulation and ellipsoidal distortion: $A_{\mathrm{atm}}=29.6_{-9.1}^{+8.3}$ ppm and $A_{\mathrm{ellip}}=28.4^{+9.6}_{-8.0}$ ppm. We also measured a weak Doppler boosting amplitude of $A_{\mathrm{Dopp}}=18.7^{+9.1}_{-8.4}$ ppm; including this term in the joint fit is not preferred by the BIC, while it yields a slight improvement to the AIC. In addition, we report marginal detections ($\Delta\mathrm{AIC}>0$, $\Delta\mathrm{BIC}>0$) of the planet's average relative brightness $\bar{f}_{p}$, from which we derive rough constraints on the dayside and nightside flux of HIP 65Ab. In Table \ref{tab:fit}, we list these and all other marginal/non-detections in parentheses.

\citet{hip65} carried out an independent fit for the phase curve components and obtained the following values: $A_{\mathrm{atm}}=57.5\pm4.7$ ppm, $A_{\mathrm{ellip}}=33.0\pm4.7$ ppm, and $A_{\mathrm{Dopp}}=15.4\pm4.5$ ppm. We note that their approach involved applying a polynomial spline to the light curve prior to fitting to remove long-term trends (without accounting for the momentum dumps and the associated flux ramps) and then phase-folding and binning the light curve, as opposed to our approach of fitting these trends simultaneously with the astrophysical model to the unbinned and unfolded light curve. Furthermore, their analysis did not consider and account for the effect of red noise on the calculated uncertainties as we have done (Section \ref{subsec:rednoise}). These are the primary reasons for the significantly smaller uncertainties on the phase curve amplitudes in their analysis. Furthermore, \citet{hip65} did not remove the two anomalous ramps in the vicinity of momentum dumps that we excised from the light curve prior to fitting (see Appendix A). When including both gravitationally induced phase curve terms in the astrophysical model, the fitted values for $A_{\mathrm{ellip}}$ and $A_{\mathrm{Dopp}}$ from our two analyses agree to well within $1\sigma$. Meanwhile, the atmospheric brightness modulation amplitudes differ by $2.7\sigma$. All other fitted astrophysical parameters agree to within $1\sigma$ with the values in \citet{hip65}.

A major difference in the data analysis methodologies employed by the two independent studies is that \citet{hip65} removed the primary transits from the light curve prior to fitting. When considering the relatively low signal-to-noise of the data as well as the significant stellar variability in the photometry of a few percent, the atmospheric brightness modulation is the most likely of the three phase curve components to be affected by transit removal, because trimming those data removes the minima of the associated photometric variation and may induce a systematic bias in the retrieved amplitude. To explore this possibility, we repeated our joint fit with the transits removed and obtained $A_{\mathrm{atm}}=35\pm11$ ppm. This value is consistent with the value derived from our fit including the transits ($29.6_{-9.1}^{+8.3}$ ppm), while being closer ($1.9\sigma$) to the value reported in \citet[$57.5\pm4.7$ ppm]{hip65}. 

This test illustrates that transit trimming and ramp removal, as well as the general treatment of systematics and stellar variability, can have notable consequences for the light-curve fits, particularly in datasets with inherently weak signals and relatively poor photometric precision. The \tess\ spacecraft will reobserve this system during the first year of the extended mission and will likely more than double the amount of available photometry, improving the precision and robustness of the phase curve fits.

\begin{figure}
\includegraphics[width=\linewidth]{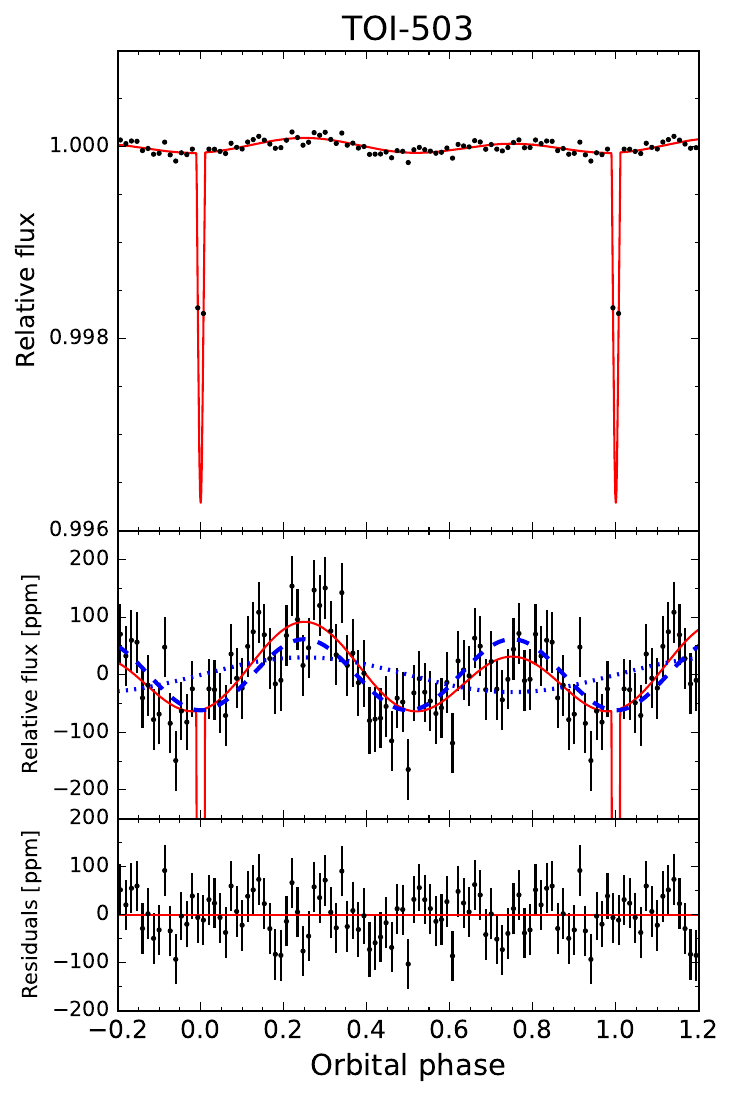}
\caption{Same as Figure \ref{fig:hip65}, but for TOI-503. The phase-folded light curve is binned in 70 minute intervals. Both ellipsoidal distortion and Doppler boosting phase curve signals were robustly detected in the \tess\ photometry, with measured semi-amplitudes of $61.6^{+9.4}_{-8.1}$ ppm and $30\pm10$ ppm, respectively. The individual components are shown in the middle panel by the dashed and dotted blue curves.}
\label{fig:toi503}
\end{figure}

\subsection{TOI-503}\label{toi503}
TOI-503b was the first brown dwarf discovered by the \tess\ mission \citep{subjak2019} and was observed during sector 7. This 54 $M_{\mathrm{Jup}}$ object, listed in the \tess\ Releases as TOI-129b, orbits a metallic-line 7650 K A-type star every 3.68 days. We detected strong phase curve signals corresponding to ellipsoidal distortion ($A_{\mathrm{ellip}}=61.6^{+9.4}_{-8.1}$ ppm) and Doppler boosting ($A_{\mathrm{Dopp}}=30\pm10$ ppm). Meanwhile, the predicted secondary eclipse depth is less than 20 ppm, and we did not measure significant values for $\bar{f_{p}}$ or $A_{\mathrm{atm}}$. The full list of fitted astrophysical parameters is given in Table \ref{tab:fit}.

For the other astrophysical parameters, which we allowed to vary unconstrained, we obtained values that agree with the measurements in \citet{subjak2019} to within $1.2\sigma$. The uncertainties in their work are up to four times smaller than ours because they incorporated other ground-based transit observations, radial velocities, the stellar spectral energy distribution, and Gaia parallax into a joint fit to better constrain the limb-darkening and orbital parameters. The light-curve plot from our phase curve analysis is provided in Figure \ref{fig:toi503}.


\begin{figure}
\includegraphics[width=\linewidth]{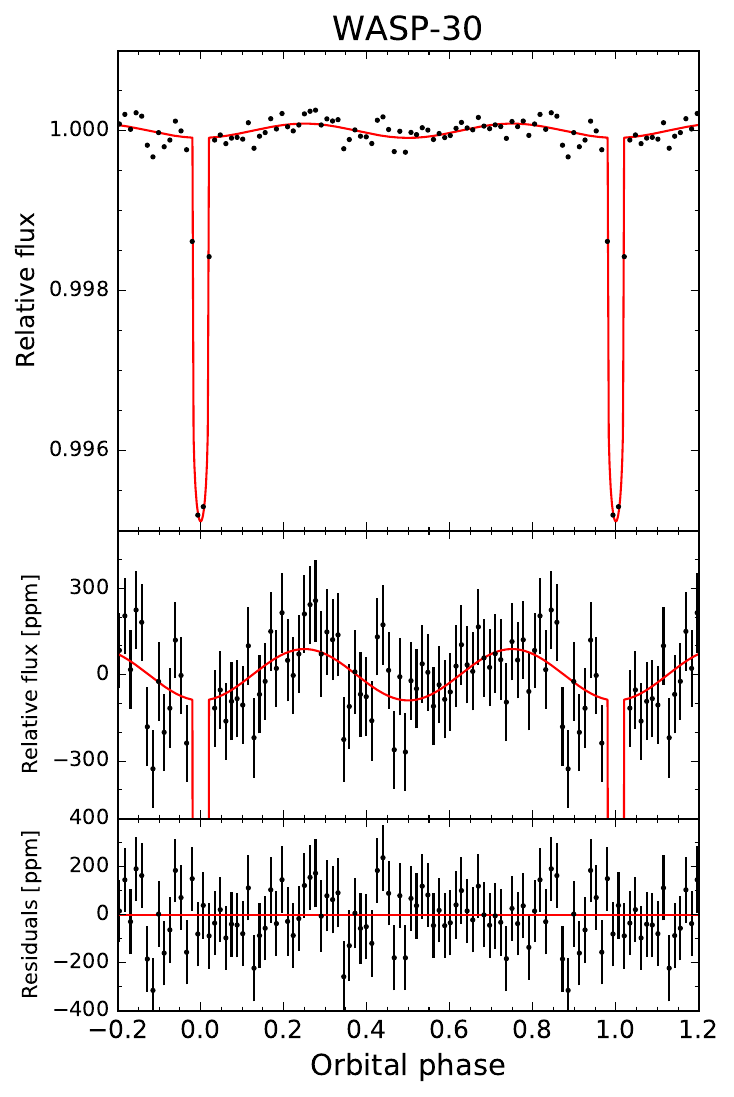}
\caption{Same as Figure \ref{fig:hip65}, but for WASP-30. The phase-folded light curve is binned in 80 minute intervals. The light curve shows a strong phase curve signal from the ellipsoidal distortion of the host star with a semi-amplitude of $97^{+24}_{-25}$ ppm.}
\label{fig:wasp30}
\end{figure}

\subsection{WASP-30}\label{wasp30}
WASP-30b is a 63 $M_{\mathrm{Jup}}$ 0.95 $R_{\mathrm{Jup}}$ brown dwarf orbiting an F8V star with an orbital period of 4.16 days \citep{wasp30,triaud2013}. The \tess\ spacecraft observed this system during sector 2. The only significant phase curve modulation detected was ellipsoidal distortion, and we measured a semi-amplitude of $97^{+24}_{-25}$ ppm. From separate joint fits, we obtained a $2\sigma$ upper limit on the Doppler boosting amplitude of 62 ppm and a marginal atmospheric brightness modulation signal with a semi-amplitude of $A_{\mathrm{atm}}=34^{+29}_{-26}$ ppm. In addition, we report weak constraints on the dayside and nightside fluxes of WASP-30b. Including any of these terms in the joint fit incurred significant increases to the BIC and no improvement to the AIC, indicating that the measured signals are not statistically robust. See Table \ref{tab:fit} for the full set of results. The corresponding best-fit phase curve model and systematics-corrected \tess\ light curve are shown in Figure \ref{fig:wasp30}.

From our PB analysis (Section \ref{subsec:rednoise}), we derived a new transit timing of $T_{0}=2458369.4316^{+0.0019}_{-0.0014}$ BJD$_{\mathrm{TDB}}$. Extrapolating the most recent literature ephemeris from \citet{triaud2013}, calculated using observations obtained in 2010, to the \tess\ epoch, we found that the transits occurred $\sim$38 minutes ($3.4\sigma$) later than predicted. This discrepancy most likely indicates significant ephemeris deprecation over the 8 yr that transpired between the \tess\ observations and the last published epoch. In Section \ref{subsec:ephem}, we provide an updated transit ephemeris for this system utilizing all previously published transit timing measurements.

This target was observed for only a single \tess\ sector and due to its relatively long orbital period, only six transits are contained in the light curve, resulting in relatively large uncertainties on the transit shape parameters. We obtained an impact parameter of $b=0.17^{+0.17}_{-0.12}$, which is consistent with the value from \citet{triaud2013}: $b=0.10^{+0.12}_{-0.10}$. The scaled semimajor axis values $a/R_{*}$ are also consistent: $8.42^{+0.16}_{-0.35}$ vs. $8.59^{+0.09}_{-0.18}$.

\subsection{WASP-72}\label{wasp72}
WASP-72b was discovered by \citet{wasp6472} using observations collected by the WASP southern station in 2006 and 2007. The host star is a bright ($V=9.6$ mag), moderately evolved F7-type star with $T_{\mathrm{eff}}=6250$ K, $M_{*}=1.4 M_{\Sun}$, and $R_{*}=2.0 R_{\Sun}$. The 1.5 $M_{\mathrm{Jup}}$ companion has an orbital period of 2.22 days, and subsequent measurements of the Rossiter--McLaughlin effect revealed a sky-projected spin-orbit angle of $-7^{+12}_{-11}$ deg \citep{addison2018}. \tess\ observed the WASP-72 system in sectors 3 and 4.

\begin{figure}
\includegraphics[width=\linewidth]{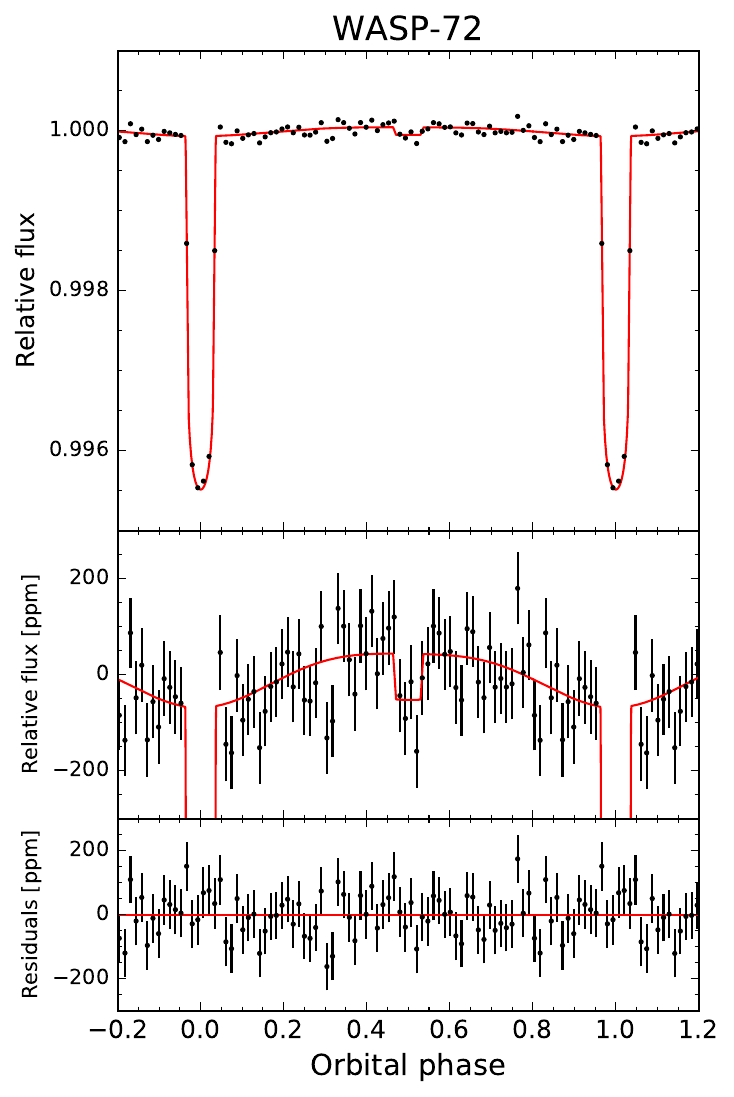}
\caption{Same as Figure \ref{fig:hip65}, but for WASP-72. The phase-folded light curve is binned in 43 minute intervals. Both the secondary eclipse ($D_{d}=113^{+39}_{-31}$ ppm) and the atmospheric brightness modulation ($A_{\mathrm{atm}}=70^{+15}_{-16}$ ppm) were robustly detected.}
\label{fig:wasp72}
\end{figure}

We did not detect a significant ellipsoidal distortion or Doppler boosting signal from the light-curve fits, and when deriving the results listed in Table \ref{tab:fit}, we applied Gaussian priors on both amplitudes based on the literature planet-to-star mass ratio (shown with square brackets in the table for this and all other similar targets). The light-curve fit is plotted in Figure \ref{fig:wasp72}. We measured a prominent $3.6\sigma$ secondary eclipse depth of $113^{+39}_{-31}$ ppm, along with a $4.4\sigma$ atmospheric brightness modulation signal with a semi-amplitude of $70^{+15}_{-16}$ ppm. When including a phase shift in the atmospheric component in the joint fit, we derived a $2\sigma$ upper limit on the eastward phase shift of $25^{\circ}$. The nightside flux is also consistent with zero at the $0.7\sigma$ level. Meanwhile, we obtained constraints on the transit shape parameters ($b=0.659^{+0.046}_{-0.070}$, $a/R_{*}=3.74^{+0.24}_{-0.18}$) and transit depth ($R_{p}/R_{*}=0.06446^{+0.00078}_{-0.00089}$) that significantly supersede the published values in \citet{wasp6472} and \citet{addison2018}, with uncertainties that are as much as three times smaller, while being consistent with the previous values to within $1\sigma$.

\subsection{WASP-100}\label{wasp100}
WASP-100b ($2.0 M_{\mathrm{Jup}}$, $1.7 R_{\mathrm{Jup}}$) orbits a solar-metallicity F2-type star ($1.6 M_{\Sun}$, $2.0 R_{\Sun}$, 6900 K) every 2.85 days \citep{wasp100}. Follow-up observations of the Rossiter--McLaughlin effect revealed that WASP-100b is on a nearly polar orbit ($\lambda=79^{+19}_{-10}$ deg; \citealt{addison2018}). Located at RA=04:35:50, decl.=$-$64:01:37, this target lies in the \tess\ CVZ and was observed during all 13 Southern Sky sectors, resulting in one year of nearly continuous photometry.

Due to the high volume of data, we did not use the uncorrected PDC light curves in the joint fit, because that would entail over 100 systematics parameters and incur forbiddingly large computational overheads. Instead, we followed a technique similar to the one utilized in \citet{wong2019koi964} and carried out smaller joint fits of each sector of \tess\ photometry, optimizing the orders of the detrending polynomials for all segments contained within the sector. We then divided out the best-fit systematics model from the corresponding segment light curves, before concatenating all of the detrended data segments together for the full 13-sector joint fits; in the final joint fits, no systematics model was included. 

The PDC light curves for this system are largely well behaved, with no significant variations due to stellar variability or uncorrected instrumental systematics. Therefore, the long-term trends could be removed prior to fitting without introducing any significant biases to the resulting fitted astrophysical parameters. As an empirical test of the reliability of this detrending method, we compared the phase curve parameters from the individual sector light-curve fits and found that the values are consistent with those from the joint fit to within roughly $2\sigma$.

\begin{figure}
\includegraphics[width=\linewidth]{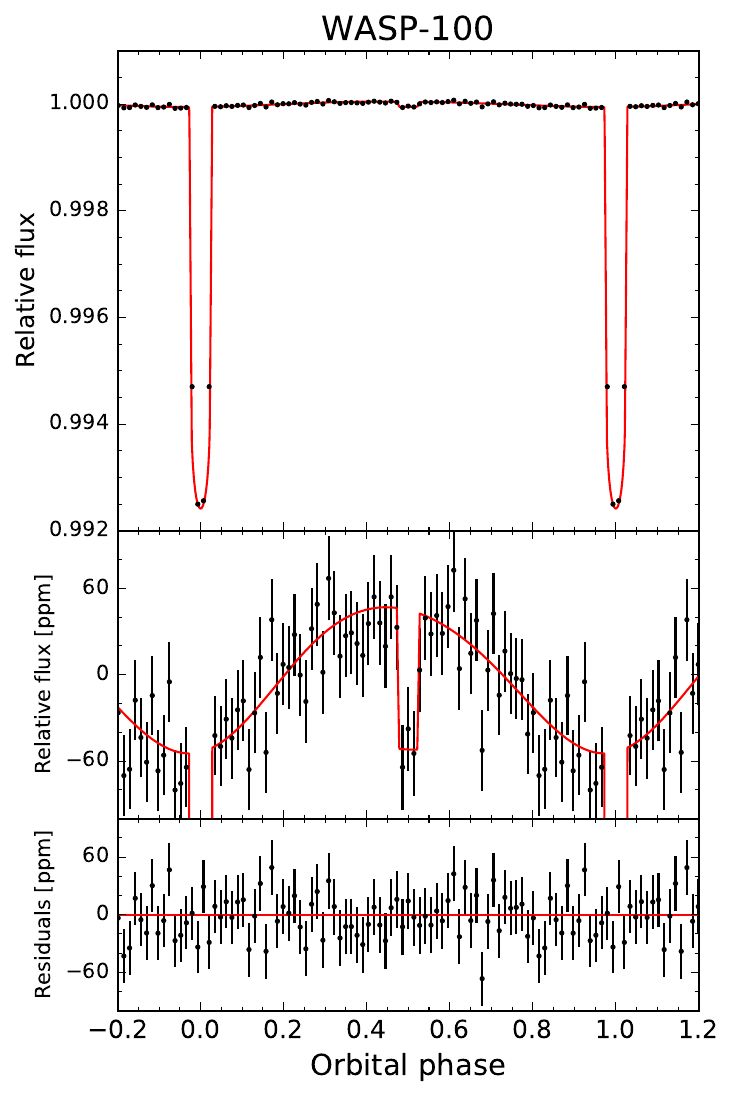}
\caption{Same as Figure \ref{fig:hip65}, but for WASP-100. The phase-folded light curve is binned in 56 minute intervals. Both the secondary eclipse ($D_{d}=97\pm17$ ppm) and atmospheric brightness modulation ($A_{\mathrm{atm}}=48.5^{+5.7}_{-5.4}$ ppm) are measured at high signal-to-noise. We also detect a slight phase shift in the atmospheric brightness modulation, with the location of maximum brightness on the dayside hemisphere shifted to the east by $\delta=12\overset{\circ}{.}0^{+6\overset{\circ}{.}3}_{-5\overset{\circ}{.}7}$.}
\label{fig:wasp100}
\end{figure}

The BIC is optimized for the model fit that includes only the secondary eclipse and atmospheric brightness modulation components; meanwhile, the AIC strongly favors the addition of a phase shift in the atmospheric brightness modulation ($\Delta\mathrm{AIC}=-7.2$ relative to the fit without phase shift). We have chosen to present the results from our lowest-AIC fit in Table \ref{tab:fit} and Figure \ref{fig:wasp100}, primarily to aid in comparison with the results of \citet[see below]{jansen2020}. 

The long-baseline photometry of the WASP-100 system yielded exquisite updated measurements of the transit ephemeris, transit shape, transit depth, and limb-darkening coefficients. All of these agree with the previous literature values \citep{wasp100} to well within $1\sigma$, with the exception of the mid-transit time $T_{0}$: our measurement (from the PB uncertainty analysis) is 7 minutes ($1.4\sigma$) earlier than the extrapolated transit timing from the most recent published ephemeris \citep{addison2018}. We measured a $5.5\sigma$ secondary eclipse depth of $97\pm17$ ppm and detected a very significant $9.0\sigma$ atmospheric brightness modulation phase curve component with a semi-amplitude of $48.5^{+5.7}_{-5.4}$ ppm. The derived nightside flux is consistent with zero, and we obtained a slight phase shift in the brightness modulation of $\delta=12\overset{\circ}{.}0^{+6\overset{\circ}{.}3}_{-5\overset{\circ}{.}7}$. As in the case of WASP-72, both ellipsoidal distortion and Doppler boosting amplitudes were constrained by Gaussian priors.

\subsubsection{Comparison with \citet{jansen2020}}\label{jansen}
\citet{jansen2020} presented an independent phase curve analysis of the full 13-sector WASP-100 dataset. Their fits of the secondary eclipses alone yielded a depth of $100\pm14$ ppm, consistent with our result to well within $1\sigma$. From fitting the phase curve, they reported an eastward offset in the region of maximum dayside brightness of $28^{\circ}\pm9^{\circ}$, in broad agreement with our measured value $\delta=12\overset{\circ}{.}0^{+6\overset{\circ}{.}3}_{-5\overset{\circ}{.}7}$.

\citet{jansen2020} reported a peak-to-peak brightness variation of $73\pm9$ ppm, whereas we measured $2A_{\mathrm{atm}}=97\pm11$ ppm, roughly $1.7\sigma$ larger. We note that in their phase curve analysis, they removed both the primary transit and the secondary eclipse (roughly $20\%$ of the time series altogether). Additionally, they phase-folded and binned the data and detrended the light curves prior to fitting, similar to the approach used in the analysis of the HIP 65A phase curve in \citet{hip65}. Trimming away the data points spanning the two conjunctions removes the regions near both the maxima and the minima of the characteristic atmospheric brightness modulation, and, as discussed in Section \ref{hip65}, this may affect the measured amplitude and phase shift, particularly given the relatively low signal-to-noise of the \tess\ photometry for this system.

Follow-up studies of WASP-100 using data from the \tess\ extended mission will help better constrain the phase shift in the atmospheric brightness component and more robustly measure the day--night brightness contrast.

\begin{figure}
\includegraphics[width=\linewidth]{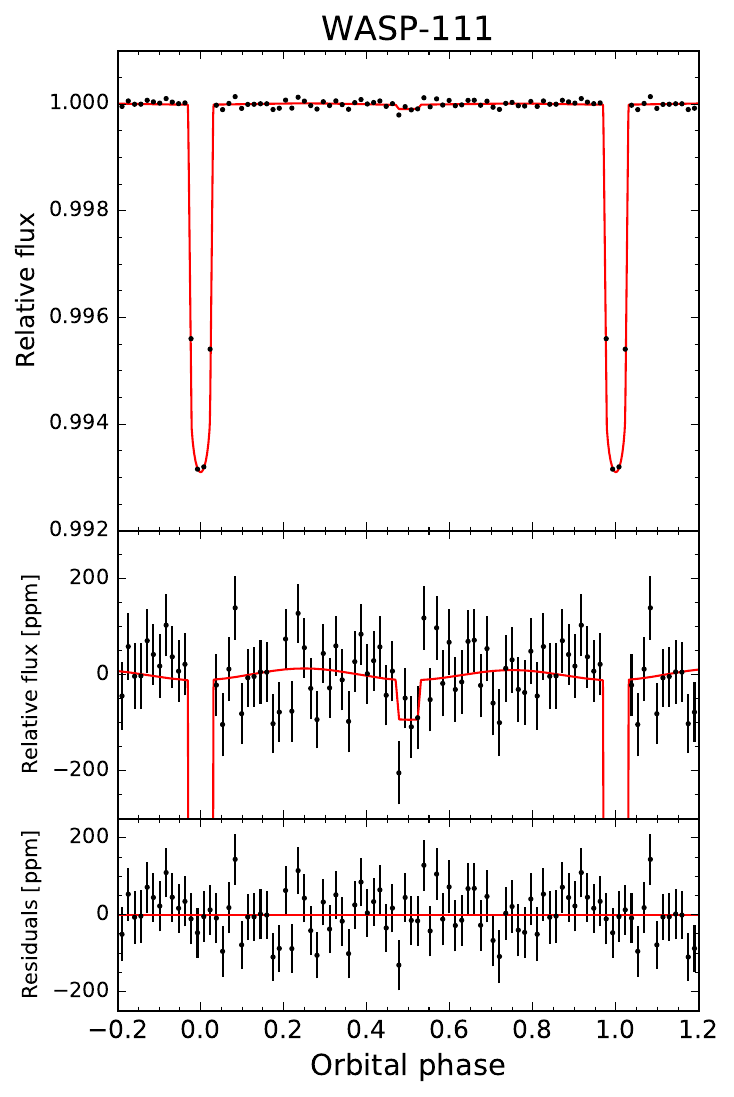}
\caption{Same as Figure \ref{fig:hip65}, but for WASP-111. The phase-folded light curve is binned in 44 minute intervals. The only significant signal detected in our fit is the secondary eclipse, which has a measured depth of $102^{+38}_{-37}$ ppm.}
\label{fig:wasp111}
\end{figure}

\begin{figure}
\includegraphics[width=\linewidth]{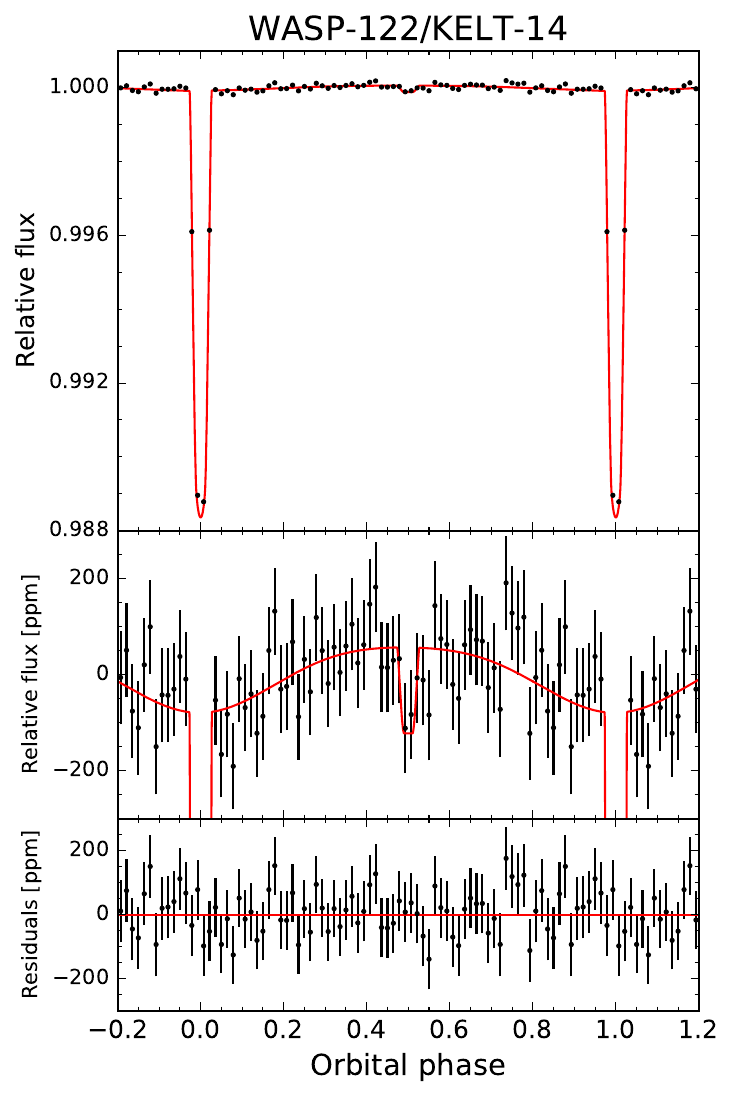}
\caption{Same as Figure \ref{fig:hip65}, but for WASP-122/KELT-14. The phase-folded light curve is binned in 33 minute intervals. Both the secondary eclipse ($D_{d}=165^{+63}_{-60}$ ppm) and atmospheric brightness modulation ($A_{\mathrm{atm}}=64\pm17$ ppm) were detected in our phase curve fit.}
\label{fig:wasp122}
\end{figure}

\subsection{WASP-111}\label{wasp111}
WASP-111b is a 1.8 $M_{\mathrm{Jup}}$, 1.4 $R_{\mathrm{Jup}}$ hot Jupiter that orbits its $V=10.3$ mag 6400 K F5 host star every 2.31 days \citep{wasp111}. Measurements of the Rossiter--McLaughlin effect indicate a prograde orbit with a sky-projected stellar obliquity of $\lambda=-5^{\circ}\pm16^{\circ}$. \tess\ observed the WASP-111 system during sector 1. No significant phase curve components were detected in our joint fits, and we placed literature-derived priors on the ellipsoidal distortion and Doppler boosting amplitudes. A marginal $2.8\sigma$ secondary eclipse depth of $102^{+38}_{-37}$ ppm was measured; we also placed a $2\sigma$ upper limit of 31 ppm on the atmospheric brightness modulation semi-amplitude. We obtained improved values for the transit depth and transit shape parameters that agree with the previously published results from the discovery paper to within $1\sigma$. The full list of fitted parameters is given in Table \ref{tab:fit}; the phase-folded, systematics-corrected light curve and best-fit model are shown in Figure \ref{fig:wasp111}.

\subsection{WASP-122/KELT-14}\label{wasp122}
This 1.2 $M_{\mathrm{Jup}}$, 1.5 $R_{\mathrm{Jup}}$ hot Jupiter on a 1.71 day orbit was discovered independently by the WASP consortium (as WASP-122b; \citealt{wasp122}) and the Kilodegree Extremely Little Telescope survey (as KELT-14b; \citealt{kelt14}). The metal-rich G4-type host star has $T_{\mathrm{eff}}=5800$ K, $M_{*}=1.2 M_{\Sun}$, and $R_{*}=1.2 R_{\Sun}$. Our analysis of the \tess\ sector 7 light curve revealed a $2.8\sigma$ secondary eclipse depth of $165^{+63}_{-60}$ ppm and a $3.8\sigma$ atmospheric brightness modulation semi-amplitude of $64\pm17$ ppm. As in the case of WASP-72b and WASP-100b, the corresponding nightside flux is consistent with zero, and no significant phase shift in the brightness modulation was detected ($|\delta|<34^{\circ}$ at $2\sigma$). All other fitted transit ephemeris and transit shape parameters are consistent to better than the $1.5\sigma$ level with the results of \citet{wasp122} and \citet{kelt14}, with comparable relative uncertainties. See Table \ref{tab:fit} and Figure \ref{fig:wasp122}.

\subsection{Reanalysis of WASP-18, WASP-19, and WASP-121}\label{previous}

\begin{figure*}
\includegraphics[width=\linewidth]{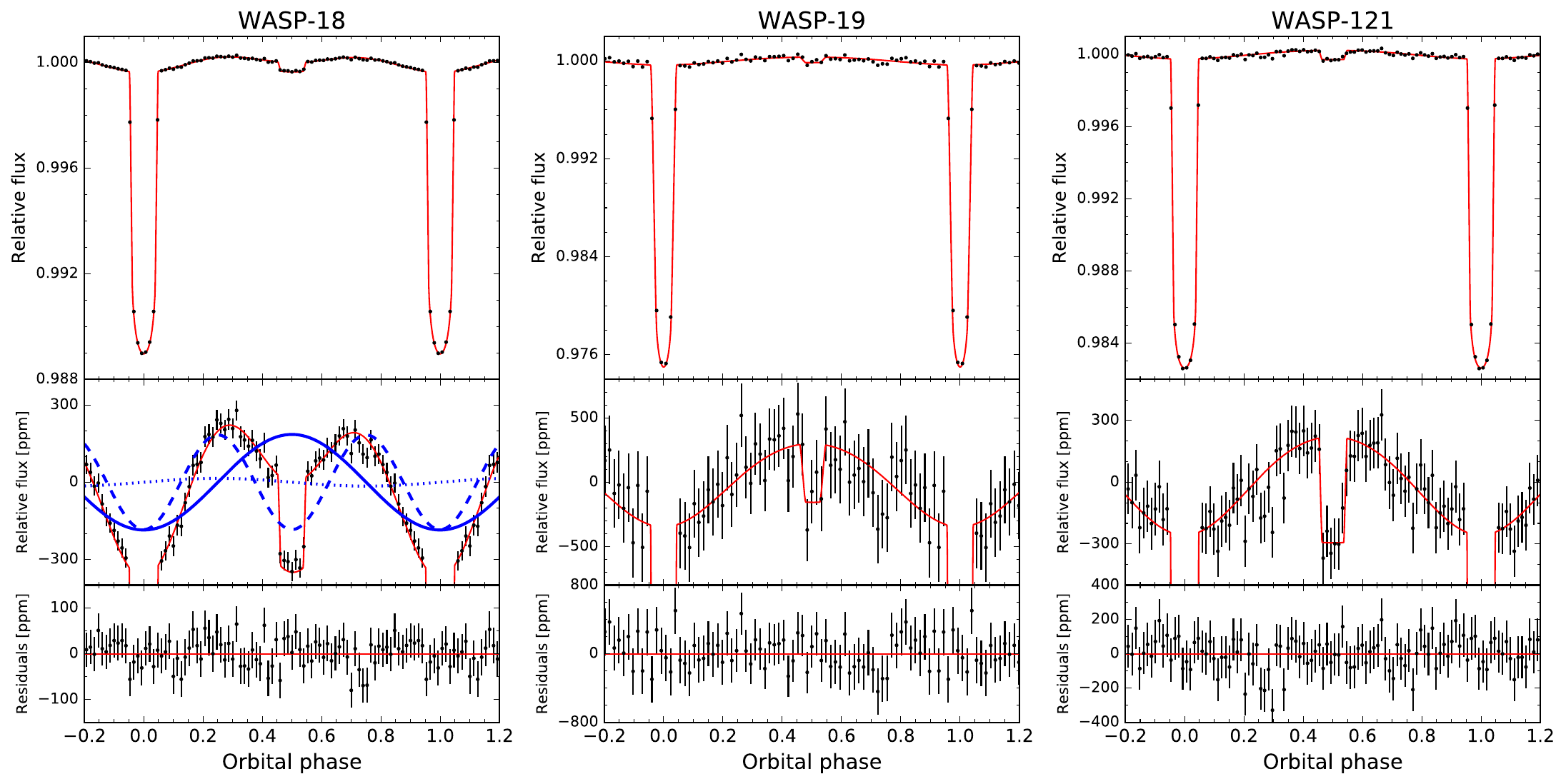}
\caption{Same as Figure \ref{fig:hip65}, but from our reanalysis of WASP-18, WASP-19, and WASP-121. The light curves are binned in 18, 15, and 24 minute intervals, respectively. WASP-19b and WASP-121b show strong atmospheric brightness modulation signals and high signal-to-noise secondary eclipses. Meanwhile, WASP-18 displays phase curve signals attributed to all three processes: atmospheric brightness modulation (solid blue curve in middle panel), ellipsoidal distortion (dashed curve), and Doppler boosting (dotted curve).}
\label{fig:previous}
\end{figure*}

We reanalyzed the WASP-18, WASP-19, and WASP-121 light curves using the same uniform analysis methodology that we applied to all of the other targets from the first year of the \tess\ mission. Phase curve analyses of these systems were previously published in \citet{shporer2019}, \citet{wong2019wasp19}, \citet{bourrier2019}, and \citet{daylan2019}. WASP-18 was observed by \tess\ in sectors 2 and 3 and consists of a massive 10.4 $M_{\mathrm{Jup}}$ hot Jupiter on a 0.94 day orbit around a $T_{\mathrm{eff}}=6400$ K F-type star \citep{wasp18}. With one of the shortest orbital periods of any gas giant exoplanet hitherto discovered, WASP-19b lies on a 0.789 day orbit around its G-type host star \citep{wasp19} and is situated in sector 9. Observations of WASP-121b ($M_{p}=1.2$ $M_{\mathrm{Jup}}$, $R_{p}=1.8$ $R_{\mathrm{Jup}}$; \citealt{wasp121}) took place during sector 7.

The prior analyses of the WASP-18 and WASP-19 phase curves were carried out using a very similar data processing and analysis framework to the one described in this paper. As such, only minor alterations were needed, primarily in the handling of red noise and by allowing the limb-darkening coefficients to vary freely. Meanwhile, the particular treatments of systematics in the WASP-121 light curve employed by \citet{bourrier2019} and \citet{daylan2019} differ considerably from our technique of simultaneous polynomial detrending: \citet{bourrier2019} pre-cleaned the photometry using median detrending with a window length equal to one planetary orbit, while \citet{daylan2019} applied a cubic spline across each spacecraft orbit's worth of data prior to fitting the light curve. Furthermore, while \citet{daylan2019} directly modeled the flux variations using sinusoidal terms, similar to our present approach, \citet{bourrier2019} fit the photometry to a physical model of the planet's temperature distribution.

The results of our reanalysis of these three systems are presented in Table \ref{tab:fit} and Figure \ref{fig:previous}. For WASP-18, we detected phase curve signals corresponding to all three major processes --- atmospheric brightness modulation ($A_{\mathrm{atm}}=181.7^{+8.2}_{-8.1}$ ppm), ellipsoidal distortion ($A_{\mathrm{ellip}}=181.9^{+8.6}_{-8.4}$ ppm), and Doppler boosting ($A_{\mathrm{Dopp}}=20.0^{+6.4}_{-6.7}$ ppm) --- as well as a secondary eclipse depth of $334\pm20$ ppm. All four measurements are consistent with the published values in \citet{shporer2019} to well within $1\sigma$. Meanwhile, most of the other astrophysical parameters agree to within $1\sigma$, with the exception of the transit depth: the value for $R_{p}/R_{*}$ we obtained from allowing the limb-darkening coefficients to vary is $1.4\sigma$ higher than the measurement in \citet{shporer2019}, derived from fixing the coefficients to the tabulated values in \citet{claret2018}.

We measured a secondary eclipse depth of $470^{+130}_{-110}$ ppm and an atmospheric brightness modulation semi-amplitude of $311^{+57}_{-54}$ ppm for WASP-19b, statistically identical to the corresponding values from \citet{wong2019wasp19} ($473^{+131}_{-106}$ ppm and $319^{+50}_{-52}$ ppm). We did not retrieve a statistically significant phase shift in the phase curve signal and placed a $2\sigma$ upper limit of $17^{\circ}$ on an eastward offset in the dayside hotspot.

Our results from the joint fit of the WASP-121 light curve are largely consistent with both sets of previously published values. For the secondary eclipse, we measured a depth of $486\pm59$ ppm, while \citet{bourrier2019} and \citet{daylan2019} reported $419^{+47}_{-41}$ ppm and $534^{+43}_{-42}$ ppm, respectively. The atmospheric brightness modulation amplitude we obtained is $214^{+25}_{-26}$ ppm, as compared to $178\pm17$ ppm and $223\pm17$ ppm. None of the analyses indicate any significant phase shift in the phase curve modulation; we found weak evidence for a small westward offset of $9\overset{\circ}{.}7\pm5\overset{\circ}{.}7$. The broad agreement between these three independent phase curve analyses provides an excellent consistency check for the respective methodologies. The notably higher uncertainties on our values are due to the presence of significant red noise on roughly hour-long timescales, which was not addressed in the previously published analyses. As a result, following the technique described in Section \ref{subsec:rednoise}, the per-point uncertainty was inflated by $52\%$ in our final fit.

\section{Discussion}\label{sec:dis}

\subsection{Mutual Gravitational Interaction}\label{subsec:grav}
We detected phase curve signals attributed to ellipsoidal distortion and Doppler boosting in four systems: HIP 65A (Section \ref{hip65}), TOI-503 (Section \ref{toi503}), WASP-18 (Section \ref{previous}), and WASP-30 (Section \ref{wasp30}). Comparisons of the observed amplitudes from our light-curve analysis with the predictions from theoretical modeling provide an empirical test of the commonly used simple models for describing those phenomena. For the remaining systems with detected phase curve signals, we placed Gaussian priors on the respective amplitudes; in those cases, the posteriors are identical to the prior constraints and are listed in square brackets in Table \ref{tab:fit}.

For ellipsoidal distortion signals, we combined measurements of $a/R_{*}$ and $i$ derived from our phase curve analysis with literature values for the mass ratio $M_{p}/M_{*}$ to compute the predicted modulation semi-amplitudes using Equation \eqref{ellip}. For the limb-darkening and gravity-darkening coefficients in the pre-factor $\alpha_{\mathrm{ellip}}$, we took the tabulated values from \citet{claret2017} for the nearest available combination of stellar parameters. To propagate the uncertainties on the system parameters to the prediction, we used a standard Monte Carlo sampling method and drew values from the individual Gaussian distributions based on the measurement uncertainties.

For HIP 65A, the predicted ellipsoidal distortion amplitude is $39\pm3$ ppm, which is consistent with the measured amplitude of $28.4^{+9.6}_{-8.0}$ ppm at the $1.1\sigma$ level. The measured effective temperature of TOI-503 is $7650\pm160$ K and lies in a region where the model-generated gravity-darkening coefficients vary significantly and non-monotonically across a relatively narrow range of temperatures. To account for the large corresponding uncertainty in $g$, we used the range of values for temperatures spanning 7500--7800 K: $0.1174<g<0.5684$. The resultant spread in the predicted ellipsoidal distortion amplitudes is 57--133 ppm. Our measured amplitude of $61.6^{+9.4}_{-8.1}$ ppm agrees well with the lower end of this range. The predicted ellipsoidal distortion amplitude for WASP-30 is $98^{+10}_{-9}$ ppm when using the values for $b$ and $a/R_{*}$ measured in our fits (Table \ref{tab:fit}); when assuming the more precise literature values from \citet{wasp30}, the predicted amplitude is $92\pm5$ ppm. Both of these predictions are consistent with our measured amplitude of $97^{+24}_{-25}$ ppm at better than the $1\sigma$ level.

The theoretical model for the Doppler boosting signal is given in Equation \eqref{doppler}. Assuming the host star is a blackbody and substituting in the expression for the RV semi-amplitude for a circular orbit, we can analytically calculate the pre-factor $\alpha_{\mathrm{Dopp}}$ (defined in Equation \eqref{alphadopp}) and express the Doppler boosting amplitude as \citep[e.g.,][]{shporer2017}
\begin{equation}\label{doppsimp}
A_{\mathrm{Dopp}}=\sin i \left\lbrack\frac{2\pi G} {Pc^{3}}\frac{q^2M_p}{(1+q)^2}\right\rbrack^{\frac{1}{3}}\left\langle \frac{xe^{x}}{e^{x}-1}\right\rangle_{\mathrm{TESS}},
\end{equation}
where $x\equiv hc/kT_{\mathrm{eff}}\lambda$, $q\equiv M_{p}/M_{*}$, and the term in the angled brackets is integrated over the \tess\ bandpass. Using the values for $M_{p}$, $M_{*}$, and $T_{\mathrm{eff}}$ from the literature and the values for $P$ we obtained in our \tess\ light-curve fits, we calculated predicted Doppler boosting amplitudes of  $10.3\pm0.3$ ppm, $41\pm1$ ppm, and $71\pm2$ ppm for HIP 65A, TOI-503, and WASP-30, respectively. For HIP 65A, we measured a marginal Doppler boosting signal with $A_{\mathrm{Dopp}}=18.7^{+9.1}_{-8.4}$ ppm, consistent with the predicted amplitude. Likewise for TOI-503, our measured amplitude of $30\pm10$ ppm agrees with the prediction at the $1.1\sigma$ level.

For WASP-30, we did not detect a statistically significant Doppler boosting signal. Our $2\sigma$ upper limit on the semi-amplitude is comparable to the predicted signal of 71 ppm. Taking into account breaks in data collection between spacecraft orbits and data points flagged by the SPOC pipeline, the \tess\ light curve of WASP-30 contains only four uninterrupted orbital cycles (transit to transit). The inclusion of more data, e.g., when the system is revisited during the \tess\ extended mission, will help better constrain the amplitude of this phase curve component and more robustly determine whether the predicted Doppler boosting signal is present in the photometry.

The comparison between the theoretical and measured amplitudes for the ellipsoidal distortion and Doppler boosting signals was previously done in the published analysis of the WASP-18 phase curve \citep{shporer2019}. Using the new measurements from our updated analysis of this system, we likewise find excellent agreement between the fitted values ($A_{\mathrm{ellip}}=181.9^{+8.6}_{-8.5}$ ppm, $A_{\mathrm{Dopp}}=20.0^{+6.4}_{-6.7}$ ppm) and predictions ($173\pm15$ ppm and $18\pm2$ ppm, respectively).

As mentioned in Section \ref{subsec:model}, there is a degeneracy between Doppler boosting and a phase shift in the atmospheric brightness modulation. There are two systems --- HIP 65A and WASP-18 --- where Doppler boosting and atmospheric brightness modulation were simultaneously detected at least marginally. By subtracting the predicted Doppler boosting signal from the overall phase curve variation at the fundamental of the orbital phase, we can place constraints on any phase shift in the atmospheric brightness modulation. For HIP 65A and WASP-18, we find $2\sigma$ upper limits on the eastward phase shift of $33^{\circ}$ and $4^{\circ}$, respectively.

We can also apply the same technique described above to retrieve the companion mass from the measured ellipsoidal distortion and Doppler boosting amplitudes. Using Equations \eqref{ellip} and \eqref{doppsimp}, we obtained the following photometric mass constraints for HIP 65Ab, TOI-503b, WASP-18b, and WASP-30b: $2.64\pm0.71$, $28-53$, $9.8\pm1.1$, and $60\pm17$ $M_{\mathrm{Jup}}$. The RV-determined masses from the literature are $3.213\pm0.078$, $53.7\pm1.2$, $10.30\pm0.69$, and $60.96\pm0.89$ $M_{\mathrm{Jup}}$, respectively. The constraints we derive from the phase curve amplitudes (also listed in Table \ref{tab:fit}) are significantly weaker than the RV mass measurement precision in most cases, though consistent to within the error bars.

For the four systems analyzed in this paper with significant phase curve contributions from the mutual star--companion gravitational interaction, the measured signals are all broadly consistent with theoretical predictions. In contrast, the published phase curve of KELT-9 shows an anomalous ellipsoidal distortion signal that is shifted in time relative to the expected phase, indicating possible secondary effects due to the stellar dynamical tide excited by the orbiting planet or additional contributions to the overall photometric modulation at the first harmonic of the orbital phase due to time-varying stellar illumination across the planet's near-polar orbit \citep{wong2019kelt9}. 

Looking more broadly at phase curve studies from the Kepler era, we find numerous systems for which the measured ellipsoidal distortion signals diverge from the corresponding theoretical predictions. Examples include KOI-964, a transiting hot white dwarf system for which the theoretical model underestimates the observed ellipsoidal distortion amplitude \citep{wong2019koi964}, and KOI-74, another transiting white dwarf system where the opposite deviation in the measured ellipsoidal distortion signal was reported \citep[e.g.,][]{vankerkwijk2010, ehrenreich2011, bloemen2012}. Many of these discrepant systems, including KELT-9, KOI-74, and KOI-964, contain hot primary stars, suggesting that the non-convective nature of the hosts may be affecting the tidal response to the orbiting companion's gravitational potential. Detailed numerical modeling of hot stars  \citep[e.g.,][]{pfahl2008,burkart2012} has revealed that the dynamical tide can induce significant deviations in the observed ellipsoidal distortion signal from the classical theoretical predictions (i.e., Equation \eqref{ellip}) that assume only the equilibrium tide.

TOI-503 joins the list of binary systems with hot primary stars that have detected phase curve modulations due to ellipsoidal distortion and Doppler boosting. Future follow-up spectroscopic studies of the host star can yield improved measurements of the stellar effective temperature, which in turn translate into better constraints on the limb- and gravity-darkening coefficients, and more precise estimates of the gravitational phase curve amplitudes. Combining these results with additional \tess\ photometry from the extended mission will allow us to determine whether this system's phase curve variability deviates significantly from theoretical predictions.


\subsection{Brightness Temperatures and Albedos}\label{subsec:temp}

\begin{deluxetable*}{lccccccccc}
\tablewidth{0pc}
\tabletypesize{\scriptsize}
\tablecaption{
    Dayside Brightness Temperatures and Optical Geometric Albedos Derived from \tess, Kepler, CoRoT, and Spitzer/IRAC 3.6 and 4.5 $\mu$m Secondary Eclipses
    \label{tab:temps}
}
\tablehead{
    \colhead{Planet} &
    \colhead{$D_{d,\mathrm{vis}}$ (ppm)\tablenotemark{\scriptsize$\mathrm{a}$}}                     &
    \colhead{$D_{d,3.6}$ (ppm)\tablenotemark{\scriptsize$\mathrm{a}$}}                     &
    \colhead{$D_{d,4.5}$ (ppm)\tablenotemark{\scriptsize$\mathrm{a}$}} &
    \colhead{$T_{b,\mathrm{vis}}$ (K)\tablenotemark{\scriptsize$\mathrm{b}$}} & 
    \colhead{$T_{b,3.6}$ (K)\tablenotemark{\scriptsize$\mathrm{b}$}} & 
    \colhead{$T_{b,4.5}$ (K)\tablenotemark{\scriptsize$\mathrm{b}$}} & 
    \colhead{$T_{\mathrm{day}}$ (K)}  &
    \colhead{$A_{g}$} &
    \colhead{Ref.}
}
\startdata
WASP-4b & $120^{+80}_{-70}$ & $3190\pm310$ & $3430\pm270$ & $1600^{+600}_{-1200}$ & $2050\pm110$ & $1880\pm90$ & $1957\pm68$ & $<0.26$\tablenotemark{\scriptsize$\mathrm{c}$} & 1, 2\\
WASP-5b  & $31_{-55}^{+73}$ & $1970\pm280$ & $2370\pm240$ & $1200\pm800$ & $2050^{+140}_{-130}$ & $2000\pm120$ & $2000\pm90$ & $<0.30$\tablenotemark{\scriptsize$\mathrm{c}$} & 1, 3\\
WASP-18b & $339\pm21$  & $3037\pm62$ & $4033\pm97$ & $2970\pm40$ & $2960\pm70$ & $3220\pm90$ & $3029\pm50$ & $< 0.03$\tablenotemark{\scriptsize$\mathrm{c}$} & 1, 4\\
WASP-19b & $470^{+130}_{-110}$ & $5015\pm175$\tablenotemark{\scriptsize$\mathrm{d}$} & $5343\pm318$\tablenotemark{\scriptsize$\mathrm{d}$} & $ 2500^{+110}_{-160}$ & $2260\pm50$ & $2070\pm80$ & $2219^{+44}_{-43}$ & $0.16\pm0.07$ & 1, 4\\
WASP-36b & $90_{-70}^{+100}$  & $914\pm578$ & $1953\pm544$ & $1500^{+700}_{-1100}$ & $1200^{+400}_{-700}$ & $1480\pm210$ & $1440^{+150}_{-160}$ & $0.16^{+0.16}_{-0.15}$ & 1, 4\\
WASP-43b & $170\pm70$ & $3773\pm138$ & $3866\pm195$ & $1800^{+200}_{-1000}$ & $1720\pm60$ & $1510\pm60$ & $1666\pm48$ & $0.12\pm0.06$ & 1, 4\\
WASP-46b & $230_{-110}^{+140}$ & $1360\pm701$ & $4446\pm589$ & $2000^{+400}_{-1400}$ & $1300^{+300}_{-700}$ & $2090\pm160$ & $1870^{+130}_{-120}$ & $0.37^{+0.27}_{-0.26}$ & 1, 4\\
WASP-64b & $230_{-110}^{+130}$ & $2859\pm270$ & $2071\pm471$ & $2100^{+400}_{-1400}$ & $2110\pm100$ & $1610\pm180$ & $1989^{+87}_{-88}$ & $0.36\pm0.26$ & 1, 4\\
WASP-77Ab & $53_{-22}^{+32}$ & $2016\pm94$ & $2487\pm127$ & $1800^{+300}_{-1200}$ & $1870\pm40$ & $1800\pm50$ & $1842^{+34}_{-33}$ & $0.05\pm0.05$ & 1, 4\\
WASP-78b & $210_{-90}^{+100}$ & $2001\pm218$ & $2013\pm351$ & $2540^{+320}_{-1330}$ & $2620\pm160$ & $2380\pm270$ & $2560\pm130$ & $<0.56$\tablenotemark{\scriptsize$\mathrm{c}$} & 1, 4\\
WASP-100b & $94\pm17$ & $1267\pm98$ & $1720\pm119$ & $2620^{+80}_{-100}$ & $2290^{+90}_{-100}$ & $2430\pm100$ & $2356^{+69}_{-67}$ & $0.20\pm0.08$ & 1, 4\\
WASP-121b &  $486\pm59$ & $3685\pm114$ &  $4684\pm121$ & $2920\pm70$ & $2550\pm50$ & $2640\pm60$ & $2596\pm43$ & $0.24\pm0.06$ & 1, 4\\
\hline
CoRoT-1b & $160\pm60$ & $4150\pm420$ & $4820\pm420$ & $2400^{+200}_{-1000}$ & $2310\pm130$ & $2240\pm130$ & $2277\pm94$ & $0.10\pm0.08$ & 5, 6\\
CoRoT-2b & $60\pm20$ & $3550\pm200$ & $5000\pm200$ & $2010^{+100}_{-270}$ & $1810\pm50$ & $1850\pm50$ & $1831^{+34}_{-33}$ & $0.07\pm0.03$ & 6, 7 \\
HAT-P-7b & $71.2^{+1.9}_{-2.2}$ & $1560\pm90$ & $1900\pm60$ & $2800\pm30$ & $2630\pm90$ & $2680\pm60$ & $2666^{+46}_{-47}$ & $0.06\pm0.02$ & 8, 9 \\
Kepler-5b & $18.6^{+5.1}_{-5.3}$ & $1030\pm170$ & $1070\pm150$ & $2290^{+90}_{-160}$ & $2070\pm160$ & $1900^{+140}_{-150}$ & $1970\pm100$ & $0.09\pm0.04$ & 8, 10 \\
Kepler-6b & $11.1^{+4.8}_{-5.3}$ & $690\pm270$ & $1510\pm190$ & $1900^{+200}_{-1100}$ & $1500^{+200}_{-400}$ & $1810^{+110}_{-120}$ & $1728\pm98$ & $0.05\pm0.03$ & 8, 10 \\
Kepler-7b & $39\pm11$ & $164\pm150$ & $367\pm221$ & $2410^{+90}_{-170}$ & $900^{+400}_{-600}$ & $1100^{+300}_{-600}$ & $990^{+210}_{-530}$ & $0.24^{+0.07}_{-0.08}$ & 8, 11 \\
Kepler-12b & $20.2^{+8.3}_{-7.6}$ & $1370\pm200$ & $1160\pm310$ & $2000^{+200}_{-1100}$ & $1670\pm100$ & $1350\pm150$ & $1554^{+79}_{-80}$ & $0.09\pm0.04$ & 8, 12\\
KOI-13b & $173.7\pm1.8$ & $1560\pm310$ & $2220\pm230$ & $3420\pm70$ & $2600\pm300$ & $2850\pm210$ & $2770\pm170$ & $0.35^{+0.04}_{-0.05}$ & 8, 13 \\
TrES-2b & $7.7^{+2.4}_{-2.6}$ & $1270\pm210$ & $2300\pm240$ & $1820^{+70}_{-280}$ & $1550\pm100$ & $1720\pm90$ & $1629^{+62}_{-63}$ & $0.02\pm0.01$ & 8, 14 \\
\enddata
\textbf{Notes.}
\vspace{-0.25cm}\tablenotetext{\textrm{a}}{Secondary eclipse depths measured in the \tess/Kepler/CoRoT bandpass and the 3.6 and 4.5 $\mu$m Spitzer/IRAC bandpasses, in parts-per-million. All of the \tess\ targets are listed in the first half of the table.}
\vspace{-0.25cm}\tablenotetext{\textrm{b}}{Blackbody brightness temperatures calculated from the individual eclipse depths, assuming zero geometric albedo. $T_{\mathrm{day}}$ and $A_{g}$ were derived from a simultaneous fit of all three eclipse depths.}
\vspace{-0.25cm}\tablenotetext{\textrm{c}}{For marginal cases, $2\sigma$ upper limits are provided.}
\vspace{-0.25cm}\tablenotetext{\textrm{d}}{Error-weighted averages of the two eclipse depths in each band from \citet{garhart2019}.}
\textbf{References.} (1) This work; (2) \citealt{beerer2011}; (3) \citealt{baskin2013}; (4) \citealt{garhart2019}; (5) \citealt{alonso1}; (6) \citealt{deming2011}; (7) \citealt{alonso2}; (8) \citealt{esteves2015}; (9) \citealt{wong2016}; (10) \citealt{desert2011}; (11) \citealt{demory2013}; (12) \citealt{fortney2011}; (13) \citealt{shporer2014}; (14) \citealt{odonovan2010}.
\end{deluxetable*}

The secondary eclipse depth at visible wavelengths contains contributions from reflected starlight as well as thermal emission. Due to the composite nature of this quantity, there is an inherent degeneracy between the dayside temperature, $T_{\mathrm{day}}$, and the geometric albedo of the atmosphere in the observed bandpass, $A_{g}$. More specifically, these parameters are negatively correlated, because a decrease in the dayside temperature lowers the contribution from thermal emission, which in turn requires a larger fraction of incident starlight to be reflected.

In previous systematic studies of secondary eclipses for transiting exoplanet systems observed by Kepler \citep[e.g.,][]{esteves2013,esteves2015,angerhausen2015}, optical geometric albedos were derived by fixing the dayside temperatures to values calculated under certain assumptions on the extent of heat redistribution across the planets' surfaces, e.g., homogeneous reradiation or dayside redistribution only. Therefore, the calculations were not fully self-consistent, and as a result, the derived albedo values do not constitute direct measurements.

To break the degeneracy between dayside temperature and optical geometric albedo, we must include secondary eclipse depth measurements at longer wavelengths, where the planet's brightness is dominated by thermal emission. \citet{garhart2019} carried out a uniform analysis of secondary eclipse light curves obtained in both the 3.6 and 4.5 $\mu$m bandpasses of the Spitzer/IRAC instrument and computed eclipse depths for several dozen exoplanets. In addition, there is a large body of other analyses of Spitzer secondary eclipses, covering well over 50 systems. 

Among the 22 systems we analyzed in this paper, 12 have eclipse depths measured in all three bandpasses (\tess, 3.6, and 4.5 $\mu$m); these include eight systems for which only marginal secondary eclipse detections were obtained (Section \ref{subsec:marginal}). These planets are listed in Table \ref{tab:temps}. To increase the body of uniformly derived albedos, we also include all targets with measured Spitzer secondary eclipses that were observed with CoRoT or Kepler. The eclipse depths and references are provided in the table.

For each system, we first utilized Equation \eqref{eclipse} to calculate the dayside blackbody brightness temperature $T_{b}$ in each bandpass, assuming zero geometric albedo. In order to straightforwardly propagate the uncertainties in stellar properties when calculating $T_{b}$, we computed the integrated stellar flux in the three bandpasses for a grid of PHOENIX stellar models \citep{husser2013} spanning the ranges $T_{\mathrm{eff}}=[3000,7500]$ K, $\log g=[3.50,5.00]$, and $[\mathrm{Fe}/\mathrm{H}]=[-1.0,+1.0]$ and constructed best-fit polynomial functions that we subsequently sampled in a Monte Carlo fashion when simultaneously fitting for the temperature. We computed the posterior distribution of $T_{b}$ using \texttt{emcee}, with $R_p/R_*$, $a/R_*$, $T_{\mathrm{eff}}$, $\log g$, and [Fe/H] as additional parameters constrained by Gaussian priors. The median and $1\sigma$ uncertainties for $R_p/R_*$ and $a/R_*$ were taken from the results of our \tess\ light-curve analysis, while constraints on $T_{\mathrm{eff}}$, $\log g$, and [Fe/H] were taken from the respective discovery papers. In the case of KOI-13, for which several mutually inconsistent stellar parameter measurements have been published, we used the values derived from the spectroscopic analysis in \citet{shporer2014}. 

The individual bandpass brightness temperatures are shown in Table \ref{tab:temps}. In most cases, the visible-band brightness temperature $T_{b,\mathrm{vis}}$ is consistent with the values in the Spitzer bandpasses. However, there are several targets for which $T_{b,\mathrm{vis}}$ is significantly higher than the temperatures in the infrared, suggesting that there is an additional contribution in the visible-light secondary eclipse depth from reflected starlight (i.e., $A_{g}>0$). 

To constrain the optical geometric albedo and further refine the dayside temperature estimate, we simultaneously fit all three eclipse depths to a single blackbody, while allowing the optical geometric albedo to vary freely. The resulting temperature and albedo measurements are listed in Table \ref{tab:temps}; in marginal cases where the derived optical geometric albedo is less than $1\sigma$ above zero, we provide $2\sigma$ upper limits.

To ensure maximum uniformity in this analysis, we only utilized available 3.6 and 4.5 $\mu$m secondary eclipse depths, even if measurements at other wavelengths were available in the literature. In the case of WASP-19b, there are two pairs of 3.6 and 4.5 $\mu$m eclipse measurements listed in \citet{garhart2019}; the independent eclipse depth measurements in each band are self-consistent to within $\sim$1.1$\sigma$, and we computed the error-weighted average eclipse depth in each band for use in our analysis. \citet{garhart2019} calculated dayside brightness temperatures separately for the 3.6 and 4.5 $\mu$m eclipse depths and found a weak correlation between increasing equilibrium temperature and increasing the brightness temperature ratio between the two Spitzer bands. This suggests a possible systematic deviation between the true emission spectra and those of isothermal blackbodies in the infrared. Nevertheless, for the majority of the planets we analyzed in this paper, all three eclipse depths lie within $1\sigma$ of the band-integrated model spectrum points and uncertainties, with better than $2\sigma$ consistency across all targets.

While most of the \tess\ targets we analyzed have optical geometric albedos that lie within $2\sigma$ of zero, WASP-19b, WASP-43b, WASP-100b, and WASP-121b show evidence for enhanced dayside atmospheric reflectivity. In addition, we confirmed the previously reported high albedos of Kepler-7b and KOI-13b \citep[e.g.,][]{demory2013,shporer2014,shporer2015}.

For WASP-43b, \citet{keating2017} likewise assumed an isothermal blackbody for the planet's emission spectrum and obtained $T_{\mathrm{day}}=1483\pm10$ K and $A_{g}=0.24\pm0.01$ from a combined analysis of Spitzer and HST/WFC3 eclipse depths; the high-precision WFC3 data points were responsible for the exquisite precision in their retrieved parameter values. We note that their analysis utilized Spitzer eclipse depths from the analysis of \citet{stevenson2017}, which measured a significantly smaller 3.6 $\mu$m eclipse depth of $3230\pm60$ ppm. This explains the higher brightness temperature and lower geometric albedo we obtained. 

Meanwhile, the albedo measurements we obtained for WASP-18b and WASP-19b are consistent with the constraints derived from previous analyses that utilized additional eclipse depths at other wavelengths and carried out more detailed modeling of the planet's atmosphere: $A_{g}<0.048$ ($2\sigma$) for WASP-18b \citep{shporer2019} and $A_{g}=0.16\pm0.04$ for WASP-19b \citep{wong2019wasp19}. For WASP-100b, \citet{jansen2020} derived the geometric albedo from the phase curve without utilizing the Spitzer secondary eclipses, instead using a thermal energy balance model to retrieve the Bond albedo and day--night heat recirculation while assuming Lambertian scattering. Their value of $A_{g}=0.16^{+0.04}_{-0.03}$ is consistent with ours at better than the $1\sigma$ level.

Only a handful of other exoplanets have direct optical geometric albedo measurements or upper limits in the literature. These include HD 189733b ($<$0.12 across 450--570 nm; \citealt{evans2013}), HD 209458b ($0.038\pm0.045$; \citealt{rowe2008}), Kepler-7b ($0.35\pm0.02$; \citealt{demory2013}), WASP-12b (97.5\% confidence upper limit at 0.064; \citealt{bell2017}), and TrES-2b ($0.014\pm0.003$; \citealt{barclay2012}). The significant body of new direct geometric albedo constraints presented here underscores the importance of \tess\ and the synergy with Spitzer in broadening the picture of exoplanet reflectivity at visible wavelengths. Future analyses of targets in the Northern Sky and repeated observation of targets during the extended mission promise to further expand upon these results and refine existing albedo values.

\begin{figure*}
    \centering
    \includegraphics[width=\linewidth]{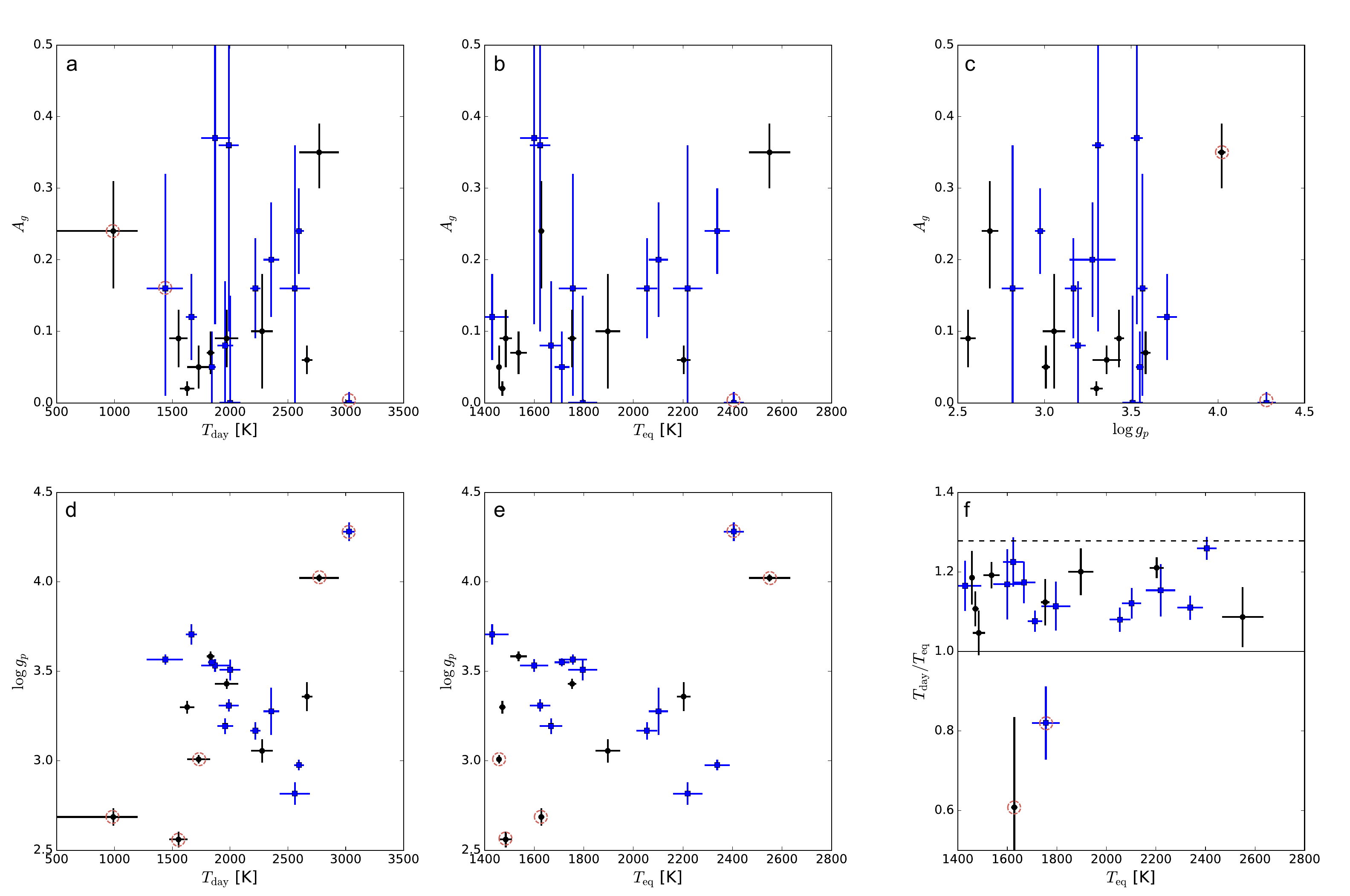}
    \caption{Two-parameter plots from our trend analysis. \tess\ targets are shown in blue, while Kepler and CoRoT targets are shown in black; outliers that were removed in subset analyses are indicated by the red circles and detailed in Table~\ref{tab:trends}. Panels (a)--(b) and (d)--(e) show the measured optical geometric albedo $A_{g}$ and planetary surface gravity $\log\, g_{p}$ as a function of dayside temperature $T_{\mathrm{day}}$ (measured from blackbody fits to secondary eclipse depths) and equilibrium temperature $T_{\mathrm{eq}}$ (theoretical calculation assuming full heat redistribution and zero Bond albedo). Panel (c) shows $A_{g}$ vs. $\log \, g_{p}$. In panel (f), the ratio between dayside and equilibrium temperature is plotted as a function of the latter; the horizontal solid and dashed lines indicate the limiting cases of uniform heat recirculation and no heat recirculation (assuming zero Bond albedo), respectively.}
    \label{fig:trends}
\end{figure*}

\subsection{Exploring Trends in Geometric Albedo}\label{subsec:trends}

By quantifying the reflectivity of a planet’s dayside hemisphere, the geometric albedo is an important diagnostic tool for probing the presence of clouds on exoplanets. The study of exoplanet clouds is a rich field, and extensive theoretical work has produced increasingly complex models for describing the microphysics of cloud formation, as well as the interplay between incident starlight, atmospheric composition, temperature--pressure profiles, and three-dimensional heat transport (see, for example, reviews by \citealt{marley2013} and \citealt{helling2019}). At the same time, a diverse range of exoplanets has been intensively observed in reflected, transmitted, and emitted light, revealing that clouds are a common feature on exoplanets \citep[e.g.,][]{sing2016}. As numerical models, observational techniques, and telescope capabilities continue to advance, the search for population-wide trends linking the presence and properties of clouds to other fundamental planetary and stellar parameters will provide crucial empirical tests of our current understanding of exoplanet clouds.

\begin{deluxetable*}{l|ccc|ccc|ccc}
\tablewidth{0pc}
\tabletypesize{\scriptsize}
\tablecaption{
    Temperature and Albedo Trend Analysis Results
    \label{tab:trends}
}

\tablehead{ \multicolumn{10}{c}{} \vspace{-0.3cm}\\ \colhead{} \vspace{-0.2cm}
&  \multicolumn{3}{c}{\underline{\tess\ targets only}} & \multicolumn{3}{c}{\underline{Kepler/CoRoT targets only}} & \multicolumn{3}{c}{\underline{All targets}} \\
    \colhead{Variables} &
    \colhead{PCC\tablenotemark{\scriptsize$\mathrm{a}$}}                    &
    \colhead{Slope ($10^{-3}$)\tablenotemark{\scriptsize$\mathrm{a}$}} &
    \colhead{Sig.\tablenotemark{\scriptsize$\mathrm{a}$}}  &
    \colhead{PCC}                    &
    \colhead{Slope ($10^{-3}$)} &
    \colhead{Sig.}  &
    \colhead{PCC}                    &
    \colhead{Slope ($10^{-3}$)} &
    \colhead{Sig.}}
\startdata
$A_{g}$ vs. $T_{\mathrm{day}}$ & $-$0.16 & $-0.099\pm0.028$ & 3.5 & 0.19 & $0.086\pm0.021$ & 4.1 & 0.03 & $0.001\pm0.012$ & 0.1 \\
$A_{g}$ vs. $T_{\mathrm{day}}$ ($1500<T_{\mathrm{day}}<3000$ K) & 0.15 & $0.174\pm0.079$ & 2.2 & 0.65 & $0.092\pm0.020$ & 4.6 & 0.38 & $0.105\pm0.019$ & 5.5 \\
\hline
$A_{g}$ vs. $T_{\mathrm{eq}}$ & $-$0.19 & $-0.129\pm0.046$ & 2.8 & 0.64 & $0.131\pm0.026$ & 5.0 & 0.19 & $0.028\pm0.017$ & 1.6 \\
$A_{g}$ vs. $T_{\mathrm{eq}}$ (no WASP-18b) & 0.02 & $0.165\pm0.085$ & 1.9 & '' & '' & '' & 0.33 & $0.146\pm0.024$ & 6.1\\
\hline
$A_{g}$ vs. $\log\,g_{p}$ & $-$0.39 & $-153\pm32$ & 4.8 & 0.31 & $102\pm37$ & 2.8 &  $-$0.05 & $-32\pm15$ & 2.1\\
$A_{g}$ vs. $\log\,g_{p}$ (no KOI-13b, WASP-18b) & $-$0.15 & $-184\pm96$ & 1.9 & $-$0.54 & $-63\pm42$ & 1.5 & $-$0.10 & $-69\pm40$ & 1.7 \\
\hline
$\log\,g_{p}$ vs. $T_{\mathrm{day}}$   & 0.00 & $-0.221\pm0.034$ & 6.5 & 0.72 & $0.985\pm0.073$ & 13.5 & 0.40 & $0.868\pm0.048$ & 18.1 \\
$\log\,g_{p}$ vs. $T_{\mathrm{day}}$  ($0.75<M_{p}<4\, M_{\mathrm{Jup}}$) & $-$0.88 & $-0.821\pm0.067$ & 12.3 & $-$0.31 & $-0.45\pm0.11$ & 4.1 & $-$0.72 & $-0.735\pm0.051$ & 14.4 \\
\hline
$\log\,g_{p}$ vs. $T_{\mathrm{eq}}$  & $-$0.12 & $-0.474\pm0.046$ & 10.3 & 0.64 & $1.022\pm0.068$ & 15.0 & 0.29 & $1.006\pm0.062$ & 16.2 \\
$\log\,g_{p}$ vs. $T_{\mathrm{eq}}$ ($0.75<M_{p}<4\, M_{\mathrm{Jup}}$) & $-$0.82 & $-1.002\pm0.089$ & 11.3 & $-$0.33 & $-0.358\pm0.094$ & 3.8 & $-$0.70 & $-0.737\pm0.049$ & 15.0\\
\hline
$T_{\mathrm{day}}/T_{\mathrm{eq}}$ vs. $T_{\mathrm{eq}}$ & 0.12 & $0.100\pm0.042$ & 2.4 & 0.15 & $0.067\pm0.044$ & 1.5 & 0.15 & $0.068\pm0.029$ & 2.3 \\
$T_{\mathrm{day}}/T_{\mathrm{eq}}$ vs. $T_{\mathrm{eq}}$ (no Kepler-7, WASP-36) & 0.01 & $0.085\pm0.041$ & 2.1 & 0.04 & $0.065\pm0.045$ & 1.4 & 0.03 & $0.057\pm0.029$ & 2.0\\
\enddata
\textbf{Notes.}
\vspace{-0.25cm}\tablenotetext{\textrm{a}}{Pearson correlation coefficient (PCC) and the derived correlation slope measurement and significance level from the MCMC linear fit to the two-parameter datasets involving dayside temperature, equilibrium temperature, \tess-band geometric albedo, and planetary surface gravity.}
\end{deluxetable*}

Having obtained a new set of uniformly derived visible geometric albedos (Table~\ref{tab:temps}), we briefly explored the possibility of emergent trends within this dataset. We carried out linear fits with MCMC to various combinations of stellar and planetary parameters with plausible relevance to cloud formation, including geometric albedo $A_{g}$, dayside brightness temperature $T_{\mathrm{day}}$, dayside equilibrium temperature assuming zero Bond albedo and uniform heat recirculation $T_{\mathrm{eq}}\equiv T_{\mathrm{eff}}\sqrt{R_{*}/2a}$, planetary surface gravity $\log\,g_{p}$, and stellar metallicity $\mathrm{[Fe/H]}$. We also computed the Pearson correlation coefficient (PCC) for each two-parameter correlation; we note that the PCC does not account for the uncertainties in the parameter estimates, and thus was used primarily to illustrate the overall sense of the correlation (i.e., positive vs. negative). For each pair of parameters, we searched for trends among the \tess\ and Kepler/CoRoT targets separately, as well as the full combined dataset. In addition, we reran the analysis on subsets of the full dataset after removing salient outliers. 

No significant correlation was found between $A_{g}$ and $\mathrm{[Fe/H]}$. Meanwhile, planetary metallicity measurements were not available for most targets on our list. For the six remaining two-parameter combinations, we plot the datasets in Figure~\ref{fig:trends}; outliers that were removed in our subset trend analyses are circled in red. Table~\ref{tab:trends} lists the complete set of results from our trend analysis. 

Condensate clouds form when the local temperature--pressure profile crosses the condensation curve of a particular cloud species. As such, the temperature across the dayside hemisphere is expected to be the primary factor in controlling the formation and extent of clouds and, by extension, the optical geometric albedo. When restricting the temperature range to $1500<T_{\mathrm{day}}<3000$ K (i.e., removing the hottest and the two coolest planets in the dataset; see panel (a) in Figure~\ref{fig:trends}), we found a marginal $
2.2\sigma$ positive correlation between $T_{\mathrm{day}}$ and $A_{g}$ for the \tess\ targets alone. The inclusion of albedo measurements for the Kepler and CoRoT targets in the same temperature range strongly reinforces this trend, and the combined dataset has a positive correlation at $5.5\sigma$ significance. An analogous $6.1\sigma$ positive correlation was found between $T_{\mathrm{eq}}$ and $A_{g}$ across all targets when the ultra-hot, ultra-massive outlier WASP-18b is removed (see panel (b)).

We also found a strong ($4.8\sigma$) correlation between $A_{g}$ and $\log\,g_{p}$ among the \tess\ targets (see panel (c)). However, this negative trend is anchored by the very low albedo of the massive planet WASP-18b, and after removing it, the significance drops to $1.9\sigma$. Likewise, when removing both supermassive hot Jupiters (KOI-13b and WASP-18b; 9.3 and 10.3 $M_{\mathrm{Jup}}$, respectively) from the full dataset, the significance of the negative correlation between $A_{g}$ and $\log\,g_{p}$ is only $1.7\sigma$. While the planet's surface gravity may be expected to play a role in determining the presence and behavior of clouds in exoplanet atmospheres, we note that this marginal trend is most likely a direct consequence of the corresponding very strong negative correlation between $\log\,g_{p}$ and $T_{\mathrm{day}}$ (or $T_{\mathrm{eq}}$) among roughly Jupiter-mass planets ($0.75<M_{p}<4 \, M_{\mathrm{Jup}}$; see panels (d) and (e)). The latter correlation reflects the well-known irradiation-radius dependence seen across the most highly irradiated hot Jupiters \citep[e.g.,][]{laughlin2011,enoch2012}. Therefore, we cannot confirm an independent fundamental trend between $A_{g}$ and $\log\,g_{p}$ from these data.

Lastly, we compared $T_{\mathrm{day}}$ and $T_{\mathrm{eq}}$ by considering the ratio between measured dayside and equilibrium temperatures across the dataset. $T_{\mathrm{day}}/T_{\mathrm{eq}}$ depends on the Bond albedo and the efficiency of  day--night heat recirculation. The two limiting cases are uniform day--night heat redistribution and no heat redistribution (i.e., instantaneous dayside reradiation and zero nightside temperature); see, for example, \citet{cowanagol}. The limiting values of $T_{\mathrm{day}}/T_{\mathrm{eq}}$ are shown in panel (f) assuming zero Bond albedo. Most of the measurements lie between the two limits, with the two outliers (Kepler-7b and WASP-36b --- two planets with the lowest measured $T_{\mathrm{day}}$) having very large error bars and lying within $2\sigma$ of the lower limit. Notably, there are no strong correlations across the dataset. The low $T_{\mathrm{day}}/T_{\mathrm{eq}}$ values for Kepler-7b and WASP-36b may be indicative of Bond albedos that are significantly higher than zero, consistent with the somewhat elevated optical geometric albedos of these two planets (see panel (a)).

Earlier studies using Kepler secondary eclipse measurements also searched for correlations between $A_{g}$ and other system parameters \citep[e.g,][]{hengalbedo,angerhausen2015,esteves2015}. \citet{esteves2015} reported a similar trend between $A_{g}$ and $T_{\mathrm{eq}}$ across a comparable range of temperatures to the one we report in Figure \ref{fig:trends}. One important distinction to make is that these earlier Kepler albedos did not self-consistently account for the contribution of the planets' thermal flux in the Kepler bandpass, given that they were computed in the absence of longer-wavelength secondary eclipse measurements. In most cases, an apparent geometric albedo was reported, which was derived without removing any thermal emission contribution from the secondary eclipse depth. In our study, we have, to first order, corrected the reported albedos for the thermal component. Although the sample size is small, the positive correlation between $A_{g}$ and $T_{\mathrm{day}}$ suggests that there may indeed be a systematic increase in the measured geometric albedos of hot Jupiters at dayside temperatures exceeding 1500 K. 

The relatively large uncertainties on the geometric albedos and the small number of data points prevent us from making any definitive claims concerning emergent albedo trends in the dataset. Looking to the near future, as \tess\ completes its observations in the Northern Sky and returns to the previously observed sectors during the extended mission, we will obtain direct geometric albedo measurements for many more systems and further refine existing values with additional data, allowing us to more effectively search for potential correlations between the atmospheric properties of exoplanets and the characteristics of their host stars. In the meantime, we briefly consider a few possible explanations for a positive correlation between geometric albedo and dayside temperature among these planets.



At temperatures above 2000 K, all of the major silicate condensates are expected to be in the vapor phase, so any enhanced reflectivity must be caused by different molecular species. Recent work combining microphysical cloud models with three-dimensional general circulation models has shown that clouds composed of some of the most refractory molecules, TiO$_{2}$ and Al$_{2}$O$_{3}$, can survive across the dayside hemisphere of exoplanets with dayside temperatures up to 1800 K \citep[][]{powell2019}. While these condensates are unable to form clouds near the substellar point for hotter planets, they may still condense near the limbs and the poles, where the temperatures are significantly cooler than the overall dayside brightness temperature. Further numerical modeling is needed to probe whether such partial cloud coverage can yield enough reflectivity to explain the high geometric albedos of planets above $T_{\mathrm{day}}=2000$ K.

Another plausible explanation for an apparent increase in the measured geometric albedo at these temperatures is enhanced thermal emission due to the presence of a additional opacity sources. Optical absorbers such as TiO and VO can strongly heat the atmosphere and induce temperature inversions, as has been seen on several ultra-hot Jupiters, such as WASP-121b \citep[e.g.,][]{parmentier2018,daylan2019}. Other possible sources of heating for the most highly irradiated planets include the continuum opacity of H$^{-}$, which forms from the dissociation of H$_{2}$ \cite[e.g.,][]{arcangeli2018}, and metallic gases composed of Fe and Mg, which are expected to be present in extremely irradiated hot Jupiters \citep{lothringer2018}. The extra heating due to these opacity sources can lead to a systematic deviation in the thermal emission spectrum at optical wavelengths from that of a simple blackbody, inflating the geometric albedo values we derived for the \tess\ bandpass.

\begin{deluxetable*}{ccccccl}
\tablewidth{0pc}
\tabletypesize{\scriptsize}
\tablecaption{
    Updated Transit Ephemerides
    \label{tab:ephem}
}
\tablehead{
    \colhead{System} \vspace{-0.3cm}&
    \colhead{$N$\tablenotemark{\scriptsize$\mathrm{a}$}}                    &
    \colhead{$\Delta t$\tablenotemark{\scriptsize$\mathrm{a}$}} &
    \colhead{$t_{0}$}  &
    \colhead{$P$} &
    \colhead{$dP/dt$\tablenotemark{\scriptsize$\mathrm{b}$}} &
    \colhead{References\tablenotemark{\scriptsize$\mathrm{c}$}}\\
    \colhead{} &
    \colhead{}                    &
    \colhead{(days)} &
    \colhead{(BJD$_{\mathrm{TDB}}-2450000$)}  &
    \colhead{(days)} &
    \colhead{(ms yr$^{-1}$)} &
    \colhead{}}
\startdata
HATS-24  & 2 & 1629 & $8667.45848\pm0.00035$ & $1.34849705\pm0.00000043$ & \dots &  \citet{hats24}\\
WASP-4 & 74 & 4285 & $6086.882662\pm0.000033$ & $1.338231530\pm0.000000027$ & $-8.3\pm1.4$ & \citet{bouma2020}\tablenotemark{\scriptsize$\mathrm{c}$}\\
WASP-5 & 25 & 3983 & $5079.106975\pm0.000098$ & $1.62842994\pm0.00000018$ & $< 21$ & \citet{bouma2019}\tablenotemark{\scriptsize$\mathrm{c}$}\\
WASP-18 & 16 & 9939 & $7255.782941\pm0.000048$ & $0.941452433\pm0.000000029$ & $<6.9$ & \citet{maxted2013}\tablenotemark{\scriptsize$\mathrm{c}$}, \citet{patra2020} \\
WASP-19 & 78 & 3767 & $6324.617983\pm0.000075$ & $0.788839008\pm0.000000058$ & $<13$ & \citet{petrucci2020}, \citet{patra2020}\\
WASP-30 & 3 & 3034 & $5443.06132\pm0.00034$ & $4.1567758\pm0.0000024$ & $< 1400$ & \citet{wasp30}, \citet{triaud2013}\\
WASP-36 & 25 & 2970 & $6656.755225\pm0.000072$ & $1.53736596\pm0.00000020$ & $< 28$ & \citet{mancini2016}, \citet{turner2016} \\
WASP-43 & 75 & 3018 & $6612.416074\pm0.000044$ & $0.813474061\pm0.000000046$ & $<5.6$ & \citet{patra2020}\\
WASP-46 & 45 & 2941 & $6407.88089\pm0.00026$ & $1.43037145\pm0.00000060$ & $< 80$ & \citet{bouma2019} \\
WASP-64 & 11 & 2911 & $5939.73877\pm0.00022$ & $1.57329034\pm0.00000045$ & $< 165$ & \citet{wasp6472}\tablenotemark{\scriptsize$\mathrm{c}$} \\
WASP-72 & 8 & 2551 & $7660.74049\pm0.00049$ & $2.2167429\pm0.0000010$ & $< 380$ & \citet{wasp6472}\tablenotemark{\scriptsize$\mathrm{c}$}, \citet{patra2020} \\
WASP-77A & 3 & 2557 & $6271.65907\pm0.00011$ & $1.360029033\pm0.000000098$ & $<72$ & \citet{wasp77}, \citet{turner2016} \\
WASP-78 & 2 & 2567 & $5882.35952\pm0.00054$ & $2.17518429\pm0.00000078$ & \dots & \citet{wasp78} \\
WASP-82 & 2 & 1515 & $8447.08368\pm0.00040$ & $2.7057831\pm0.0000016$ & \dots & \citet{smith2015} \\
WASP-100 & 2 & 2237 & $8509.10542\pm0.00014$ & $2.8493823\pm0.0000012$ & \dots & \citet{wasp100} \\
WASP-111 & 2 & 2061 & $8337.13711\pm0.00034$ & $2.31096978\pm0.00000060$ & \dots & \citet{wasp111} \\
WASP-121 & 2 & 1868 & $6635.708314\pm0.000011$ & $1.27492476\pm0.00000012$ &\dots & \citet{wasp121}\\
WASP-122 & 9 & 1838 & $7114.96893\pm0.00029$ & $1.71005340\pm0.00000051$ & $<300$ & \citet{wasp122}, \citet{kelt14}\tablenotemark{\scriptsize$\mathrm{c}$} \\
WASP-142 & 2 & 1519 & $7007.77866\pm0.00041$ & $2.0528714\pm0.0000015$ & \dots & \citet{wasp142} \\
WASP-173A & 2 & 1079 & $7288.85930\pm0.00021$ & $1.38665333\pm0.00000042$ & \dots & \citet{wasp173} \\
\enddata
\textbf{Notes.}
\vspace{-0.25cm}\tablenotetext{\textrm{a}}{$N$: number of published transit timing measurements included in fit; $\Delta t$: time baseline spanned by timing measurements.}
\vspace{-0.25cm}\tablenotetext{\textrm{b}}{Period derivative. In the case of non-detections, the $2\sigma$ upper limit on the absolute value is given.}
\vspace{-0.25cm}\tablenotetext{\textrm{c}}{Publications that contain the full list of transit timing measurements used in our fits. Due to missing ingress or egress, some epochs were removed. For WASP-4, we removed epochs --1085 and --818 as listed in \citet{bouma2020}. For WASP-5, epochs --653 and --476 as listed in \citet{bouma2019} were not included. To the list of transit timings provided in \citet{patra2020} for WASP-18, we added the transit times derived from analyses of WASP and Exoplanet Transit Database transits in \citet{maxted2013}. For WASP-64 and WASP-72, we removed epochs 21 and (0, 143), respectively, as listed in \citet{wasp6472}. We also removed the fifth transit time presented in \citet{kelt14}.}
\end{deluxetable*}
\subsection{Updated Transit Ephemerides}\label{subsec:ephem}
From our phase curve fits, we derived new transit timing estimates, and we used these measurements to calculate updated transit ephemerides. We excluded HIP 65A and TOI-503, which were discovered from the \tess\ light curves. Many of the targets had not been revisited for many years prior to the \tess\ observations, and our transit measurements substantially improve these deprecated ephemerides and can benefit the scheduling of future observations.

We considered both a linear (i.e., constant period) transit ephemeris model and a quadratic model with a steadily changing orbital period:
\begin{align}
t_{\mathrm{lin}}(E) &= t_{0}+PE, \label{linear}\\
t_{\mathrm{quad}}(E) &= t_{0}+PE+\frac{P}{2}\frac{dP}{dt}E^2. \label{quad}
\end{align}
Here, $E$ denotes the transit epoch, $t_{0}$ is the zeroth epoch mid-transit time, $P$ is the orbital period, and $dP/dt$ is the period derivative. We set the zeroth epoch to the published transit epoch closest to the weighted average of all available transit timings.

Following the methodology of \citet{bouma2019} and \citet{patra2020}, we considered all timing measurements from the peer-reviewed literature that (i) have a clearly denoted time system (UTC versus TDB),\footnote{For almost all WASP discovery papers, Heliocentric Julian Date (HJD) was used, and while later papers invariably state that the time system is UTC, several of the earlier papers do not; in these cases, we assumed consistency and applied the UTC system to unspecified timings.} (ii) were derived from transit light curves that include both ingress and egress, (iii) were fit with the mid-transit time as an unconstrained free parameter, and (iv) were not affected by unmodeled starspot crossings. When applicable, we utilized published timings derived from uniform, global reanalyses of previous measurements. The literature references that contain the full body of transit timing measurements used in our analysis are provided in Table \ref{tab:ephem}; see the Table Notes for details about removed epochs.

We combined the literature transit timing measurements with the transit times derived from our \tess\ light-curve fits (Tables \ref{tab:transitfit} and \ref{tab:fit}). For the most conservative uncertainties on the ephemerides, we utilized the transit times that we obtained using the residual permutation (PB) analysis: $T_{0,\mathrm{PB}}$. We calculated the transit ephemeris parameters using a standard MCMC procedure. Only the WASP-4 system shows evidence for a varying orbital period (see dedicated analyses in \citealt{bouma2019} and \citealt{bouma2020}). For the remaining systems with more than two transit timing measurements, we placed $2\sigma$ upper limits on $|dP/dT|$.

For most systems, the reduced chi-squared $\chi_r^2$ of the best-fit solution is less than 1.5. The other cases have $\chi_r^2>3$. Of these, WASP-19, WASP-43, WASP-46, and WASP-64 have attested photometric variability due to stellar activity (see Appendix B and respective discovery papers). To account for the scatter in the transit timings induced by activity, we followed, for example, the previous transit timing analysis of WASP-46 in \citet{petrucci2018} and added an additional scatter term $\sigma_{t}$ in quadrature to the listed transit timing uncertainties, which we allowed to vary freely in the MCMC fit to ensure $\chi_r^2=1$. For WASP-19, WASP-43, WASP-46, and WASP-64, we obtained best-fit scatter terms of $\sigma_t=45$ s, $\sigma_t=18$ s, $\sigma_t=2.3$ minutes, and $\sigma_t=33$ s, respectively. For the remaining cases with $\chi_r^2>3$ --- WASP-4, WASP-36, and WASP-122 --- we applied a similar method to the one used in our phase curve fits (see also \citealt{bouma2020}) and scaled up the transit timing uncertainties by the factor $\sqrt{\chi_r^2}$ to ensure $\chi_r^2=1$.

The full set of updated transit ephemerides is listed in Table \ref{tab:ephem}. In Appendix D, we show observed minus calculated ($O-C$) transit timing residual plots for the 12 systems with more than two transit timing measurements. 

The most notable result is the robust detection of a decreasing orbital period in the WASP-4 system. We obtained a rate of $dP/dt=-8.3\pm1.4$ ms yr$^{-1}$, which is statistically identical to the value presented in the independent analysis by \citet{bouma2020}: $-8.6\pm1.3$ ms yr$^{-1}$. Comparing the BIC for fits assuming the linear and nonlinear transit ephemeris models, we find $\Delta\mathrm{BIC}=29$ in favor of the nonlinear model. From the $O-C$ plot for WASP-4, we see that the two high-precision transit timings at orbit numbers 366 and 626, as well as the first two epochs, strongly anchor the nonlinear shape of the overall transit ephemeris. However, even when removing those four transit timings, the BIC still strongly favors the nonlinear model ($\Delta\mathrm{BIC}=8.1$). Long-baseline RV monitoring of WASP-4 has confirmed that the system is accelerating toward Earth at a speed that is consistent with the observed orbital period decay rate \citep{bouma2020}.

For WASP-18 and WASP-43, we derived tight constraints on orbital decay, with $2\sigma$ upper limits less than 10 ms yr$^{-1}$. For the other targets, the precision of the orbital periods is significantly improved --- up to an order of magnitude in some cases. \tess\ will reobserve all of these systems during the first year of the extended mission, and the additional $\sim$2 yr of time baseline will yield even more refined ephemerides.

\section{Summary}\label{sec:con}

We have presented the results from our systematic study of \tess\ phase curves of known short-period transiting systems with substellar companions observed in sectors 1--13, encompassing the first year of the \tess\ primary mission. After selecting for targets with likely detectable phase curve signals, we carried out a uniform analysis of 22 systems. The main findings of our work are summarized below:
\begin{itemize}
\item No significant phase curve signals were found for 12 systems: HATS-24, WASP-4, WASP-5, WASP-36, WASP-43, WASP-46, WASP-64, WASP-77A, WASP-78, WASP-82, WASP-142, and WASP-173A. We reported marginal secondary eclipse depth and atmospheric brightness modulation amplitude measurements for these planets in Table \ref{tab:bad} and presented the results of our fits to the transit light curves in Table \ref{tab:transitfit}.
\item We detected statistically significant secondary eclipse depths for WASP-18b ($339\pm21$ ppm), WASP-19b ($470^{+130}_{-110}$ ppm), WASP-72b ($113^{+39}_{-31}$ ppm), WASP-100b ($94\pm17$ ppm), WASP-111b ($102^{+38}_{-37}$ ppm), WASP-121b ($486\pm59$ ppm), and WASP-122b/KELT-14b ($165^{+63}_{-60}$ ppm). 
\item Of these seven planets, WASP-18b, WASP-19b, WASP-72b, WASP-100b, WASP-121b, and WASP-122b/KELT-14b show atmospheric brightness modulation, with measured semi-amplitudes of $181.7^{+8.2}_{-8.1}$ ppm, $311^{+57}_{-54}$ ppm, $70^{+15}_{-16}$ ppm, $48.5^{+5.7}_{-5.4}$ ppm, $214^{+25}_{-26}$ ppm, and $64\pm17$ ppm, respectively. We also recovered this phase curve signal for the grazing hot Jupiter system HIP 65A, with a semi-amplitude of $29.6_{-9.1}^{+8.3}$ ppm.
\item A marginal detection of a phase shift in the atmospheric brightness modulation signal was reported for WASP-100b, indicating an eastward offset in the region of maximum brightness on the dayside hemisphere of $\delta=12\overset{\circ}{.}0^{+6\overset{\circ}{.}3}_{-5\overset{\circ}{.}7}$. The results for this system are in good agreement with those reported in the independent analysis by \citet{jansen2020}.
\item HIP 65A, TOI-503, WASP-18, and WASP-30 display photometric variability associated with ellipsoidal distortion of the host star, with semi-amplitudes of $28.4^{+9.6}_{-8.0}$ ppm, $61.6^{+9.4}_{-8.1}$ ppm, $181.9^{+8.6}_{-8.4}$ ppm, and $97^{+24}_{-25}$ ppm, respectively. HIP 65A, TOI-503, and WASP-18 also show Doppler boosting signals with semi-amplitudes of $18.7^{+9.1}_{-8.4}$ ppm, $30\pm10$ ppm, and $20.0^{+6.4}_{-6.7}$ ppm, respectively. The amplitudes of these measured signals are generally consistent with the predictions from theoretical models of the corresponding physical processes.
\item We combined the measured \tess\ secondary eclipse depths for all systems with available Spitzer 3.6 and 4.5 $\mu$m secondary eclipse data to simultaneously constrain the dayside brightness temperature and optical geometric albedo (Table \ref{tab:temps}). Of the targets analyzed in this work, WASP-19b ($A_{g}=0.18\pm0.08$), WASP-43b ($A_{g}=0.14\pm0.06$), WASP-100b ($A_{g}=0.22\pm0.09$), and WASP-121b ($A_{g}=0.29^{+0.07}_{-0.08}$) show enhanced geometric albedos.
\item Using the geometric albedos measured for the \tess\ targets analyzed in this paper, we found a weak positive correlation between dayside temperature and optical geometric albedo for planets with dayside temperatures between 1500 and 3000 K. This trend is reinforced when including targets with Kepler and CoRoT secondary eclipse measurements and suggests that planets with $T_{\mathrm{day}}>2000$ K may have systematically higher atmospheric reflectivity due to high-temperature condensates and/or opacity sources contributing additional emission at visible wavelengths. Additional analyzed systems and refined secondary eclipse measurements are needed to definitively confirm this trend.
\item We computed updated transit ephemerides using the mid-transit times measured in our light-curve fits and literature values. We confirmed the decreasing orbital period in the WASP-4 system and placed upper-limit constraints on the orbital period variations for 11 other systems.
\end{itemize}

This work is the first systematic study of orbital phase curves provided by the \tess\ mission. These efforts will be expanded in the coming year with an analogous analysis of targets observed by \tess\ in the northern ecliptic hemisphere. Looking past the primary mission, most of the sky observed during the first two years of the \tess\ mission will be revisited during the approved extended mission, along with much of the ecliptic to achieve almost full sky coverage, with possible further extensions in the future.

The availability of additional photometry will greatly refine existing phase curve fits, as well as recover statistically significant secondary eclipses and phase curve signals for many of the currently marginal cases. By narrowing the constraints on the amplitude and phase shift of the atmospheric brightness modulation, we will be able to carry out more detailed characterization of the temperature distribution and probe the possibility of inhomogeneous clouds, as have been detected on Kepler-7b, for example \citep{demory2013,shporer2015}. Simultaneously, the long time baseline spanned by these repeated observations will enable numerous other scientific objectives of interest, such as the search for orbital decay in short-period hot Jupiter systems \citep[e.g.,][]{yee2020} or the detection of temporal variability in exoplanet atmospheres (i.e., weather), which has been predicted by some recent modeling work \citep[e.g.,][]{komacek2020}.

The tentative trends in the visible-light geometric albedo values reported in Section \ref{subsec:trends} provide particularly fertile ground for follow-up study. With the end of the Spitzer era, we can look to near-future facilities such as James Webb Space Telescope to continue space-based infrared observations, which, when combined with \tess-band secondary eclipses, will expand the set of direct albedo measurements and help us better understand the emergent trends. This, alongside other intensive atmospheric characterization campaigns, will yield new insights into the formation and properties of exoplanet clouds.

\acknowledgments
$\,$

Funding for the \tess\ mission is provided by NASA’s science mission directorate. This paper includes data collected by the \tess\ mission, which are publicly available from the Mikulski Archive for Space Telescopes (MAST). Resources supporting this work were provided by the NASA High-End Computing (HEC) Program through the NASA Advanced Supercomputing (NAS) Division at Ames Research Center for the production of the SPOC data products. We acknowledge Peter Gao, Tiffany Jansen, and David Kipping for helpful discussions. We also thank an anonymous reviewer for helpful comments that greatly improved the manuscript. I.W. is supported by a Heising-Simons \textit{51 Pegasi b} postdoctoral fellowship. T.D. acknowledges support from MIT’s Kavli Institute as a Kavli postdoctoral fellow. 


\appendix 
\restartappendixnumbering

\section{List of Light-curve Segments}

Table \ref{tab:segments} lists all of the data segments extracted from the \tess\ light curves for the 22 systems analyzed in this paper. In the second column, each segment is referred to by a three-number sequence separated by dashes: the first number denotes the \tess\ sector, the second number indicates the spacecraft orbit (two per sector), and the last digit is a sequential data segment number. Only data segments with time series spanning longer than 1 day are included in the table and were considered in our analysis; for example, during \tess\ sectors 1 and 2, a momentum dump was scheduled within the last day of observation during each spacecraft orbit, and we discarded the short segment of photometry between the final momentum dump and the break in data collection for downlink. The number of data points before and after flagged data removal, outlier trimming, and ramp trimming are indicated, as well as the start and final time stamps of each data segment. The sixth column lists the order of the polynomial detrending function used for the respective segment (see Section \ref{sec:ana}). The final column indicates whenever data was removed to alleviate flux ramps and other uncorrectable systematic artifacts. Several of the segments showed severe short-term (i.e., on timescales comparable to or shorter than the orbital period) photometric variations due to uncorrected instrumental systematics. These are also indicated under the ``Comments'' column and were removed from the light curve prior to our joint fits. We have also indicated the two systems --- WASP-19 and WASP-121 --- for which the SAP photometry was utilized instead of the PDCSAP light curves.

\startlongtable
\begin{deluxetable}{cccccccc}
\tablewidth{0pc}
\tabletypesize{\scriptsize}
\tablecaption{
    Summary of Light-curve Segments
    \label{tab:segments}
}
\tablehead{
    \colhead{Target} &
    \colhead{Segment\tablenotemark{\scriptsize$\mathrm{a}$}}                     &
    \colhead{$n_{\mathrm{raw}}$\tablenotemark{\scriptsize$\mathrm{b}$}}                     &
    \colhead{$n_{\mathrm{trimmed}}$\tablenotemark{\scriptsize$\mathrm{b}$}} &
    \colhead{$T_{\mathrm{start}}$\tablenotemark{\scriptsize$\mathrm{c}$}}  &
    \colhead{$T_{\mathrm{end}}$\tablenotemark{\scriptsize$\mathrm{c}$}} & 
    \colhead{Order\tablenotemark{\scriptsize$\mathrm{d}$}} & 
    \colhead{Comments}
}
\startdata
HATS-24 & 13-1-1 & 2470 & 2413 & 653.925 & 657.339  & 3 & \\
        & 13-1-2 & 2430 & 2339 & 657.350 & 660.714  & 2 & \\
        & 13-1-3 & 2430 & 2380 & 660.725 & 664.089  & 3 & \\
        & 13-1-4 & 2589 & 2491 & 664.100 & 667.695  & 3 & \\
        & 13-2-1 & 2473 & 2412 & 668.631 & 672.047  & 2 & \\
        & 13-2-2 & 2430 & 2357 & 672.058 & 675.422  & 4 & \\
        & 13-2-3 & 2430 & 2390 & 675.433 & 678.797  & 2 & \\
        & 13-2-4 & 2430 & 2385 & 678.808 & 682.172  & 0 & \\
\hline
HIP 65A & 1-1-1 & 1839 & 1774 &  325.299 &  327.838 & 4 & \\
& 1-1-2 & 1800 & 1390 &  327.855 &  329.841 & 1 & trimmed 0.50 d from end\\
& 1-1-3 & 1800 & 1706 &  330.355 &  332.831 & 3 & \\
& 1-1-4 & 1800 & 1727 &  332.855 &  335.330 & 4 & \\
& 1-1-5 & 1800 & 1730 &  335.354 &  337.834 & 2 & \\
& 1-2-1 & 1828 & 1774 &  339.665 &  342.183 & 2 & \\
& 1-2-2 & 1800 & 1749 &  342.197 &  344.683 & 3 & \\
& 1-2-3 & 1800 & 1761 &  344.697 &  347.183 & 3 & \\
& 1-2-4 & 1800 & 875 &  347.445 &  349.683 & 3 & trimmed 0.25 d from start\\
& 1-2-5 & 1800 & 1748 &  349.697 &  352.183 & 2 & \\
& 2-1-1 & 1827 & 1778 &  354.113 &  356.632 & 3 & \\
& 2-1-2 & 1800 & 1741 &  356.651 &  359.132 & 3 & \\
& 2-1-3 & 1800 & 1742 &  359.151 &  361.631 & 3 & \\
& 2-1-4 & 1800 & 1739 &  361.651 &  364.131 & 3 & \\
& 2-1-5 & 1800 & 1736 &  364.151 &  366.631 & 3 & \\
& 2-2-1 & 1827 & 1766 &  368.606 &  371.124 & 3 & \\
& 2-2-2 & 1800 & 1750 &  371.144 &  373.624 & 3 & \\
& 2-2-3 & 1800 & 1741 &  373.644 &  376.124 & 2 & \\
& 2-2-4 & 1800 & 1713 &  376.144 &  378.602 & 3 & \\
& 2-2-5 & 1800 & 1718 &  378.644 &  381.124 & 2 & \\
\hline
TOI-503 & 7-1-1 & 2288 & 2070 &  491.887 &  494.798 & 0 & trimmed 0.25 d from start \\
& 7-1-2 & 2250 & 2207 &  494.810 &  497.923 & 1 & \\
& 7-1-3 & 2250 & 2223 &  497.935 &  501.049 & 2 & \\
& 7-1-4 & 1429 & 1403 &  501.060 &  503.043 & 0 & \\
& 7-2-1 & 2280 & 2066 &  504.961 &  507.861 & 0 & trimmed 0.25 d from start\\
& 7-2-2 & 2250 & 2207 &  507.872 &  510.986 & 0 & \\
& 7-2-3 & 2250 & 2039 &  510.997 &  514.111 & 0 & trimmed 511.75---512.00 \\ 
& 7-2-4 & 1419 & 1394 &  514.122 &  516.092 & 0 & \\
\hline
WASP-4 & 2-1-1 & 1827 & 1768 &  354.114 &  356.632 & 2 & \\
& 2-1-2 & 1800 & 1743 &  356.652 &  359.132 & 0 & \\
& 2-1-3 & 1800 & 1726 &  359.152 &  361.632 & 1 & \\
& 2-1-4 & 1800 & 1747 &  361.652 &  364.132 & 0 & \\
& 2-1-5 & 1800 & 1399 &  364.152 &  366.132 & 0 & trimmed 0.50 d from end\\
& 2-2-1 & 1827 & 1593 &  368.859 &  371.125 & 2 & trimmed 0.25 d from start\\
& 2-2-2 & 1800 & 1758 &  371.145 &  373.625 & 1 & \\
& 2-2-3 & 1800 & 1739 &  373.645 &  376.125 & 1 & \\
& 2-2-4 & 1800 & 1714 &  376.145 &  378.603 & 0 & \\
& 2-2-5 & 1800 & 1725 &  378.645 &  381.125 & 2 & \\
\hline
WASP-5 & 2-1-1 & 1827 & 1760 &  354.114 &  356.632 & 2 & \\
& 2-1-2 & 1800 & 1747 &  356.652 &  359.132 & 1 & \\
& 2-1-3 & 1800 & 1736 &  359.152 &  361.632 & 3 & \\
& 2-1-4 & 1800 & 1742 &  361.652 &  364.132 & 0 & \\
& 2-1-5 & 1800 & 1735 &  364.152 &  366.632 & 0 & \\
& 2-2-1 & 1827 & 1775 &  368.607 &  371.125 & 0 & \\
& 2-2-2 & 1800 & 1743 &  371.145 &  373.625 & 0 & \\
& 2-2-3 & 1800 & 1732 &  373.645 &  376.125 & 2 & \\
& 2-2-4 & 1800 & 1715 &  376.145 &  378.603 & 1 & \\
& 2-2-5 & 1800 & 1718 &  378.645 &  381.125 & 1 & \\
\hline
WASP-18 & 2-1-1 & 1827 & 1779 &  354.113 &  356.631 & 1 & \\
& 2-1-2 & 1800 & 1745 &  356.650 &  359.131 & 2 & \\
& 2-1-3 & 1800 & 1742 &  359.151 &  361.630 & 3 & \\
& 2-1-4 & 1800 & 1732 &  361.651 &  364.131 & 0 & \\
& 2-1-5 & 1800 & 1742 &  364.151 &  366.631 & 1 & \\
& 2-2-1 & 1827 & 1592 &  368.856 &  371.124 & 2 & trimmed 0.25 d from start\\
& 2-2-2 & 1800 & 1748 &  371.144 &  373.624 & 0 & \\
& 2-2-3 & 1800 & 1739 &  373.644 &  376.124 & 1 & \\
& 2-2-4 & 1800 & \dots &  376.144 &  378.602 & \dots & severe systematics \\
& 2-2-5 & 1800 & 1725 &  378.644 &  381.124 & 1 & \\
& 3-1-1 & 1406 & 1101 &  385.952 &  387.499 & 2 & trimmed 0.25 d from end\\
& 3-1-2 & 1800 & 1741 &  387.769 &  390.249 & 1 & \\
& 3-1-3 & 1800 & 1724 &  390.269 &  392.749 & 2 & \\
& 3-1-4 & 1800 & \dots &  392.769 &  395.249 & \dots & severe systematics \\
& 3-2-1 & 1586 & 1454 &  396.638 &  398.709 & 1 & \\
& 3-2-2 & 1440 & 1390 &  398.728 &  400.709 & 3 & \\
& 3-2-3 & 1440 & 868 &  400.728 &  401.959 & 0 & trimmed 0.75 d from end\\
& 3-2-4 & 1440 & 702 &  402.971 &  403.959 & 0 & trimmed 0.25 d from start\\
& & & & & & & and 0.75 d from end\\
& 3-2-5 & 1222 & \dots &  404.728 &  406.216 & \dots & severe systematics\\
\hline
WASP-19 & 9-1-1 & 1669 & 1140 &  544.771 &  546.380 & 3 & trimmed 0.50 d from start\\
(SAP) & 9-1-2 & 2250 & 1856 &  546.889 &  549.505 & 2 & trimmed 0.50 d from start\\
& 9-1-3 & 2250 & 1854 &  550.014 &  552.630 & 1 & trimmed 0.50 d from start\\
& 9-1-4 & 2092 & 1710 &  553.139 &  555.545 & 1 & trimmed 0.50 d from start\\
& 9-2-1 & 1854 & 945 &  558.035 &  559.899 & 1 & trimmed 0.50 d from start\\
& 9-2-2 & 2250 & 1861 &  560.410 &  563.025 & 3 & trimmed 0.50 d from start\\
& 9-2-3 & 2250 & 1857 &  563.537 &  566.150 & 3 & trimmed 0.50 d from start\\
& 9-2-4 & 1669 & 1291 &  566.662 &  568.478 & 2 & trimmed 0.50 d from start\\
\hline
WASP-30 & 2-1-1 & 1827 & 1761 &  354.115 &  356.633 & 0 & \\
& 2-1-2 & 1800 & 1394 &  356.652 &  359.133 & 0 & trimmed 358.1---358.6\\
& 2-1-3 & 1800 & 1743 &  359.153 &  361.633 & 0 & \\
& 2-1-4 & 1800 & 1732 &  361.653 &  364.133 & 1 & \\
& 2-1-5 & 1800 & 1742 &  364.153 &  366.633 & 0 & \\
& 2-2-1 & 1827 & 1783 &  368.608 &  371.126 & 0 & \\
& 2-2-2 & 1800 & 1742 &  371.146 &  373.626 & 0 & \\
& 2-2-3 & 1800 & 1750 &  373.646 &  376.126 & 0 & \\
& 2-2-4 & 1800 & 1711 &  376.146 &  378.604 & 0 & \\
& 2-2-5 & 1800 & 1728 &  378.646 &  381.126 & 1 & \\
\hline
WASP-36 & 8-1-1 & 2283 & 1851 &  517.902 &  520.506 & 0 & trimmed 0.50 d from start\\
& 8-1-2 & 2250 & 2205 &  520.518 &  523.631 & 1 & \\
& 8-1-3 & 2250 & 2214 &  523.643 &  526.756 & 0 & \\
& 8-1-4 & 1659 & 1637 &  526.768 &  529.070 & 0 & \\
& 8-2-1 & 1121 & \dots &  535.009 &  536.548 & \dots & severe systematics \\
& 8-2-2 & 2250 & 2214 &  536.559 &  539.673 & 2 & \\
& 8-2-3 & 1672 & 1645 &  539.684 &  542.005 & 2 & \\
\hline
WASP-43 & 9-1-1 & 1548 & 1371 &  544.441 &  546.382 & 3 & \\
& 9-1-2 & 2250 & 2213 &  546.393 &  549.507 & 2 & \\
& 9-1-3 & 2250 & 2218 &  549.518 &  552.632 & 4 & \\
& 9-1-4 & 2092 & 2057 &  552.643 &  555.547 & 2 & \\
& 9-2-1 & 1627 & 912 &  558.105 &  559.902 & 2 & trimmed 0.25 d from start \\
& 9-2-2 & 2250 & 2207 &  559.913 &  563.027 & 5 & \\
& 9-2-3 & 2250 & 2219 &  563.038 &  566.152 & 1 & \\
& 9-2-4 & 1669 & 1640 &  566.163 &  568.480 & 2 & \\
\hline
WASP-46 & 1-1-1 & 1839 & 1766 &  325.300 &  327.839 & 3 & \\
& 1-1-2 & 1800 & 1685 &  327.856 &  330.328 & 2 & \\
& 1-1-3 & 1800 & 1692 &  330.356 &  332.832 & 3 & \\
& 1-1-4 & 1800 & 1723 &  332.856 &  335.331 & 1 & \\
& 1-1-5 & 1800 & 1738 &  335.354 &  337.835 & 0 & \\
& 1-2-1 & 1828 & 1770 &  339.665 &  342.183 & 3 & \\
& 1-2-2 & 1800 & 1746 &  342.197 &  344.683 & 2 & \\
& 1-2-3 & 1800 & 1753 &  344.697 &  347.183 & 0 & \\
& 1-2-4 & 1800 & 1038 &  347.197 &  349.683 & 0 & \\
& 1-2-5 & 1800 & 1755 &  349.697 &  352.183 & 3 & \\
\hline
WASP-64 & 6-1-1 & 2250 & 2211 &  468.390 &  471.504 & 2 & \\
& 6-1-2 & 2250 & 2200 &  471.515 &  474.629 & 1 & \\
& 6-1-3 & 1717 & 1691 &  474.640 &  477.024 & 2 & \\
& 6-2-1 & 2288 & 2058 &  478.370 &  481.275 & 0 & trimmed 0.25 d from start \\
& 6-2-2 & 2250 & 2206 &  481.286 &  484.400 & 0 & \\
& 6-2-3 & 2250 & 2216 &  484.411 &  487.525 & 2 & \\
& 6-2-4 & 1809 & 1780 &  487.536 &  490.048 & 2 & \\
& 7-1-1 & 2288 & 2248 &  491.635 &  494.796 & 0 & \\
& 7-1-2 & 2250 & 2203 &  494.807 &  497.921 & 1 & \\
& 7-1-3 & 2250 & 2218 &  497.932 &  501.046 & 0 & \\
& 7-1-4 & 1429 & 1411 &  501.057 &  503.041 & 1 & \\
& 7-2-1 & 2280 & 2232 &  504.709 &  507.858 & 2 & \\
& 7-2-2 & 2250 & 2217 &  507.870 &  510.983 & 0 & \\
& 7-2-3 & 2250 & 2204 &  510.996 &  514.108 & 1 & \\
& 7-2-4 & 1419 & 1401 &  514.119 &  516.089 & 1 & \\
\hline
WASP-72 & 3-1-1 & 1405 & 1270 &  385.952 &  387.750 & 0 & \\
& 3-1-2 & 1800 & 1738 &  387.769 &  390.250 & 0 & \\
& 3-1-3 & 1800 & 1726 &  390.269 &  392.750 & 0 & \\
& 3-1-4 & 1800 & 1736 &  392.769 &  395.250 & 0 & \\
& 3-2-1 & 1585 & 1277 &  396.889 &  398.710 & 0 & trimmed 0.25 d from start \\
& 3-2-2 & 1440 & 1386 &  398.729 &  400.710 & 0 & \\
& 3-2-3 & 1440 & 1393 &  400.729 &  402.710 & 2 & \\
& 3-2-4 & 1440 & 1388 &  402.729 &  404.710 & 2 & \\
& 3-2-5 & 1221 & 1045 &  404.729 &  406.218 & 0 & \\
& 4-1-1 & 2191 & 1604 &  410.907 &  413.189 & 0 & trimmed 0.75 d from the end\\
& 4-1-2 & 2160 & 2123 &  413.958 &  416.939 & 1 & \\
& 4-1-3 & 1110 & 1060 &  416.958 &  418.492 & 0 & \\
& 4-1-4 & 1249 & 1219 &  421.218 &  422.939 & 0 & \\
& 4-2-1 & 2186 & 2136 &  424.560 &  427.584 & 1 & \\
& 4-2-2 & 2160 & 2117 &  427.604 &  430.584 & 2 & \\
& 4-2-3 & 2160 & 2117 &  430.604 &  433.584 & 2 & \\
& 4-2-4 & 2160 & 2120 &  433.604 &  436.584 & 2 & \\
\hline
WASP-77A & 4-1-1 & 2191 & 2119 &  410.908 &  413.940 & 2 & \\
& 4-1-2 & 2160 & 2107 &  413.959 &  416.940 & 3 & \\
& 4-1-3 & 1110 & \dots &  416.959 &  418.493 & \dots & severe systematics\\
& 4-1-4 & 1249 & \dots &  421.219 &  422.940 & \dots & severe systematics \\
& 4-2-1 & 2190 & 2134 &  424.561 &  427.586 & 4 & \\
& 4-2-2 & 2160 & 2107 &  427.605 &  430.586 & 3 & \\
& 4-2-3 & 2160 & 2109 &  430.605 &  433.586 & 2 & \\
& 4-2-4 & 2160 & 2108 &  433.605 &  436.586 & 3 & \\
\hline
WASP-78 & 5-1-1 & 2200 & 2155 &  437.996 &  441.026 & 2 & \\
& 5-1-2 & 2160 & 2126 &  441.037 &  444.026 & 1 & \\
& 5-1-3 & 2160 & 2130 &  444.037 &  447.026 & 0 & \\
& 5-1-4 & 2160 & 2130 &  447.037 &  450.026 & 0 & \\
& 5-2-1 & 2193 & 2159 &  451.560 &  454.589 & 0 & \\
& 5-2-2 & 2160 & 2117 &  454.600 &  457.589 & 0 & \\
& 5-2-3 & 2160 & 2135 &  457.600 &  460.589 & 1 & \\
& 5-2-4 & 2160 & 2117 &  460.600 &  463.589 & 1 & \\
\hline
WASP-82 & 5-1-1 & 2200 & 2161 &  437.997 &  441.027 & 0 & \\
& 5-1-2 & 2160 & 1945 &  441.038 &  443.779 & 1 & trimmed 0.25 d from end \\
& 5-1-3 & 2160 & 1943 &  444.287 &  447.027 & 1 & trimmed 0.25 d from start\\
& 5-1-4 & 2160 & 2116 &  447.039 &  450.027 & 1 & \\
& 5-2-1 & 2193 & 1802 &  451.561 &  454.091 & 1 & trimmed 0.50 d from end\\
& 5-2-2 & 2160 & \dots &  454.601 &  457.590 & \dots& severe systematics\\
& 5-2-3 & 2160 & \dots &  457.601 &  460.590 & \dots & severe systematics\\
& 5-2-4 & 2160 & 1948 &  460.601 &  463.340 & 2 & trimmed 0.25 d from end\\
\hline
WASP-100 & 1-1-1 & 1839 & 1418 &  325.796 &  327.835 & 0 & trimmed 0.50 d from start\\
& 1-1-2 & 1800 & 1692 &  327.852 &  330.324 & 0 & \\
& 1-1-3 & 1800 & 1707 &  330.352 &  332.828 & 0 & \\
& 1-1-4 & 1800 & 1729 &  332.852 &  335.327 & 0 & \\
& 1-1-5 & 1800 & 1737 &  335.350 &  337.831 & 0 & \\
& 1-2-1 & 1828 & 1417 &  340.162 &  342.180 & 0 & trimmed 0.50 d from start\\
& 1-2-2 & 1800 & 1392 &  342.692 &  344.680 & 1 & trimmed 0.50 d from start\\
& 1-2-3 & 1800 & 1771 &  344.694 &  347.180 & 0 & \\
& 1-2-4 & 1800 & 1033 &  347.194 &  349.680 & 0 & \\
& 1-2-5 & 1800 & 1736 &  349.694 &  352.180 & 0 & \\
& 2-1-1 & 1827 & 1776 &  354.110 &  356.628 & 1 & \\
& 2-1-2 & 1800 & 1751 &  356.648 &  359.128 & 0 & \\
& 2-1-3 & 1800 & 1736 &  359.148 &  361.628 & 0 & \\
& 2-1-4 & 1800 & 1748 &  361.648 &  364.129 & 0 & \\
& 2-1-5 & 1800 & 1746 &  364.148 &  366.629 & 2 & \\
& 2-2-1 & 1827 & 1774 &  368.604 &  371.122 & 0 & \\
& 2-2-2 & 1800 & 1750 &  371.141 &  373.622 & 0 & \\
& 2-2-3 & 1800 & 1749 &  373.641 &  376.122 & 1 & \\
& 2-2-4 & 1800 & 1713 &  376.141 &  378.599 & 1 & \\
& 2-2-5 & 1800 & 1725 &  378.641 &  381.122 & 1 & \\
& 3-1-1 & 1405 & 1272 &  385.949 &  387.747 & 0 & \\
& 3-1-2 & 1800 & 1736 &  387.766 &  390.247 & 0 & \\
& 3-1-3 & 1800 & 1739 &  390.266 &  392.747 & 1 & \\
& 3-1-4 & 1800 & 1724 &  392.766 &  395.247 & 2 & \\
& 3-2-1 & 1586 & 1459 &  396.636 &  398.706 & 0 & \\
& 3-2-2 & 1440 & 1385 &  398.726 &  400.706 & 1 & \\
& 3-2-3 & 1440 & 1399 &  400.726 &  402.706 & 0 & \\
& 3-2-4 & 1440 & 1044 &  402.726 &  404.206 & 0 & trimmed 0.50 d from end\\
& 3-2-5 & 1222 & 1049 &  404.726 &  406.215 & 0 & \\
& 4-1-1 & 2191 & 2111 &  410.904 &  413.935 & 1 & \\
& 4-1-2 & 2160 & 2115 &  413.955 &  416.935 & 0 & \\
& 4-1-3 & 1110 & 1080 &  416.955 &  418.488 & 0 & \\
& 4-2-1 & 2186 & 2136 &  424.556 &  427.581 & 1 & \\
& 4-2-2 & 2160 & 2113 &  427.601 &  430.581 & 1 & \\
& 4-2-3 & 2160 & 2117 &  430.601 &  433.581 & 0 & \\
& 4-2-4 & 2160 & 1757 &  433.601 &  436.081 & 1 & trimmed 0.50 d from end\\
& 5-1-1 & 2200 & 1792 &  438.494 &  441.023 & 0 & trimmed 0.50 d from start\\
& 5-1-2 & 2160 & 2116 &  441.034 &  444.023 & 0 & \\
& 5-1-3 & 2160 & 2110 &  444.034 &  447.023 & 0 & \\
& 5-1-4 & 2160 & 2123 &  447.034 &  450.023 & 0 & \\
& 5-2-1 & 2193 & 1798 &  452.057 &  454.585 & 2 & trimmed 0.50 d from start\\
& 5-2-2 & 2160 & 2115 &  454.596 &  457.585 & 0 & \\
& 5-2-3 & 2160 & 2130 &  457.596 &  460.585 & 1 & \\
& 5-2-4 & 2160 & 2118 &  460.596 &  463.585 & 0 & \\
& 6-1-1 & 2250 & \dots &  468.388 &  471.501 & \dots& severe systematics\\
& 6-1-2 & 2250 & \dots &  471.513 &  474.626 & \dots& severe systematics\\
& 6-1-3 & 1717 & \dots &  474.637 &  477.021 & \dots& severe systematics\\
& 6-2-1 & 2288 & 1344 &  479.371 &  481.272 & 2 & trimmed 1.25 d from start\\
& 6-2-2 & 2250 & 2210 &  481.283 &  484.397 & 0 & \\
& 6-2-3 & 2250 & 2211 &  484.408 &  487.522 & 2 & \\
& 6-2-4 & 1809 & 1790 &  487.533 &  490.044 & 0 & \\
& 7-1-1 & 2288 & 1892 &  492.133 &  494.793 & 0 & trimmed 0.50 d from start\\
& 7-1-2 & 2250 & 2209 &  494.804 &  497.918 & 0 & \\
& 7-1-3 & 2250 & 2203 &  497.929 &  501.043 & 0 & \\
& 7-1-4 & 1429 & 1407 &  501.054 &  503.037 & 1 & \\
& 7-2-1 & 2280 & 1876 &  505.211 &  507.855 & 0 & trimmed 0.50 d from start\\
& 7-2-2 & 2250 & 2212 &  507.866 &  510.980 & 1 & \\
& 7-2-3 & 2250 & 2207 &  510.991 &  514.105 & 0 & \\
& 7-2-4 & 1419 & 1400 &  514.116 &  516.085 & 0 & \\
& 8-1-1 & 2283 & 2190 &  517.397 &  520.501 & 1 & \\
& 8-1-2 & 2250 & 2200 &  520.512 &  523.626 & 0 & \\
& 8-1-3 & 2250 & 2201 &  523.637 &  526.751 & 1 & \\
& 8-1-4 & 1659 & 1630 &  526.762 &  529.064 & 2 & \\
& 8-2-1 & 1121 & \dots &  535.003 &  536.542 & \dots& severe systematics \\
& 8-2-2 & 2250 & 2207 &  536.553 &  539.667 & 2 & \\
& 8-2-3 & 1672 & 1649 &  539.678 &  541.999 & 1 & \\
& 9-1-1 & 1227 & 885 &  545.134 &  546.375 & 1 & trimmed 0.25 d from start \\
& 9-1-2 & 2250 & 2215 &  546.387 &  549.500 & 3 & \\
& 9-1-3 & 2250 & 2225 &  549.511 &  552.624 & 0 & \\
& 9-1-4 & 2092 & 2054 &  552.636 &  555.541 & 2 & \\
& 9-2-1 & 1313 & \dots &  558.284 &  559.896 & \dots & severe systematics\\
& 9-2-2 & 2250 & 2205 &  559.907 &  563.021 & 0 & \\
& 9-2-3 & 2250 & 2212 &  563.032 &  566.146 & 2 & \\
& 9-2-4 & 1669 & 1638 &  566.157 &  568.474 & 2 & \\
& 10-1-1 & 1623 & 1454 &  570.560 &  572.604 & 1 & \\
& 10-1-2 & 2250 & 1678 &  573.364 &  575.729 & 2 & trimmed 0.75 d from start\\
& 10-1-3 & 2250 & 2211 &  575.741 &  578.854 & 1 & \\
& 10-1-4 & 2102 & 2070 &  578.866 &  581.784 & 2 & \\
& 10-2-1 & 1324 & 986 &  584.538 &  585.917 & 0  & trimmed 0.25 d from start\\
& 10-2-2 & 2250 & 2211 &  585.928 &  589.042 & 0 & \\
& 10-2-3 & 2250 & 1681 &  589.802 &  592.167 & 0 & trimmed 0.75 d from start\\
& 10-2-4 & 2522 & 2126 &  592.677 &  595.680 & 2 & trimmed 0.50 d from start\\
& 11-1-1 & 2250 & 2161 &  599.952 &  603.063 & 1 & \\
& 11-1-2 & 2250 & 2190 &  603.074 &  606.188 & 0 & \\
& 11-1-3 & 2517 & 2476 &  606.199 &  609.694 & 2 & \\
& 11-2-1 & 1044 & \dots &  612.698 &  613.938 & \dots & severe systematics \\
& 11-2-2 & 2250 & 2030 &  614.199 &  617.063 & 0 & trimmed 0.25 d from start\\
& 11-2-3 & 2250 & 2035 &  617.323 &  620.188 & 1 & trimmed 0.25 d from start\\
& 11-2-4 & 2659 & 2624 &  620.199 &  623.891 & 3 & \\
& 12-1-1 & 2280 & \dots &  624.962 &  628.105 & \dots & severe systematics \\
& 12-1-2 & 2250 & 2141 &  628.119 &  631.230 & 0 & \\
& 12-1-3 & 2250 & 2184 &  631.244 &  634.355 & 2 & \\
& 12-1-4 & 2250 & 2189 &  634.366 &  637.480 & 1 & \\
& 12-1-5 & 1084 & 1010 &  637.491 &  638.995 & 1 & \\
& 12-2-1 & 2280 & \dots &  640.038 &  643.188 & \dots & severe systematics\\
& 12-2-2 & 2250 & 2203 &  643.199 &  646.313 & 0 & \\
& 12-2-3 & 2250 & 2209 &  646.325 &  649.438 & 1 & \\
& 12-2-4 & 2479 & 2438 &  649.450 &  652.891 & 2 & \\
& 13-1-1 & 2470 & 2414 &  653.920 &  657.334 & 2 & \\
& 13-1-2 & 2430 & 2337 &  657.346 &  660.709 & 2 & \\
& 13-1-3 & 2430 & 2375 &  660.721 &  664.085 & 2 & \\
& 13-1-4 & 2589 & 2494 &  664.096 &  667.690 & 2 & \\
& 13-2-1 & 2473 & 2405 &  668.626 &  672.043 & 2 & \\
& 13-2-2 & 2430 & 2360 &  672.054 &  675.418 & 1 & \\
& 13-2-3 & 2430 & 2389 &  675.429 &  678.793 & 1 & \\
& 13-2-4 & 2430 & 2389 &  678.806 &  682.168 & 2 & \\
\hline
WASP-111 & 1-1-1 & 1839 & 1765 &  325.301 &  327.840 & 2 & \\
& 1-1-2 & 1800 & 1699 &  327.857 &  330.329 & 0 & \\
& 1-1-3 & 1800 & 1699 &  330.357 &  332.833 & 1 & \\
& 1-1-4 & 1800 & 1731 &  332.857 &  335.332 & 1 & \\
& 1-1-5 & 1800 & 1742 &  335.356 &  337.836 & 0 & \\
& 1-2-1 & 1828 & 1762 &  339.667 &  342.185 & 2 & \\
& 1-2-2 & 1800 & 1743 &  342.199 &  344.685 & 2 & \\
& 1-2-3 & 1800 & 1758 &  344.699 &  347.185 & 3 & \\
& 1-2-4 & 1800 & 1029 &  347.199 &  349.685 & 0 & \\
& 1-2-5 & 1800 & 1742 &  349.699 &  352.185 & 1 & \\
\hline
WASP-121 & 7-1-1 & 2288 & 2240 &  491.635 &  494.796 & 0 & \\
(SAP) & 7-1-2 & 2250 & 2211 &  494.807 &  497.921 & 0 & \\
& 7-1-3 & 2250 & 2209 &  497.932 &  501.046 & 2 & \\
& 7-1-4 & 1429 & 1413 &  501.057 &  503.040 & 1 & \\
& 7-2-1 & 2280 & 2242 &  504.708 &  507.858 & 3 & \\
& 7-2-2 & 2250 & 2210 &  507.869 &  510.983 & 4 & \\
& 7-2-3 & 2250 & 2213 &  510.994 &  514.108 & 2 & \\
& 7-2-4 & 1419 & 1398 &  514.119 &  516.089 & 4 & \\
\hline
WASP-122/KELT-14 & 7-1-1 & 2288 & 2234 &  491.634 &  494.795 & 2 & \\
& 7-1-2 & 2250 & 2211 &  494.806 &  497.920 & 0 & \\
& 7-1-3 & 2250 & 2208 &  497.932 &  501.045 & 1 & \\
& 7-1-4 & 1429 & 1403 &  501.057 &  503.040 & 0 & \\
& 7-2-1 & 2280 & 1890 &  505.208 &  507.858 & 2 & trimmed 0.50 d from start\\
& 7-2-2 & 2250 & 2215 &  507.869 &  510.983 & 3 & \\
& 7-2-3 & 2250 & 2222 &  510.994 &  514.108 & 0 & \\
& 7-2-4 & 1419 & 1399 &  514.119 &  516.088 & 0 & \\
\hline
WASP-142 & 8-1-1 & 2283 & 1851 &  517.901 &  520.506 & 2 & trimmed 0.50 d from start \\
& 8-1-2 & 2250 & 2215 &  520.518 &  523.631 & 0 & \\
& 8-1-3 & 2250 & 2222 &  523.642 &  526.756 & 1 & \\
& 8-1-4 & 1659 & 1632 &  526.767 &  529.070 & 0 & \\
& 8-2-1 & 1121 & \dots &  535.009 &  536.547 & \dots & severe systematics \\
& 8-2-2 & 2250 & 2202 &  536.559 &  539.672 & 0 & \\
& 8-2-3 & 1672 & 1653 &  539.684 &  542.004 & 1 & \\
\hline
WASP-173A & 2-1-1 & 1827 & 1773 &  354.115 &  356.633 & 4 & \\
& 2-1-2 & 1800 & 1742 &  356.652 &  359.133 & 3 & \\
& 2-1-3 & 1800 & 1736 &  359.152 &  361.631 & 4 & \\
& 2-1-4 & 1800 & 1755 &  361.652 &  364.133 & 3 & \\
& 2-1-5 & 1800 & 1749 &  364.152 &  366.633 & 3 & \\
& 2-2-1 & 1827 & 1771 &  368.608 &  371.126 & 3 & \\
& 2-2-2 & 1800 & 1750 &  371.145 &  373.626 & 2 & \\
& 2-2-3 & 1800 & 1744 &  373.645 &  376.126 & 3 & \\
& 2-2-4 & 1800 & 1713 &  376.145 &  378.603 & 3 & \\
& 2-2-5 & 1800 & 1733 &  378.645 &  381.126 & 4 & \\
\enddata
\vspace{0.15cm}\textbf{Notes.}
\vspace{-0.25cm}\tablenotetext{\textrm{a}}{The numbers indicate the TESS sector, spacecraft orbit (two per sector), and segment number, respectively.}
\vspace{-0.25cm}\tablenotetext{\textrm{b}}{Number of data points contained in each data segment before and after removing flagged points, filtering out outliers, and trimming ramps (as detailed under Comments).}
\vspace{-0.25cm}\tablenotetext{\textrm{c}}{Start and end times of each data segment, in units of $\mathrm{BJD}_{\mathrm{TDB}}-2458000$.}
\vspace{-0.25cm}\tablenotetext{\textrm{d}}{Order of the polynomial systematics detrending model used in the final joint fits.}
\end{deluxetable}

\section{Raw and Corrected Light Curves}
In Figures \ref{fig:raw1} and \ref{fig:raw2}, we present a compilation of light-curve plots for all 22 systems included in our phase curve analysis. The top panels show the raw light curves as contained in the \tess\ data products. The vertical blue lines indicate the locations of the scheduled momentum dumps. The corresponding outlier-removed, trimmed, and systematics-corrected light curves are provided in the bottom panels. These light curves contain only the data segments used in the joint fits and listed in Appendix A. The raw light curves of four systems (WASP-87A, WASP-120, WASP-167, and K2-237) that were ignored due to strong short-term photometric variability are also included. All targets with attested stellar variability in the literature and/or the \tess\ light curves are indicated by asterisks next to the labels.

\begin{figure*}
    \centering
    \includegraphics[width=\linewidth]{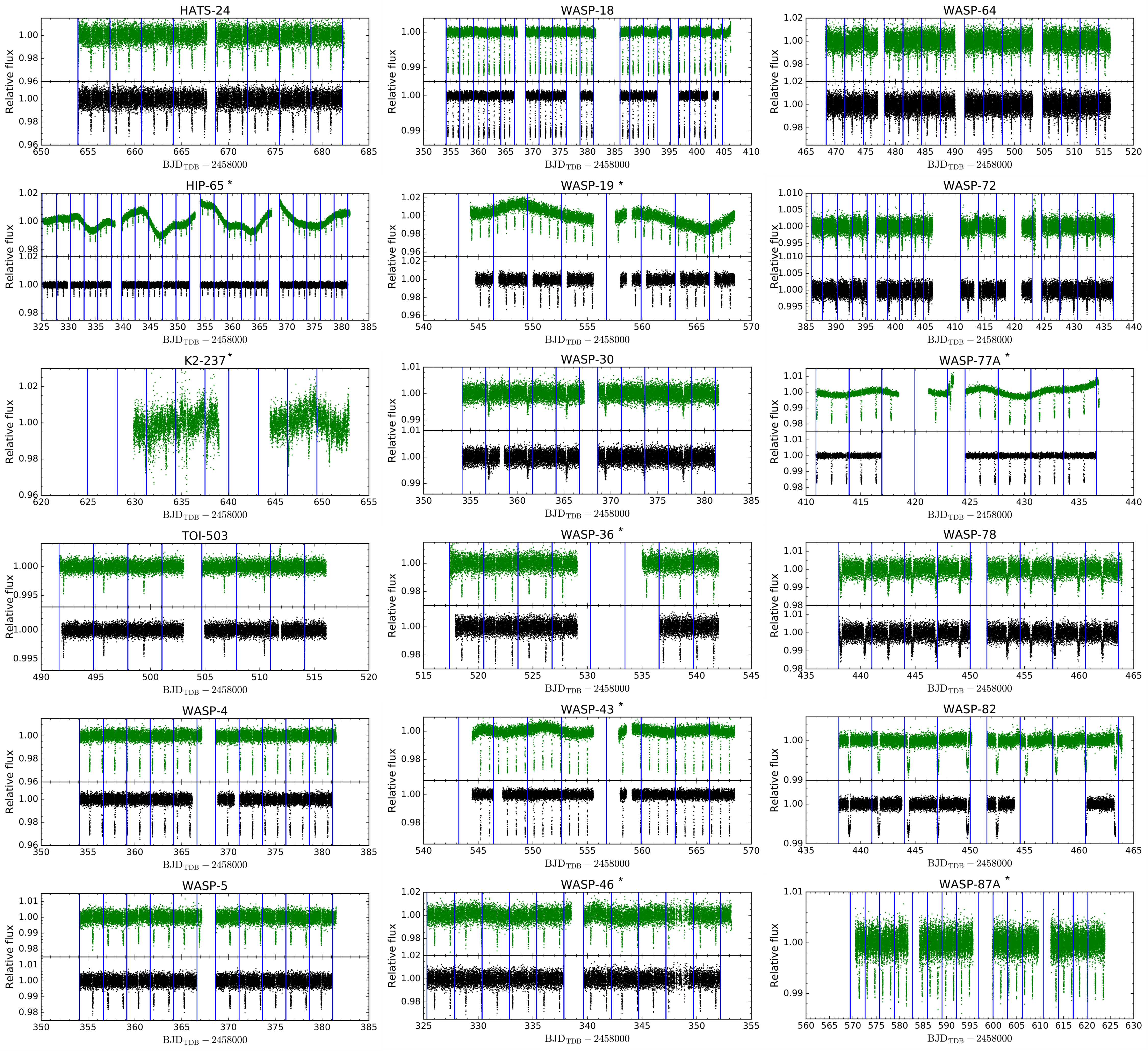}
    \caption{Raw and corrected light curves for 18 of the targets discussed in this paper. The top panels show the raw \tess\ light curves, and the bottom panels show the outlier-removed, trimmed, and systematics-corrected light curves. The vertical blue lines indicate the scheduled momentum dumps. Systems marked with asterisks show stellar variability.}
    \label{fig:raw1}
\end{figure*}

\begin{figure*}
    \centering
    \includegraphics[width=0.75\linewidth]{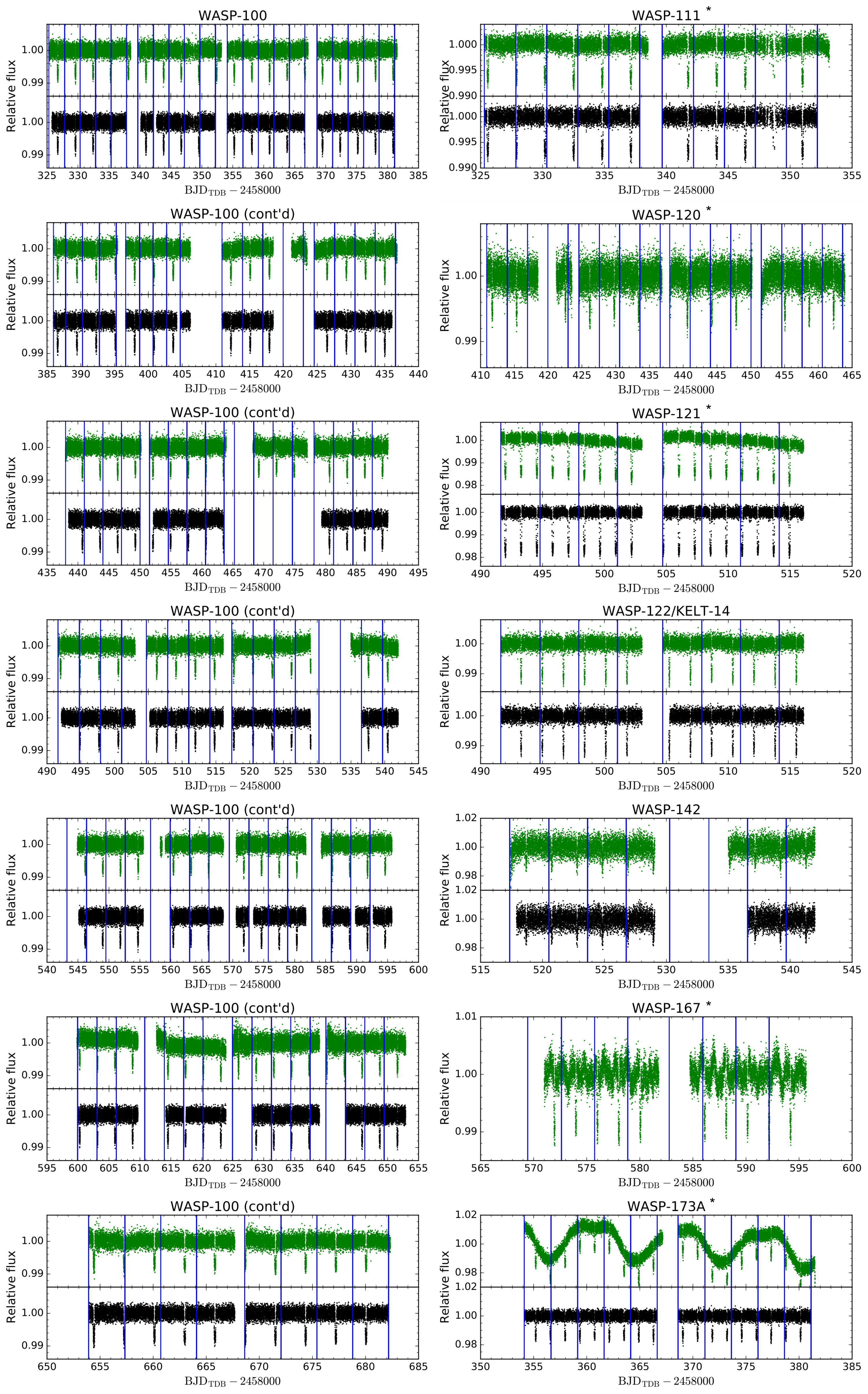}
    \caption{Continuation of Figure \ref{fig:raw1}. The 13-sector light curve of WASP-100 is divided into seven plots.}
    \label{fig:raw2}
\end{figure*}

\section{Full-orbit Light Curves: Marginal Detections and Non-detections}
\begin{figure*}
    \centering
    \includegraphics[width=0.85\linewidth]{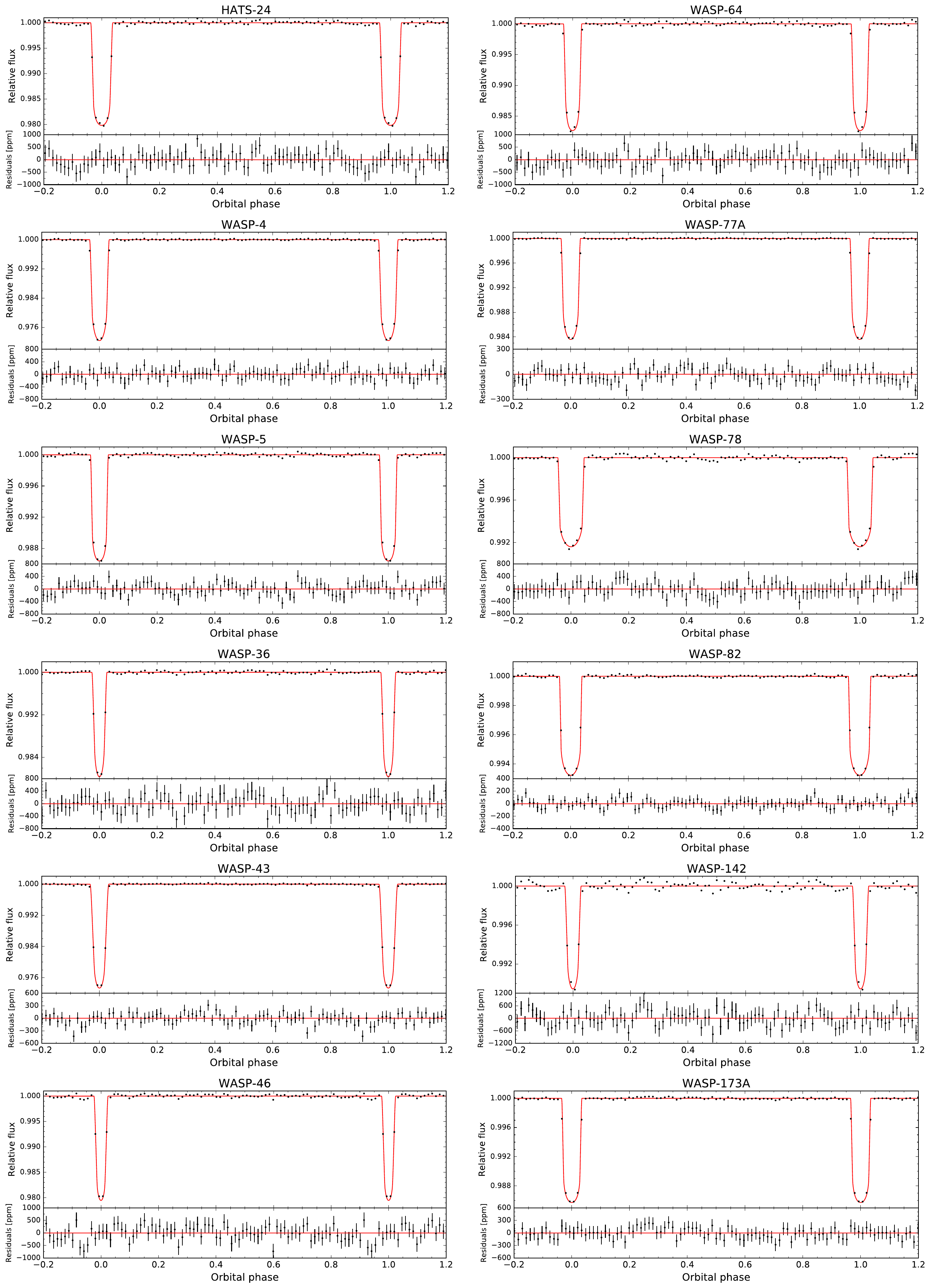}
    \caption{Phase-folded, systematics-corrected, and binned light curves for all targets that did not yield statistically significant secondary eclipse depths or phase curve signals.}
    \label{fig:nondetect}
\end{figure*}

Figure \ref{fig:nondetect} shows the full phase-folded, systematics-corrected, and binned light curves for the 12 targets where we did not measure any significant phase curve components or secondary eclipse; the corresponding residuals from the best-fit transit-only model are given in the bottom panels. The bin size is set to ensure roughly 75 bins per orbit. The results from the transit-only fits are provided in Table \ref{tab:transitfit}.
\restartappendixnumbering

\section{Multi-epoch Transit Ephemeris Fits}
\begin{figure*}
    \centering
    \includegraphics[width=1.0\linewidth]{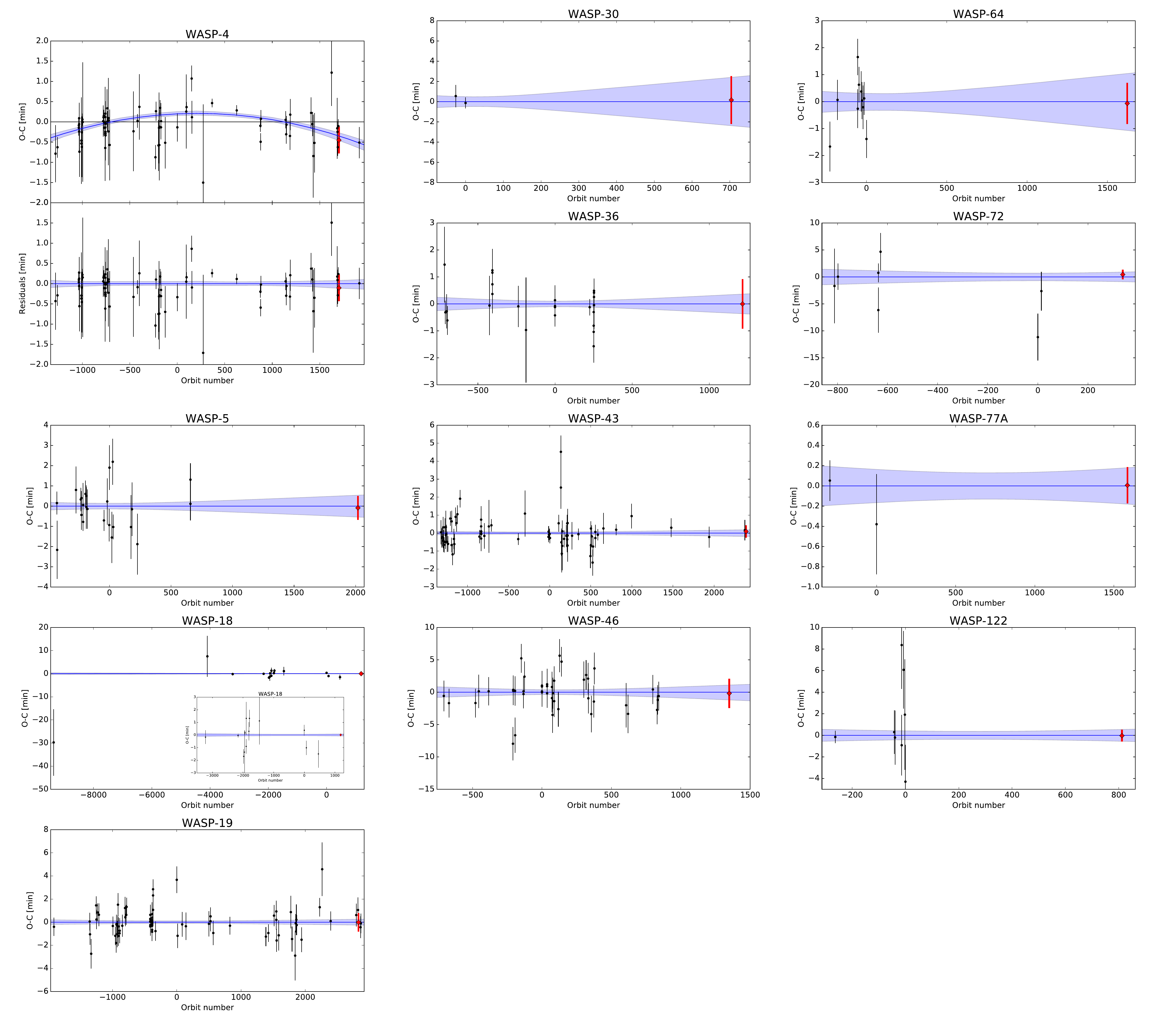}
    \caption{Observed minus calculated plots for the 12 systems analyzed in this paper with more than two published transit timings. The \tess\ measurements are shown in red. The blue shaded areas show the $1\sigma$ confidence region.}
    \label{fig:ephem}
\end{figure*}

\restartappendixnumbering
The newly derived transit ephemeris fits for the 12 systems with more than two published timing measurements are shown in Figure \ref{fig:ephem}. The $O-C$ values are computed relative to the best-fit linear transit ephemeris. Previously published transit timings are shown in black, and the \tess-epoch mid-transit time is shown with the red diamonds. The full list of updated ephemerides is given in Table \ref{tab:ephem}. For several systems, the timing uncertainties have been adjusted following the prescription described in Section \ref{subsec:ephem}. Only WASP-4 shows significant evidence for a nonlinear ephemeris. For that system, we include an additional panel showing the residuals from the best-fit nonlinear transit model.

\begin{thebibliography}{}
\tighten
\footnotesize

\bibitem[Addison et al.(2018)]{addison2018}
Addison, B. C., Wang, S., Johnson, M. C., et al. 2018, \aj, 156, 197

\bibitem[Alonso et al.(2009a)]{alonso1}
Alonso, R., Alapini, A., Aigrain, S., et al. 2009a, \aap, 506, 353

\bibitem[Alonso et al.(2009b)]{alonso2}
Alonso, R., Guillot, T., Mazeh, T., et al. 2009b, \aap, 501, L23

\bibitem[Anderson et al.(2014)]{wasp111}
Anderson, D. R., Brown, D. J. A., Collier Cameron, A., et al. 2014, \mnras, submitted  (arXiv:1410.3449)

\bibitem[Anderson et al.(2012)]{wasp46}
Anderson, D. R., Collier Cameron, A., Gillon, M., et al. 2012, \mnras, 422, 1988

\bibitem[Anderson et al.(2011)]{wasp30}
Anderson, D. R., Collier Cameron, A., Hellier, C., et al. 2011, \apjl, 726, L19

\bibitem[Anderson et al.(2008)]{wasp5}
Anderson, D. R., Gillon, M., Hellier, C., et al. 2008, \mnras, 387, L4

\bibitem[Angerhausen et al.(2015)]{angerhausen2015}
Angerhausen, D., DeLarme, E., \& Morse, J. A. 2015, \pasp, 127, 1113

\bibitem[Arcangeli et al.(2018)]{arcangeli2018}
Arcangeli, J., D{\'e}sert, J.-M., Line, M. R., et al. 2018, \apjl, 855, L30

\bibitem[Barclay et al.(2012)]{barclay2012}
Barclay, T., Huber, D., Rowe, J. F., et al. 2012, \apj, 761, 53

\bibitem[Baskin et al.(2013)]{baskin2013}
Baskin, N. J., Knutson, H. A., Burrows, A., et al. 2013, \apj, 773, 124

\bibitem[Beatty et al.(2019)]{beatty2019}
Beatty, T. G., Marley, M. S., Gaudi, B. S., et al. 2019, \aj, 158, 166

\bibitem[Beerer et al.(2011)]{beerer2011}
Beerer, I. M., Knutson, H. A., Burrows, A., et al. 2011, \apj, 727, 23




\bibitem[Bell et al.(2017)]{bell2017}
Bell, T. J., Nikolov, N., Cowan, N. B., et al. 2017, \apjl, 847, L2

\bibitem[Benneke et al.(2019)]{benneke2019}
Benneke, B., Knutson, H. A., Lothringer, J., et al. 2019, NatAs, 3, 813

\bibitem[Bento et al.(2017)]{hats24}
Bento, J., Schmidt, B., Hartman, J. D., et al. 2017, \mnras, 468, 835

\bibitem[Bloemen et al.(2012)]{bloemen2012}
Bloemen, S., Marsh, T. R., Degroote, P., et al. 2012, \mnras, 422, 2600 

\bibitem[Bouma et al.(2019)]{bouma2019}
Bouma, L. G., Winn, J. N., Baxter, C., et al. 2019, \aj, 157, 217

\bibitem[Bouma et al.(2020)]{bouma2020}
Bouma, L. G., Winn, J. N., Howard, A. W., et al. 2020, \apjl, 893, L29

\bibitem[Bourrier et al.(2019)]{bourrier2019}
Bourrier, V., Kitzmann, D., Kuntzer, T., et al. 2019, \aap, 637, A36

\bibitem[Burkart et al.(2012)]{burkart2012}
Burkart, J., Quataert, E., Arras, P., \& Weinberg, N. N. 2012, \mnras, 421, 983

\bibitem[Carmichael et al.(2019)]{carmichael2019}
Carmichael, T. W., Latham, D. W., \& Vanderburg, A. M. 2019, \aj, 158, 38

\bibitem[Claret(2017)]{claret2017}
Claret, A. 2017, \aap, 600, A30

\bibitem[Claret(2018)]{claret2018}
Claret, A. 2018, \aap, 618, A20

\bibitem[Cowan \& Agol(2011)]{cowanagol}
Cowan, N. B., \& Agol, E. 2011, \apj, 729, 54

\bibitem[Cowan et al.(2012)]{cowan2012}
Cowan, N. B., Machalek, P., Croll, B., et al. 2012, \apj, 747, 82

\bibitem[Daylan et al.(2019)]{daylan2019}
Daylan, T., G{\" u}nther, M. N., Mikal-Evans, T., et al. 2019, \aj, in revision (arXiv:1909.03000)

\bibitem[Delrez et al.(2016)]{wasp121}
Delrez, L., Santerne, A., Almenara, J.-M., et al. 2016, \mnras, 458, 4025

\bibitem[Deming et al.(2011)]{deming2011}
Deming, D., Knutson, H., Agol, E., et al. 2011, \apj, 726, 95

\bibitem[Demory et al.(2013)]{demory2013}
Demory, B.-O., de Wit, J., Lewis, N., et al. 2013, \apjl, 776, L25

\bibitem[D{\'e}sert et al.(2011b)]{desert2011}
D{\'e}sert, J.-M., Charbonneau, D., Fortney, J. J., et al. 2011, ApJS, 197, 11

\bibitem[Ehrenreich et al.(2011)]{ehrenreich2011}
Ehrenreich, D., Lagrange, A.-M., Bouchy, F., et al. 2011, \aap, 525, A85

\bibitem[Enoch et al.(2012)]{enoch2012}
Enoch, B., Collier Cameron, A., \& Horne, K. 2012, \aap, 540, A99

\bibitem[Esteves et al.(2013)]{esteves2013}
Esteves, L. J., De Mooij, E. J. W., \& Jayawardhana, R.\ 2013, \apj, 772, 51 

\bibitem[Esteves et al.(2015)]{esteves2015}
Esteves, L. J., De Mooij, E. J. W., \& Jayawardhana, R.\ 2015, \apj, 804, 150 

\bibitem[Evans et al.(2013)]{evans2013}
Evans, T. M., Pont, F., Sing, D. K., et al. 2013, \apjl, 772, L16 

\bibitem[Faigler \& Mazeh(2015)]{faigler2015}
Faigler, S., \& Mazeh, T. 2015, \apj, 800, 73 

\bibitem[Foreman-Mackey et al.(2013)]{emcee}
Foreman-Mackey, D., Hogg, D. W., Lang, D., \& Goodman, J. 2013, \pasp, 125, 306

\bibitem[Fortney et al.(2011)]{fortney2011}
Fortney, J. J., Demory, B.-O., D{\'e}sert, J.-M., et al. 2011, ApJS, 197, 9

\bibitem[Garhart et al.(2019)]{garhart2019}
Garhart, E., Deming, D., Mandell, A., et al. 2019, \aj, 159, 137

\bibitem[Gelman \& Rubin(1992)]{gelmanrubin}
Gelman, A., \& Rubin, D. B. 1992, StaSc, 7, 457

\bibitem[Gillon et al.(2013)]{wasp6472}
Gillon, M., Anderson, D. R., Collier-Cameron, A., et al. 2013, \aap, 552, A82

\bibitem[Gillon et al.(2009)]{gillon2009}
Gillon, M., Smalley, B., Hebb, L., et al. 2009, \aap, 267, 259

\bibitem[Hebb et al.(2010)]{wasp19}
Hebb, L., Collier-Cameron, A., Triaud, A. H. M. J., et al. 2010, \apj, 708, 224

\bibitem[Hellier et al.(2019)]{wasp173}
Hellier, C., Anderson, D. R., Bouchy, F., et al. 2019, \mnras, 482, 1379

\bibitem[Hellier et al.(2014)]{wasp100}
Hellier, C., Anderson, D. R., Cameron, A. C., et al. 2014, \mnras, 440, 1982

\bibitem[Hellier et al.(2009)]{wasp18}
Hellier, C., Anderson, D. R., Collier Cameron, A., et al. 2009, Natur, 460, 1098

\bibitem[Hellier et al.(2011)]{wasp43}
Hellier, C., Anderson, D. R., Collier Cameron, A., et al. 2011, \aap, 535, L7

\bibitem[Hellier et al.(2017)]{wasp142}
Hellier, C., Anderson, D. R., Collier Cameron, A., et al. 2017, \mnras, 465, 3693

\bibitem[Helling(2018)]{helling2019}
Helling, Ch. 2019, AREPS, 47, 583

\bibitem[Heng \& Demory(2013)]{hengalbedo}
Heng, K., \& Demory, B.-O. 2013, \apj, 777, 100

\bibitem[Holman et al.(2006)]{holman2006}
Holman, M. J., Winn, J. N., Latham, D. W., et al. 2006, \apj, 652, 1715

\bibitem[Husser et al.(2013)]{husser2013}
Husser, T.-O., Wende-von Berg, S., Dreizler, S., et al. 2013, \aap, 553, A6

\bibitem[Jansen \& Kipping(2020)]{jansen2020}
Jansen, T., \& Kipping, D. 2020, \mnras, 494, 4077

\bibitem[Jenkins et al.(2016)]{jenkins2016}
Jenkins, J. M., Twicken, J. D., McCauliff, S., et al. 2016, Proc. SPIE, 9913, 99133E

\bibitem[Keating \& Cowan(2017)]{keating2017}
Keating, D., \& Cowan, N. B. 2017, \apjl, 849, L5

\bibitem[Komacek \& Showman(2020)]{komacek2020}
Komacek, T. D., \& Showman, A. P. 2020, \apj, 888, 2

\bibitem[Kreidberg(2015)]{kreidberg2015}
Kreidberg, L. 2015, \pasp, 127, 1161

\bibitem[Laughlin et al.(2011)]{laughlin2011}
Laughlin, G., Crismani, M., \& Adams, F. C. 2011, \apjl, 729, L7

\bibitem[Loeb \& Gaudi(2003)]{loeb2003}
Loeb, A., \& Gaudi, B. S. 2003, \apjl, 588, L117 

\bibitem[Lothringer et al.(2018)]{lothringer2018}
Lothringer, J., Barman, T., \& Koskinen, T. 2018, \apj, 866, 27

\bibitem[Mancini et al.(2016)]{mancini2016}
Mancini, L., Kemmer, J., Southworth, J., et al. 2016, \mnras, 459, 1393

\bibitem[Marley et al.(2013)]{marley2013}
Marley, M. S., Ackerman, A. S., Cuzzi, J. N., \& Kitzmann, D. 2013, in Comparative Climatology of Terrestrial Planets, ed. S. J. Mackwell, A. A. Simon-Miller, J. W. Harder, \& M. A., Bullock (Tucson, AZ: University of Arizona Press), 367

\bibitem[Maxted et al.(2013a)]{wasp77}
Maxted, P. F. L., Anderson, D. R., Collier Cameron, A., et al. 2013, \pasp, 125, 48

\bibitem[Maxted et al.(2013b)]{maxted2013}
Maxted, P. F. L., Anderson, D. R., Doyle, A. P., et al. 2013, \mnras, 428, 2645

\bibitem[Mayorga et al.(2019)]{mayorga2019}
Mayorga, L. C., Batalha, N. E., Lewis, N. K., \& Marley, M. S. 2019, \aj, 158, 66

\bibitem[Mazeh(2008)]{mazeh2008}
Mazeh, T. 2008, EAS Publications Series, 29, 1 

\bibitem[Mazeh \& Faigler(2010)]{mazeh2010}
Mazeh, T., \& Faigler, S. 2010, \aap, 521, L59

\bibitem[Mireles et al.(2020)]{mireles2020}
Mireles, I., Shporer, A., Grieves, N., et al. 2020, \aj, 160, 133

\bibitem[Morris(1985)]{morris1985}
Morris, S. L. 1985, \apj, 295, 143 

\bibitem[Morris \& Naftilan(1993)]{morris1993}
Morris, S. L., \& Naftilan, S. A. 1993, \apj, 419, 344 

\bibitem[Nielsen et al.(2020)]{hip65}
Nielsen, L. D., Brahm, R., Bouchy, F., et al. 2020, \aap, 639, A76

\bibitem[O'Donovan et al.(2010)]{odonovan2010}
O'Donovan, F. T., Charbonneau, D., Harrington, J., et al. 2010, \apj, 710, 1551

\bibitem[Parmentier \& Crossfield(2017)]{parmentier2017}
Parmentier, V., \& Crossfield, I. J. M. 2017, in Handbook of Exoplanets, ed. H. J. Deeg \& J. A. Belmonte (Cham: Springer), 116

\bibitem[Parmentier et al.(2018)]{parmentier2018}
Parmentier, V., Line, M. R., Bean, J. L., et al. 2018, \aap, 617, A110

\bibitem[Patra et al.(2020)]{patra2020}
Patra, K. C., Winn, J. N., Holman, M. J., et al. 2020, \aj, 159, 150

\bibitem[Perez-Becker \& Showman(2013)]{perezbecker}
Perez-Becker, D., \& Showman, A. P. 2013, \apj, 776, 134

\bibitem[Perna et al.(2012)]{perna2012}
Perna, R., Heng, K., \& Pont, F. 2012, \apj, 751, 59

\bibitem[Petrucci et al.(2018)]{petrucci2018}
Petrucci, R., Jofr{\'e}, E., Ferrero, L. V., et al. 2018, \mnras, 473, 5126

\bibitem[Petrucci et al.(2020)]{petrucci2020}
Petrucci, R., Jofr{\'e}, E., G{\'o}mez Maqueo Chew, Y., et al. 2020, \mnras, 491, 1243

\bibitem[Pfahl et al.(2008)]{pfahl2008}
Pfahl, E., Arras, P., \& Paxton, B. 2008, \apj, 679, 783 

\bibitem[Pont et al.(2006)]{pont2006}
Pont, F., Zucker, S., \& Queloz, D. 2006, \mnras, 373, 231

\bibitem[Powell et al.(2019)]{powell2019}
Powell, D., Louden, T., Kreidberg, L., et al. 2019, \apj, 887, 170

\bibitem[Rodriguez et al.(2016)]{kelt14}
Rodriguez, J. E., Col{\' o}n, K. D., Stassun, K. G., et al. 2016, \apj, 151, 138

\bibitem[Rowe et al.(2008)]{rowe2008}
Rowe, J. F., Matthews, J. M., Seager, S., et al. 2008, \apj, 689, 1345

\bibitem[Shakura \& Postnov(1987)]{shakura1987}
Shakura, N. I., \& Postnov, K. A. 1987, \aap, 183, L21 

\bibitem[Shporer(2017)]{shporer2017}
Shporer, A. 2017, \pasp, 129, 072001 

\bibitem[Shporer \& Hu(2015)]{shporer2015}
Shporer, A., \& Hu, R. 2015, \aj, 150, 112

\bibitem[Shporer et al.(2014)]{shporer2014}
Shporer, A., O'Rourke, J. G., Knutson, H. A., et al. 2014, \apj, 788, 92

\bibitem[Shporer et al.(2019)]{shporer2019}
Shporer, A., Wong, I., Huang, C. X., et al. 2019, \aj, 157, 178

\bibitem[Sing et al.(2016)]{sing2016}
Sing, D. K., Fortney, J. J., Nikolov, N., et al. 2016, Natur, 529, 59

\bibitem[Smalley et al.(2012)]{wasp78}
Smalley, B., Anderson, D. R., Collier-Cameron, A., et al. 2012, \aap, 547, A61

\bibitem[Smith(2015)]{smith2015}
Smith, A. M. S. 2015, AcA, 65, 117

\bibitem[Smith et al.(2012)]{wasp36}
Smith, A. M. S., Anderson, D. R., Collier Cameron, A., et al. 2012, \aj, 143, 81

\bibitem[Smith et al.(2012)]{smith2012}
Smith, J. C., Stumpe, M. C., Van Cleve, J. E., et al. 2012, \pasp, 124, 1000

\bibitem[Stassun et al.(2017)]{stassun2017}
Stassun, K. G., Collins, K. A., \& Gaudi, B. S. 2017, \aj, 153, 136

\bibitem[Stassun et al.(2018)]{stassun2018}
Stassun, K. G., Oelkers, R. J., Pepper, J., et al. 2018, \aj, 156, 102 

\bibitem[Stevenson et al.(2017)]{stevenson2017}
Stevenson, K. B., Line, M. R., Bean, J. L., et al. 2017, \aj, 153, 68

\bibitem[Stumpe et al.(2014)]{stumpe2014}
Stumpe, M. C., Smith, J. C., Catanzarite, J. H., et al. 2014, \pasp, 126, 100

\bibitem[{\v S}ubjak et al.(2019)]{subjak2019}
{\v S}ubjak, J., Sharma, R., Carmichael, T. W., et al. 2019, \aj, 159, 151

\bibitem[Sullivan et al.(2015)]{sullivan2015}
Sullivan, P. W., Winn, J. N., Berta-Thompson, Z. K., et al. 2015, \apj, 809, 77

\bibitem[Triaud et al.(2013)]{triaud2013}
Triaud, A. H. M. J., Hebb, L., Anderson, D. R., et al. 2013, \aap, 549, A18

\bibitem[Turner et al.(2016)]{turner2016}
Turner, J. D., Pearson, K. A., Biddle, L. I., et al. 2016, \mnras, 459, 789

\bibitem[Turner et al.(2016)]{wasp122}
Turner, O. D., Anderson, D. R., Collier Cameron, A., et al. 2016, \pasp, 128, 064401

\bibitem[van Kerkwijk et al.(2010)]{vankerkwijk2010}
van Kerkwijk, M. H., Rappaport, S. A., Breton, R. P., et al. 2010, \apj, 715, 51 

\bibitem[West et al.(2016)]{wasp82}
West, R. G., Hellier, C., Almenara, J.-M., et al. 2016, \aap, 585, A126

\bibitem[Wilson et al.(2008)]{wasp4}
Wilson, D. M., Gillon, M., Hellier, C., et al. 2008, \apjl, 675, L113

\bibitem[Wong et al.(2020a)]{wong2019}
Wong, I., Benneke, B., Gao, P., et al. 2020a, \aj, 159, 234

\bibitem[Wong et al.(2020b)]{wong2019wasp19}
Wong, I., Benneke, B., Shporer, A., et al. 2020b, \aj, 159, 104

\bibitem[Wong et al.(2016)]{wong2016}
Wong, I., Knutson, H. A., Kataria, T., et al. 2016, \apj, 823, 122

\bibitem[Wong et al.(2020c)]{wong2019koi964}
Wong, I., Shporer, A., Becker, J. C., et al. 2020c, \aj, 159, 29

\bibitem[Wong et al.(2020d)]{wong2019kelt9}
Wong, I., Shporer, A., Kitzmann, D., et al. 2020d, \aj, 160, 88

\bibitem[Yee et al.(2020)]{yee2020}
Yee, S. W., Winn, J. N., Knutson, H. A., et al. 2020, \apjl, 888, L5

\bibitem[Zucker et al.(2007)]{zucker2007}
Zucker, S., Mazeh, T., \& Alexander, T. 2007, \apj, 670, 1326 

\end{thebibliography}
\end{document}